\pdfoutput=1
\documentclass[english,preprint,aps,prb,superscriptaddress,longbibliography]{revtex4-2}
\usepackage[T1]{fontenc}
\usepackage{textcomp}
\usepackage[latin9]{inputenc}
\usepackage{color}
\usepackage{placeins}
\usepackage{amsbsy}
\usepackage{amstext}
\usepackage{amssymb}
\usepackage{graphicx}

\makeatletter

\providecommand{\tabularnewline}{\\}

\makeatother

\usepackage{babel}
\begin{document}
\title{Tuning the optoelectronic properties of graphene quantum dots by BN-ring
doping: A density functional theory study}
\author{Samayita Das }
\email{samayitadas5@gmail.com}

\author{Alok Shukla}
\email{shukla@iitb.ac.in}

\affiliation{Department of Physics, Indian Institute of Technology Bombay, Powai,
Mumbai 400076, India}
\begin{abstract}
Graphene monolayer is a material with zero band gap, because of which
its applications in optoelectronics are limited. The question arises,
can we modify the optoelectronic properties of graphene by doping
it with other atoms? Synthesis of 2D monolayer of graphene doped with
heteroatoms such as boron and nitrogen, and a few computational studies
of their structural and electronic properties were previously reported.
In this work, we aim to answer this question for graphene quantum
dots (GQDs), i.e., how can we tune their optoelectronic performance
by replacing their carbon rings with (BN)\textsubscript{3} (borazine)
hexagonal rings? In this work, we have studied in detail the geometry,
electronic structure, and optical absorption spectra of fourteen different
borazine-ring doped diamond-shaped GQDs, using a computational methodology
based on the first-principles density functional theory (DFT). We
refer to the doped structures as BN-GQDs, which differ from each other
in the location, orientation and the number of borazine rings. We
computed their optical absorption spectra using the time-dependent
density-functional theory (TDDFT), and carefully examined: (a) for
the single-ring doped BN-GQDs the influence of the location of the
ring on their optical properties, and (b) for the double ring doped
systems, the influence of the location, mutual distance and orientation
of the rings on their absorption spectra. Frontier molecular orbitals
of all the structures considered here are studied in detail in order
to understand the nature of low-lying optical excitations. We also
analyzed the point-group symmetries of considered BN-GQDs, and performed
a group-theoretic analysis of the influence of their reduced symmetries,
on their optical properties. Our results indicate that\textcolor{blue}{{}
}BN-ring doping can achieve a significant control over the optical
properties of GQDs. The comparison of the optical absorption spectra
of the BN-GQDs with the parent GQD shows remarkable spectral broadening
with optical gap spanning over infrared to visible region of the spectrum.
Thus, systematic BN-ring doping provides easy tunability of the electronic
and optical properties of BN-GQDs to a considerable extent which is
very promising for their optoelectronic applications.
\end{abstract}
\maketitle

\section{Introduction }

Graphene quantum dots with hetero-atom doping have emerged as a new
class of $\pi$-conjugated molecules that are highly useful in the
field of optoelectronics \citep{ma2015general,zuo2017functionalization,yang2018exploring}.
Because of their high degree of tunability, structural flexibility,
low costs of synthesis and miniaturization, they can easily replace
the conventional Si-based technology that has several drawbacks. In
particular, the silicon based devices exhibit structural stiffness,
moderate tunability, low carrier mobility, and high production costs.
It is widely believed that these problems can be easily solved by
shifting to graphene-based materials that have high carrier mobility
\citep{graphene-synthesis-2004}, thermal and chemical stability \citep{galashev2014mechanical}
and most importantly electronic and optical tunability that can be
controlled by doping \citep{liu2011graphene,singh2013electrically}.
Although graphene has many exceptional properties, its zero band gap
\citep{graphene-synthesis-2004} makes it unsuitable for many optoelectronic
applications. Quantum confinement is one of the ways to overcome this
problem by working with reduced dimensional nanostructures such as
1D graphene nanoribbons (GNRs) \citep{cai2010atomically,jiao2010facile}
and 0D graphene quantum dots (GQDs) \citep{li2010colloidal,huang2011graphene,bacon2014graphene}.
GQDs exhibit higher band gap as compared to GNRs depending on their
size and shape \citep{kim2012anomalous}, which makes GQDs popular
for applications such as organic light emitting diodes \citep{geffroy2006organic,chen2018liquid,liu2018all},
solar cells \citep{facchetti2011pi,zhu2014efficiency}, and organic
field-effect transistors (OFETs) \citep{wang2012semiconducting,sirringhaus201425th,lussem2016doped}.
Quantum confinement is surely an efficient way of introducing band
gap in graphene, but controllable tuning of its electronic and optical
properties is very important for advanced optoelectronic applications.
Another effective way of controlling the band gap is through hetero-atom
doping, which normally reduces the symmetry of the system. For graphene,
various doping techniques have been explored in the past \citep{wehling2008molecular,liu2011chemical,kawai2015atomically};
however, a natural way to dope it is by substituting $sp^{2}$ hybridized
carbon atom pairs by isoelectronic BN units that also participate
in the planar $sp^{2}$ bonding, giving rise to stable $\pi$-conjugated
structures just like the parent material but with distinct electronic
and optical properties.

In the past, Xu \emph{et al}. \citep{xu2010density} performed large-scale
systematic first-principles density-functional theory (DFT) calculations
on graphene superlattices doped by BN units and studied the evolution
of their electronic and magnetic properties. They reported that the
band gap of the superlattice increases with the size of the BN nanodots
regardless of the shape of the dopants. On the other hand, Ajeel \emph{et
al}. \citep{bn-dimer-doped-gqds-2024} considered a coronene molecule
(C\textsubscript{24}H\textsubscript{12}) and co-doped it by single
B and N atoms (paired or separated) at various places, and studied
the evolution of its electronic and vibrational properties using the
first-principles DFT approach. They found that the HOMO-LUMO gap decreases
upon doping as compared to the pristine GQD. As far as experimental
works are concerned, BN-doped GQDs (BN-GQDs) were synthesized by Fei
\emph{et al.} \citep{fei2014boron} and were used as catalysts for
the oxygen reduction reaction. The authors concluded that BN doping
leads to a far superior catalytic performance. Li \emph{et al.} \citep{bn-codoped-gqds-for-bioimaging-synthesis-2017}
synthesized and studied the optical properties of BN-GQDs with the
aim of their applications in bio-imaging. They found that the BN-doped
GQDs emitted in the yellow with a high quantum yield. Yang \emph{et
al}. \citep{bn-doped-gqds-for-sensors-synthesis-2019} also synthesized
BN-GQDs, and studied their optical properties for their possible applications
as fluorescence sensors for the detection of Hg\textsuperscript{2+}and
F\textsuperscript{-} ions. They also reported high fluorescence quantum
yields and life times for the BN-doped GQDs. Budak \emph{et al.} \citep{BUDAK-bn-doped-dots-synthesis-2021}
synthesized BN-GQDs using a bottom-up synthesis approach, and reported
boron-regulated dual emission in their nanostructures with possible
applications in white light-emitting diodes (LEDs) and solar cells.
From a synthesis point of view, the field of BN-doped $\pi$-conjugated
molecules was comprehensively reviewed recently by Chen \emph{et al}.
\citep{multi-bn-doped-pi-conjugated-systems-review-2022}.

However, the carbon-based structures discussed above were either doped
with a bonded BN pair, or separated B and N atoms, not doped with
(BN)\textsubscript{3}-ring structures. The ring-based (BN)\textsubscript{3}-doped
graphene structures were successfully synthesized by S\'anchez-S\'anchez
\emph{et al.} \citep{doi:10.1021/acsnano.5b03895}, and Chen \emph{et
al}. \citep{doi:10.1021/acsami.9b10582}, while Herrera-Reinoza \textit{et
al.} \citep{herrera2021atomically} synthesized nanographenes doped
by h-BN clusters consisting of multiple (BN)\textsubscript{3} rings,
and reported the opening of a band gap in the range 1.4--1.6 eV for
a doping concentration of about 17\%. By means of a first-principles
DFT approach, Caputo\emph{ et al}. \citep{PhysRevMaterials.6.114001}
studied the influence of doping graphene with hexagonal (BN)\textsubscript{3}
rings on its electronic and structural properties. They found that
for all the considered cases, the structure remained planar, and demonstrated
significant variation in the band gap with the changing doping concentration
and pattern. Inspired by their work, we decided to explore the influence
of doping finite graphene fragments, i.e, GQDs by the (BN)\textsubscript{3}
rings with the aim of studying the evolution of their electronic and
optical properties. 

It is well known that doping a quantum dot introduces dopants as scattering
sites for the charge carriers. As the charge carriers move through
the material, the scattering events reduce their mean free path and
consequently, the mobility. The change in mobility, which has a significant
impact on the transport properties, can be controlled by different
doping strategies \citep{doi:10.1021/acs.jpclett.1c01284,doi:10.1021/acs.nanolett.5c02560,linh2025electronic}.
However, in this work we have mainly focused on tuning the optoelectronic
properties of the BN-GQDs in terms of systematic BN-ring doping. For
the purpose, we have employed a first-principles DFT-based methodology
in which the optical absorption spectra have been computed using the
TDDFT approach. We have chosen a diamond-shaped GQD (C\textsubscript{30}H\textsubscript{14})
of $D_{2h}$ symmetry as the host for the doping, and have demonstrated
a high degree of tunability in its electro-optical properties depending
on the number, location and orientation of the (BN)\textsubscript{3}
rings.

The remainder of the paper is organized as follows. In Sec. \ref{sec:Computational-details},
we briefly describe our computational methodology, followed by a detailed
presentation and discussion of our results in Sec. \ref{sec:Results-and-discussions}.
Finally, we present our conclusions in Sec. \ref{sec:Conclusions}.

\section{Computational details}

\label{sec:Computational-details}

We start our study with the geometry optimization of the considered
structures, for which we employed a first-principles all-electron
density functional theory (DFT) approach, along with the atom-centered
Cartesian Gaussian basis functions, as implemented in the Gaussian16
package \citep{g16}. Because of the all-electron nature of the calculations,
no pseudopotentials were employed in these calculations. As far as
the exchange-correlation functional is concerned, unless otherwise
specified, we used the B3LYP \citep{lee1988development,becke1992density,becke1993density}
hybrid functional both for the ground-state properties as well as
for the calculations of the optical absorption spectra employing the
time-dependent density functional theory (TDDFT) \citep{runge-gross-tddft-1984,marques2004time}
approach. These calculations were performed using valence double zeta
basis set 6-31++G(d,p) which includes both the polarization and diffuse
functions \citep{hehre1972self,hariharan1973influence}. The presence
of polarization function in basis set, ensures flexibility of chemical
bond formation, while the small-exponent diffuse functions help in
accurately predicting quantities such as the dipole moments. The convergence
criteria of $10^{-8}$ Hartree was set to solve the Kohn-Sham equations
\citep{kohn1965self} self consistently, while for the geometry optimization,
the threshold values of gradient forces between each constituent atoms,
average (RMS) force, maximum displacement and average (RMS) displacement
were set at 0.000450 Hartree/Bohr, 0.00030 Hartree/Bohr, 0.00180 Bohr,
and 0.00120 Bohr, respectively. The stability of the computed structures
was confirmed by performing vibrational frequency analysis. Based
on these optimized structures, the optical absorption spectra of various
BN-GQDs were calculated using the TDDFT approach, incorporating the
same exchange-correlation functional and the basis set in Gaussian16
software which employs an adiabatic frequency-space-based implementation
\citep{BAUERNSCHMITT1996454,scalmani2006geometries,furche2002adiabatic,VANCAILLIE1999249,VANCAILLIE2000159,stratmann1998efficient,casida1998molecular}.
Recently using the same functional and basis set we obtained excellent
agreement on optical absorption of some GQDs with the experimental
results \citep{samayita-jpca}. However, here for comparison purpose
we have calculated the optical absorption spectra using HSE06 functional
also and the results obtained are in excellent agreement with the
corresponding B3LYP ones, as shown in the Fig. S2 within the Supplemental
Material (SM) \citep{supplemental}. Here we have also presented results
of total density of states (TDOS) and partial density of states (PDOS)
of the BN-GQDs, and for that purpose we have used Multiwfn software
\citep{lu2012multiwfn}. 

\section{Results and discussion}

\label{sec:Results-and-discussions}

\subsection{Optimized structures and stability}

The dangling bonds of the edge atoms in all the structures considered
in this work were saturated with the hydrogen atoms. First we have
optimized the geometry of the diamond-shaped pristine GQD containing
thirty carbon atoms (C\textsubscript{30}H\textsubscript{14}), and
its optimized structure is presented in Fig. \ref{fig:Optimized-geometry-pristine}.
This GQD contains nine aromatic hexagonal rings, and it is a polycyclic
aromatic hydrocarbon named dibenzo{[}bc,kl{]}coronene of the D\textsubscript{2h}
symmetry. Next, we dope the pristine GQD by replacing the hexagonal
carbon rings (C\textsubscript{6}) with the corresponding (BN)\textsubscript{3}
rings, obtaining 14 distinct structures with different symmetries.
In all the cases, the optimized geometries were found to be planar
2D structures, i.e., no buckling was observed. Subsequently, we performed
the vibrational-frequency analysis on all the optimized structures
(pristine as well as doped), and no imaginary frequencies were found
indicating that the considered structures are dynamically stable.
For the pristine structure (dibenzo{[}bc,kl{]}coronene), the minimum
and maximum C$-$C bond lengths are 1.358 $\text{\AA}$ (C29$-$C30)
and 1.445 $\text{\AA}$ (C13$-$C14) respectively, the parentheses following
the bond lengths indicate the carbon atoms involved in the bonds,
according to the atom numbering scheme displayed in Fig. \ref{fig:Optimized-geometry-pristine}.
The edge C$-$H bond lengths ranged from 1.086 to 1.088 $\text{\AA}$,
while for the BN-GQDs in which the edge C atoms were substituted by
B/N atoms, the B-H bond lengths were close to 1.19 $\text{\AA}$, and
the N-H bond lengths were nearly 1.01 $\text{\AA}$ in all the cases.
For BN-GQDs, the optimized geometries have the values of C$-$C, C$-$H
bond lengths very close to the optimized pristine structure, and other
bond lengths (B$-$N, B$-$C, N$-$C) present in the doped structures
have average value close to the ideal value 1.4 $\text{\AA}$. All the
optimized bond lengths indicating their minima and maxima are presented
in Table S1 within the SM \citep{supplemental}. 
\begin{center}
\begin{figure}[!ht]
\begin{centering}
\includegraphics[scale=0.2]{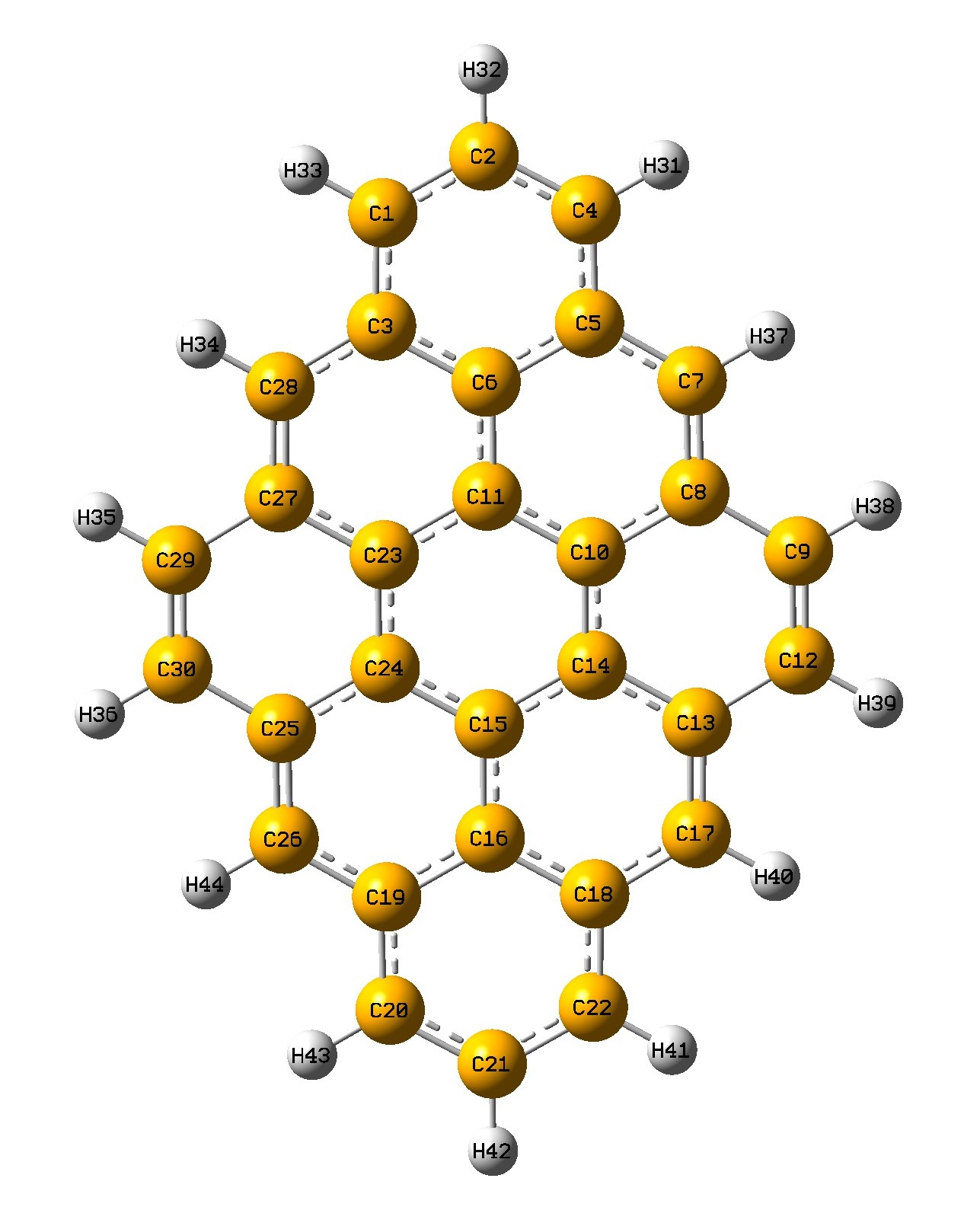}
\par\end{centering}
\caption{Optimized geometry of pristine diamond-shaped GQD (dibenzo{[}bc,kl{]}coronene,
C\protect\textsubscript{30}H\protect\textsubscript{14}).\protect\label{fig:Optimized-geometry-pristine}}
\end{figure}
\par\end{center}

As far as the pristine GQD is concerned, the optimized minimum and
maximum bond angles are 118.0$^\circ$ (C12-C13-C14) and 122.0$^\circ$ (C13-C17-C18),
respectively. The average bond lengths and angles in the pristine
and doped structures are close to the standard values of 1.4 $\text{\AA}$
and 120$^\circ$, respectively. For the doped structures, all the optimized
angles indicating their minima and maxima are shown in Table S2 within
the SM \citep{supplemental}. 

We have categorized the doped structures into three types based on
the number, location, and mutual orientation of the dopants (borazine
rings). The first category corresponds to doping with a single (BN)\textsubscript{3}
ring, for which four unique structures are possible depending on the
locations of the dopants as shown in Fig. \ref{fig:Different-(BN)ring-doped}(a).
The next category corresponds to doping with two fused (BN)\textsubscript{3}
rings, for which also four possibilities are there depending on their
locations as shown in Fig. \ref{fig:Different-(BN)ring-doped}(b).
Finally, we have substituted two separated benzene rings with two
(BN)\textsubscript{3} rings in the pristine GQD, which resulted in
three possible structures; furthermore, for each of these three structures
there are two possibilities depending on the mutual orientation of
the two rings. The structures in which the two rings have parallel
orientation are denoted by two up arrows $\uparrow\uparrow$, while
those with opposite orientation are indicated by up and down arrows
$\uparrow\downarrow$. Based on this characterization scheme, six
model structures are possible as depicted in Fig. \ref{fig:Different-(BN)ring-doped}(c).
Thus, in this work, we have considered a total of 14 different (BN)\textsubscript{3}-ring
doped diamond shaped graphene quantum dots. 
\begin{center}
\begin{figure}[!ht]
\begin{centering}
\includegraphics[scale=1.35]{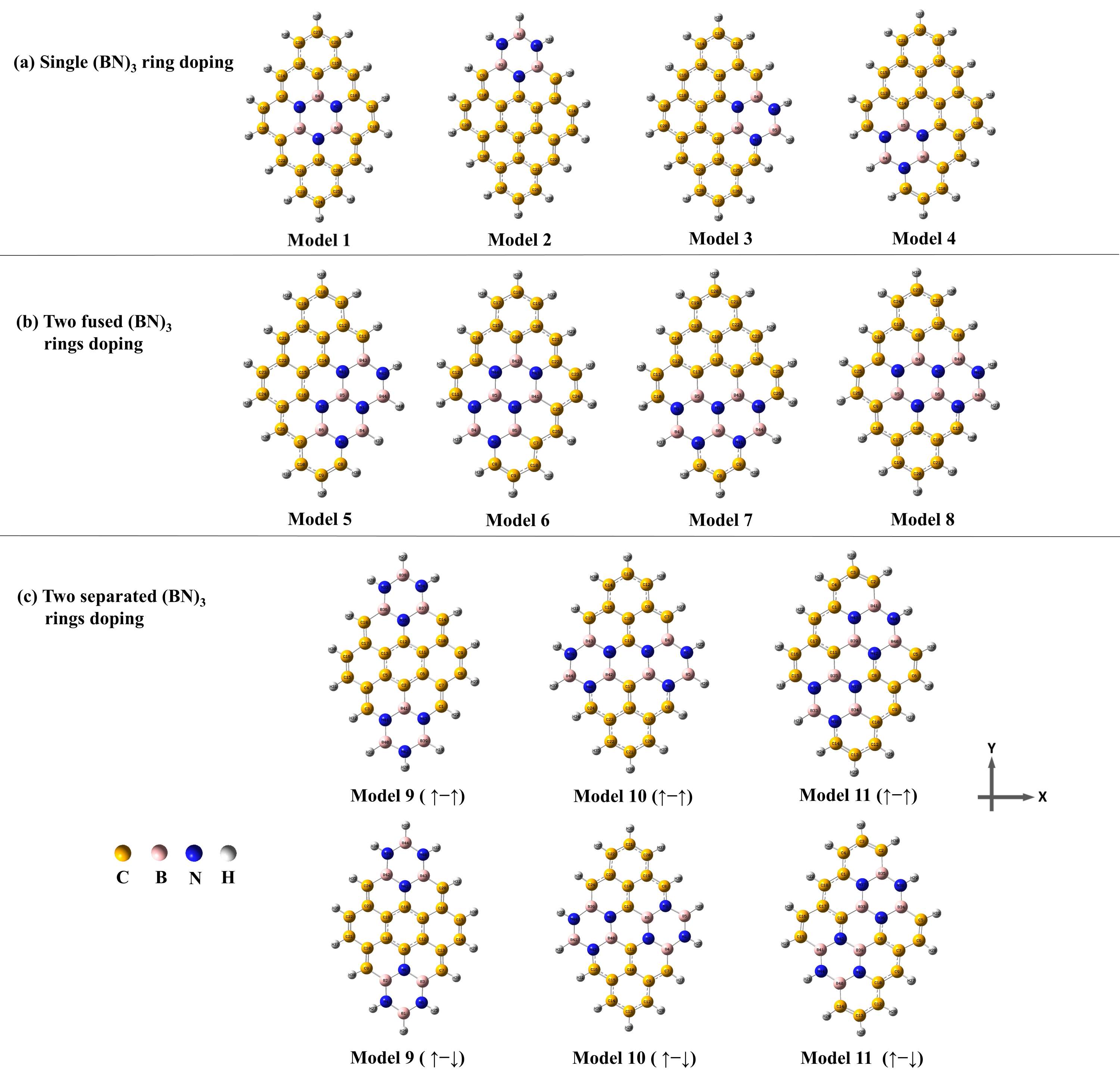}
\par\end{centering}
\caption{Diamond-shaped BN-GQDs with different doping schemes. In the last
category of two separated (BN)\protect\textsubscript{3} rings doped
structures (models 9-11), up-up and up-down arrows inside the parentheses
represent the parallel and antiparallel orientations of the two BN
rings with respect to each other.\protect\label{fig:Different-(BN)ring-doped}}
\end{figure}
\par\end{center}

Because of the hetero-atomic nature of the dopant rings, we wanted
to quantify the induced ionicity in the doped structures. Therefore,
we performed Mulliken charge analysis on all the BN-GQDs and found
that the N atoms acquire negative charges, while B atoms acquire positive
charges. The C atoms, on the other hand, exhibit both positive and
negative charges based on their proximity to B/N atoms. This behavior
can be understood quantitatively in terms of the electronegativities
of the atoms involved. Since the B atoms have the least electronegativity
(2.04) as compared to the N and C atoms, they will lose electrons
easily and acquire positive charges, while the N atoms have the highest
electronegativity (3.04) among the three, they easily attract electrons
and become negatively charged. But, the electronegativity of C atoms
(2.55) is in between of B and N atoms, so they exhibit both positive
and negative charges based on which atom they are bonded with. Electronegativity
of H atoms (2.20) is higher than B atoms but lower compared to the
C and N atoms, therefore, the H atoms connected to C and N atoms acquire
positive charges, while those connected with the B atoms have negative
charges. However, given the electrical neutrality of the considered
systems, the sum of all the induced Mulliken charges on every BN-GQD
is zero. 

\subsection{Cohesive energies}

Energetic stability of all the structures considered in this work
can be analyzed based on the cohesive energy (or binding energy) per
atom calculated using the formula, $E_{coh}=\frac{1}{N_{total}}\left(E_{total}-n_{C}E_{C}-n_{H}E_{H}-n_{B}E_{B}-n_{N}E_{N}\right)$,
where $N_{total}$, $n_{C}$, $n_{H}$, $n_{B}$, $n_{N}$ are, respectively,
the total number of atoms, and the numbers of $C$, $H$, $B$, and
$N$ atoms present in the BN-GQD structures. The calculated optimized
energies of the BN-GQDs are denoted by $E_{total}$, while $E_{C}$,
$E_{H}$, $E_{B}$, $E_{N}$ are, respectively, the calculated total
energies of the isolated $C$, $H$, $B$, $N$ atoms. The computed
cohesive energies are presented in Table \ref{tab:cohesive} from
which the following trends are obvious: (a) all the cohesive energies
are negative implying that all the structures are energetically stable;
(b) the BN-GQDs with the lowest magnitudes of cohesive energies ($\approx$
4.4 eV) correspond to models 5, 6, 7, and 8, each of which is doped
by two fused (BN)\textsubscript{3} rings; (c) the rest of the structures,
whether doped with a single (BN)\textsubscript{3} ring or with two
separated rings, have significantly larger cohesive energies in excess
of 6.5 eV/atom; and (d) comparatively speaking, the structures doped
by single ring have cohesive energies $\approx$ 0.2 eV higher than
the ones doped by two separated rings.
\begin{center}
\begin{table}[!ht]
\begin{centering}
\setlength{\tabcolsep}{10pt}
\begin{tabular}{cc}
\hline\hline 
Quantum dots & Cohesive energy/atom (eV)\tabularnewline
\hline
\hline
Pristine & $-$6.945\tabularnewline
Model 1 & $-$6.730\tabularnewline
Model 2 & $-$6.752\tabularnewline
Model 3 & $-$6.726\tabularnewline
Model 4 & $-$6.724\tabularnewline
Model 5 & $-$4.398\tabularnewline
Model 6 & $-$4.398\tabularnewline
Model 7 & $-$4.374\tabularnewline
Model 8 & $-$4.383\tabularnewline
Model 9 ($\uparrow-\uparrow$) & $-$6.548\tabularnewline
Model 9 ($\uparrow-\downarrow$) & $-$6.558\tabularnewline
Model 10 ($\uparrow-\uparrow$) & $-$6.504\tabularnewline
Model 10 ($\uparrow-\downarrow$) & $-$6.512\tabularnewline
Model 11 ($\uparrow-\uparrow$) & $-$6.514\tabularnewline
Model 11 ($\uparrow-\downarrow$) & $-$6.528\tabularnewline
\hline\hline
\end{tabular}
\par\end{centering}
\centering{}\caption{Cohesive energy per atom for all the structures considered in this
work. \protect\label{tab:cohesive}}
\end{table}
\par\end{center}

\subsection{Electronic structure}

\subsubsection{Frontier molecular orbitals}

\begin{figure}
\begin{centering}
\includegraphics[scale=0.5]{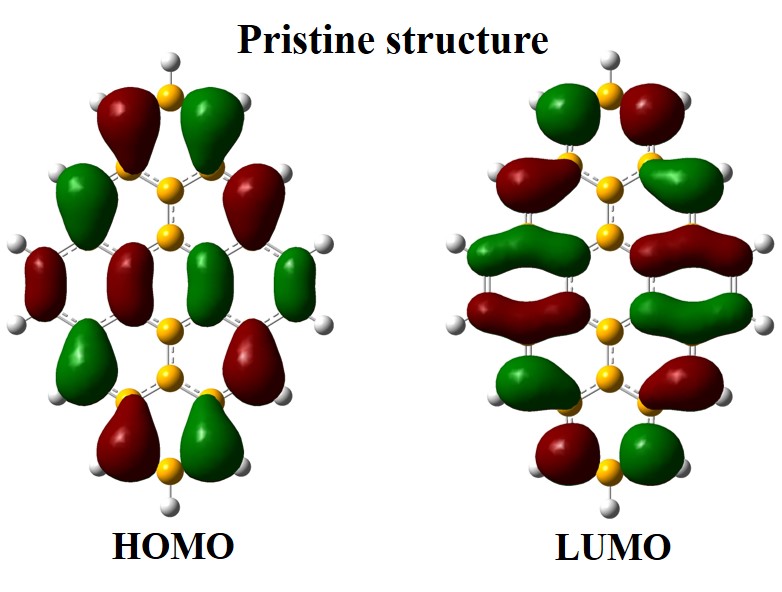}\caption{Plots of HOMO (H) and LUMO (L) orbitals of pristine diamond shaped
graphene quantum dot.\protect\label{fig:HL-GQD}}
\par\end{centering}
\end{figure}

\begin{figure}[!ht]

\centering{}\includegraphics[scale=0.41]{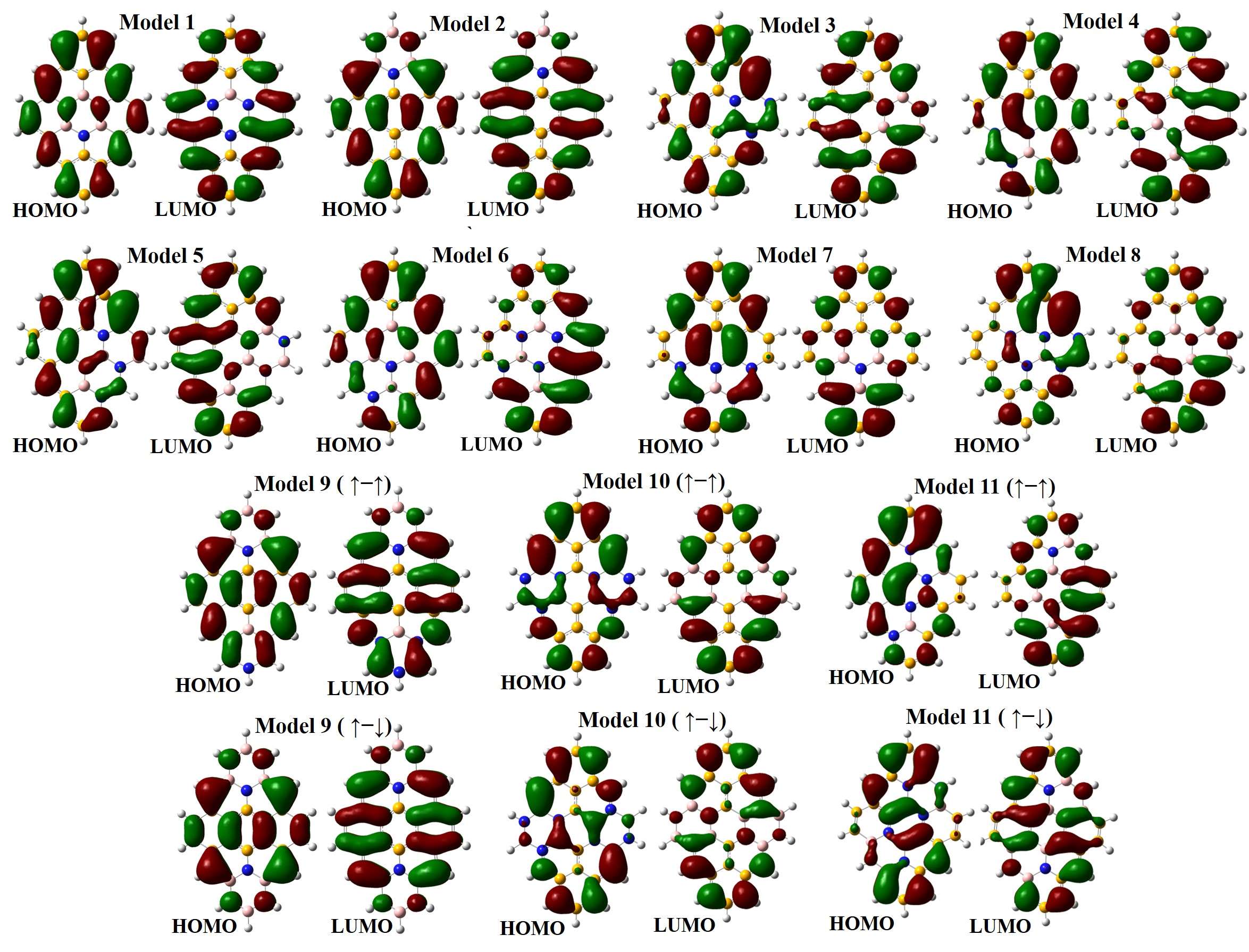}\caption{Plots of HOMO (H) and LUMO (L) orbitals of different BN-GQDs. The
doping model is written on top of each figure.\protect\label{fig:HL-BNGQD}}
\end{figure}

In order to understand the optical and electronic properties of the
BN-GQDs considered here, we need to understand the nature of their
frontier molecular orbitals (MOs), i.e., HOMO (highest occupied molecular
orbital) and LUMO (lowest unoccupied molecular orbital), and their
energies. The plots of these orbitals for the pristine GQD are presented
in Fig. \ref{fig:HL-GQD} while those for various doped structures
are shown in Fig. \ref{fig:HL-BNGQD}. It is obvious from the figures
that HOMO/LUMO orbital for both the pristine and doped structures
are of $\pi/\pi^{*}$ type, delocalized over C, B, and N atoms, but
with negligible charge densities on the passivating H atoms. Although,
we have not plotted the MOs further away from the Fermi level, there
are several $\pi/\pi^{*}$ type orbitals in the lower energy range
in spite of doping. This implies that in addition to C atoms, both
B and N atoms also contribute to the $\pi$ conjugation in a way similar
to polycyclic aromatic hydrocarbons (PAHs) which the parent pristine
molecule is. As a result, we expect the low-lying optical excitations
of these materials to be dominated by $\pi\rightarrow\pi^{*}$ transitions. 

\FloatBarrier

\subsubsection{Density of states}

In Fig. \ref{fig:dos} we present both the total density of states
(TDOS) along with the atom projected density of states (PDOS) for
the pristine GQD as well as a few doped BN-GQDs. We note the following
trends: (a) as expected, DOS is negligible in the region between the
HOMO-LUMO energies; (b) on the occupied side of the energy, the maximum
contribution is from carbon atoms; and (c) in the unoccupied region
of the orbital energies, in addition to carbon, other atoms also contribute
significantly away from the LUMO. However, for all the systems considered,
near the HOMO/LUMO orbital energies, the dominant contributions to
the DOS are from the carbon atoms. Corresponding orbital projected
DOS plots are presented in Fig. S1 within the SM \citep{supplemental}.
From the DOS plots it is again confirmed that the MOs near Fermi level
are of $\pi$ and $\pi^{*}$ type.
\begin{center}
\begin{figure}[!ht]
\begin{centering}
\includegraphics[scale=1.37]{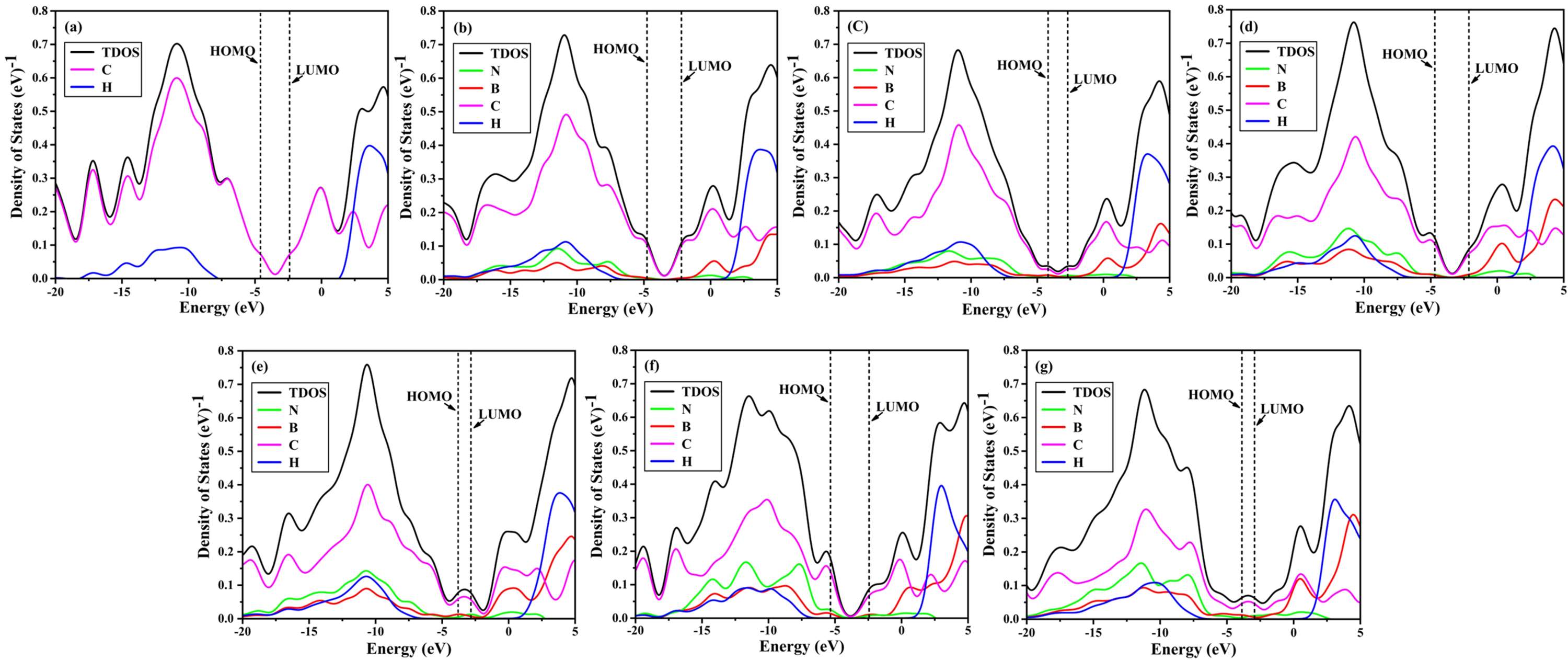}
\par\end{centering}
\caption{Total and atom projected DOS plots for: (a) pristine GQD, (b) single
BN-ring doping (model 1), (c) single BN-ring doping (model 3), (d)
fused double BN-ring doping (model 6), (e) fused double BN-ring doping
(model 7), (f) separated double BN-ring doping {[}model 9 ($\uparrow-\downarrow$){]},
and (g) separated double BN-ring doping {[}model 10 ($\uparrow-\uparrow$){]}.\protect\label{fig:dos}}
\end{figure}
\par\end{center}

\subsubsection{HOMO-LUMO Gaps}

In Table \ref{tab:H/L=000020energies}, we list the values of the
orbital energies of the H/L orbitals, along with the corresponding
H-L gaps.

From the table it is obvious that models 1, 2, 6 and 9 have larger
H-L gaps as compared to the pristine structure, while the rest of
the structures have lower values. This is quite a remarkable result
considering the fact that h-BN monolayer is a wide band gap insulator
while graphene is gapless. Therefore, one would think that doping
of GQDs with BN rings will widen the gap, however, as our results
suggest, it is happening in a small number of geometries, while for
most of the structures, the gap actually reduces on doping. To further
understand this, we examine the behavior of HOMO/LUMO orbital energies
on doping. We note that for the models 1, 2, 6 and 9, as compared
to the pristine case, the HOMO energies become lower, while the LUMO
energies, except for model 2 and model 9 ($\uparrow-\downarrow$),
move a bit higher. 
\begin{center}
\begin{table}[!ht]
\caption{The calculated B3LYP values of the orbital energies corresponding
to HOMO ($E_{HOMO}$), LUMO ($E_{LUMO}$), and the H-L gaps ($E_{HL}$)
for the pristine and BN-GQDs. \protect\label{tab:H/L=000020energies}}

\centering{}%
\setlength{\tabcolsep}{7pt}
\begin{tabular}{cccc}
\hline\hline
Quantum dots & $E_{HOMO}$ (eV) & $E_{LUMO}$ (eV) & $E_{HL}$ (eV)\tabularnewline
\hline
\hline
Pristine & $-$4.86 & $-$2.68 & 2.18\tabularnewline
Model 1 & $-$5.04 & $-$2.48 & 2.56\tabularnewline
Model 2 & $-$5.23 & $-$2.73 & 2.50\tabularnewline
Model 3 & $-$4.46 & $-$2.99 & 1.47\tabularnewline
Model 4 & $-$4.66 & $-$2.71 & 1.96\tabularnewline
Model 5 & $-$4.48 & $-$2.88 & 1.61\tabularnewline
Model 6 & $-$4.97 & $-$2.43 & 2.54\tabularnewline
Model 7 & $-$4.08 & $-$3.13 & 0.95\tabularnewline
Model 8 & $-$4.25 & $-$3.13 & 1.12\tabularnewline
Model 9 ($\uparrow-\uparrow$) & $-$5.08 & $-$2.23 & 2.86\tabularnewline
Model 9 ($\uparrow-\downarrow$) & $-$5.64 & $-$2.76 & 2.88\tabularnewline
Model 10 ($\uparrow-\uparrow$) & $-$4.17 & $-$3.23 & 0.94\tabularnewline
Model 10 ($\uparrow-\downarrow$) & $-$4.25 & $-$3.12 & 1.13\tabularnewline
Model 11 ($\uparrow-\uparrow$) & $-$4.46 & $-$2.95 & 1.51\tabularnewline
Model 11 ($\uparrow-\downarrow$) & $-$4.56 & $-$3.10 & 1.45\tabularnewline
\hline\hline
\end{tabular}
\end{table}
\par\end{center}
\FloatBarrier
For the remaining models, we note that the LUMO
energies move significantly lower as compared to the pristine case,
while the HOMO energies move up, leading to the reduced H/L gaps.
Recently, Caputo \emph{et al}. \citep{PhysRevMaterials.6.114001},
using a first-principles DFT-based approach, studied the influence
of doping graphene monolayer with single and double B$_{3}$N$_{3}$
rings on its electronic properties. For the purpose, they considered
diamond-shaped supercells similar to our pristine GQD, but of varying
sizes, and reported that in all considered cases, a finite band gap
opens. They reported GGA-PBE level band gaps of single BN ring doped
structures in the range 0.31-1.54 eV, for different doping concentrations;
while our results of B3LYP H-L gaps for the single BN ring doped GQDs
(models 1-4) are in the range 1.47-2.56 eV. Clearly, the higher H-L
gaps for BN-GQDs are from quantum confinement effects because of their
finite sizes. The noteworthy point is that when graphene is doped
with BN rings, the gap always increases compared to the parent system.
However, as discussed above, when the parent material is a GQD, the
H-L gap increases in a few cases, while it decreases in a majority
of the cases.

\subsubsection{Optical gaps}

Optical gap is defined as the lowest energy transition from ground
state to the first dipole allowed excited state corresponding to the
first peak in the optical absorption spectra. We have performed electron-correlated
calculations of the optical gaps and the absorption spectra of all
the QDs considered in this work using the TDDFT approach, with the
B3LYP and HSE06 functionals, as implemented in the Gaussian16 program
\citep{g16}. In the TDDFT method, the many-electron wave function
of a system is considered to be a linear combination of the Kohn-Sham
(KS) ground state, along with all possible electron-hole (e-h) type
singly-excited configurations obtained from the KS state, thereby
including the excitonic effects. This inclusion of the excitonic effects
leads to a change in the energy gaps as compared to the KS calculations,
which is evident in our calculations as well, in which a majority
of optical gaps listed in Table \ref{tab:Optical-gap-values} are
larger than the corresponding H-L gaps listed in Table \ref{tab:H/L=000020energies}.
However, for models 1, 6, 9 ($\uparrow-\uparrow$), and 9 ($\uparrow-\downarrow$),
we note slight reductions in the optical gaps as compared to the H-L
gaps.

From the TDDFT values of the optical gaps presented in Table \ref{tab:Optical-gap-values},
we see that there are insignificant quantitative differences in the
gaps computed using two functionals. Furthermore, we note that in
comparison to the pristine GQD, the optical gaps of the doped structures
follow the same trends as the H-L gaps, i.e., for models 1, 2, 6,
and 9, the optical gaps are larger than that of the pristine structure,
while for the rest of the structures they are smaller. This is because
the TDDFT wave functions of the excited states of the BN-GQDs corresponding
to their optical gaps are dominated by HOMO-LUMO single-particle transition.
Therefore, we conclude that even after accounting for the electron
correlations, the qualitative behavior of optical gaps remains the
same as of H-L gaps with respect to the doping configurations.

As far as the accuracy of our TDDFT calculations is concerned, we
note that for the pristine GQD (dibenzo{[}bc,kl{]}coronene, C\textsubscript{30}H\textsubscript{14})
experimental value of its optical gap was reported by Clar \textit{et
al.} \citep{CLAR19783219} to be 2.55 eV, which is $\approx$ 0.2
eV higher than our calculated values. Therefore, we expect similar
agreement for our calculated values of the optical gaps of BN-GQDs
with the experiments, as and when they are performed. 
\begin{table}[!ht]
\caption{The values of the optical gaps of pristine and BN-ring doped quantum dots calculated using the TDDFT approach, employing B3LYP and HSE06 functionals. The point-group symmetries of various BN-GQDs are also indicated. \label{tab:Optical-gap-values}}
\centering
\setlength{\tabcolsep}{5pt}
\begin{tabular}{lcccc}
\hline \hline
Types of QDs & Doping configurations & Symmetry & \multicolumn{2}{c}{TDDFT values of optical gaps (eV)} \\
\cline{4-5} 
 & & & \makebox[2.7cm] [c] {B3LYP} & \makebox[2.7cm] [c] {HSE06} \\
\hline 
Pristine & --- & $D_{2h}$ & 2.31 & 2.34 \\
\hline 
Doped with single & Model 1 & $C_{2v}$ & 2.52 & 2.54 \\
(BN)$_{3}$ ring & Model 2 & $C_{2v}$ & 2.53 & 2.57 \\
 & Model 3 & $C_{s}$ & 1.71 & 1.74 \\
 & Model 4 & $C_{s}$ & 2.09 & 2.13 \\
\hline 
Doped with two fused & Model 5 & $C_{s}$ & 1.77 & 1.79 \\
(BN)$_{3}$ rings & Model 6 & $C_{s}$ & 2.44 & 2.45 \\
 & Model 7 & $C_{2v}$ & 1.17 & 1.19 \\
 & Model 8 & $C_{s}$ & 1.23 & 1.24 \\
\hline 
Doped with two & Model 9 ($\uparrow-\uparrow$) & $C_{2v}$ & 2.75 & 2.78 \\
separated & Model 9 ($\uparrow-\downarrow$) & $D_{2h}$ & 2.79 & 2.79 \\
(BN)$_{3}$ rings & Model 10 ($\uparrow-\uparrow$) & $C_{2v}$ & 1.27 & 1.29 \\
 & Model 10 ($\uparrow-\downarrow$) & $C_{2h}$ & 1.46 & 1.49 \\
 & Model 11 ($\uparrow-\uparrow$) & $C_{s}$ & 1.59 & 1.61 \\
 & Model 11 ($\uparrow-\downarrow$) & $C_{2h}$ & 1.73 & 1.75 \\
\hline \hline
\end{tabular}
\end{table}

From Table \ref{tab:Optical-gap-values}, we can see that the changes
in the optical gaps, as compared to the pristine GQD, are sensitive
to the doping pattern. Next, we aim to arrive at a broad understanding
of this in terms of orbital localization (or delocalization) as compared
to the pristine GQD. From the spatial distribution of frontier molecular
orbitals (see Fig. \ref{fig:HL-GQD}) we can see that for the case
of the pristine GQD, HOMO and LUMO orbitals are delocalized with electron
density evenly spreading over the carbon framework. But doping with
BN ring (because of the differing electronegativities of B and N,
compared to C) disrupts the uniform electron distribution of the pristine
GQD, causing changes in the charge distributions of the frontier orbitals
(see Fig. \ref{fig:HL-BNGQD}), leading to significant changes in
the H-L energy gaps, and consequently in the optical gaps. Model 7
with fused double BN-rings exhibits the lowest value of the optical
gap. Its HOMO/LUMO orbital plots reveal that the doping significantly
delocalizes the HOMO charge distribution, thus raising its energy,
while the LUMO is highly localized around a few atoms resulting in
a lower energy. Thus, the H-L and optical gaps in this model get significantly
redshifted as compared to the pristine GQD. On the other hand, model
9 doped with two separated and oppositely oriented BN-rings {[}model
9 ($\uparrow-\downarrow$){]}, has the highest value of the optical
gap. In this unique configuration, doping stabilizes the HOMO by lowering
its energy and, as compared to the pristine GQD, localizing its charge
distribution. However, the charge distribution, and hence the energy
of the LUMO, remains relatively unchanged, leading to a wider H-L
gap and a blue-shift in the absorption spectrum. As far as other doped
models are concerned, we can use similar arguments to explain the
changes in their optical gaps as compared to the pristine GQD.

\subsection{Optical absorption spectra }

The optical gap contains information about the lowest optical excitation
(H-L transition in this case), but the extended systems such as the
ones considered here have similar excitations at higher energies as
well, involving transitions among orbitals further away from the Fermi
level. Therefore, the aim of this section is to study the absorption
spectra of the pristine and doped structures beyond their optical
gap, extending to higher energies. The spectra were computed using
the well-known sum-over-states (SOS) formula,
\begin{equation}
\sigma(\omega)=4\pi\alpha\sum_{i}\frac{\omega_{i0}|\left\langle i|\hat{e}.\boldsymbol{r}|0\right\rangle |^{2}\gamma^{2}}{(\omega_{i0}-\omega)^{2}+\gamma^{2}},\label{eq:opt-spec}
\end{equation}

where $|0\rangle$ ($|i\rangle$) denotes the ground (excited) state
TDDFT wave function, $\hbar\omega_{i0}=E_{i}-E_{0}$ is the energy
difference between the two states, $\left\langle i|\hat{e}.\boldsymbol{r}|0\right\rangle $
represents the corresponding transition dipole matrix element for
an incident photon of energy $\hbar\omega$ polarized along the $\hat{e}$
direction, $\alpha$ is the fine-structure constant, and $\gamma$
is the uniform line width assumed for all excited states. For each
QD, we considered 30 excited states in the SOS formula, and confirmed
the convergence in a few cases by performing calculations with 40
excited states (see Fig. S3 within the SM \citep{supplemental}).
Additionally, we also performed calculations of the absorption spectra
using the HSE06 functional for the pristine case as well as a few
doped cases, whose results are presented in Fig. S2 within the SM
\citep{supplemental}. We note that the spectra computed using the
two functionals are qualitatively similar with the HSE06 spectrum
being slightly blue-shifted with respect to the B3LYP one. Additionally,
in Tables S3--S17 within the SM \citep{supplemental}, we present
detailed information regarding the TDDFT wave functions of the excited
states contributing to various peaks of the optical absorption spectra
computed using the B3LYP functional for all the QDs considered in
this work. 

\subsubsection{Point-group symmetries}

The point-group symmetries of the QDs considered in this work are
$D_{2h}$, $C_{2h}$, $C_{2v}$, and $C_{s}$ (see Table \ref{tab:Optical-gap-values}),
which we briefly discuss below.

The point-group symmetry of the pristine GQD (Fig. \ref{fig:Optimized-geometry-pristine})
is $D_{2h}$, and its closed-shell singlet ground state belongs to
the irreducible representation (irrep) $A_{g}$, while the dipole-allowed
excited states belong to irreps $B_{1u}$ ($z$ polarized), $B_{2u}$
($y$ polarized), and $B_{3u}$ ($x$ polarized), respectively. Out
of these, the excited states of $B_{1u}$ irrep occur at much higher
energy as compared to those of $B_{2u}$ and $B_{3u}$; therefore,
we have ignored them in the present work. We note that with our choice
of axes, long axis is along the $y$ axis, while the short axis is
along the $x$ axis.

The BN-GQDs in which two separated BN rings are placed in such a manner
that the inversion symmetry is obeyed, the point-group symmetry is
$C_{2h}$. For these structures, the closed-shell ground state is
of irrep $A_{g}$, while the dipole allowed excited states have the
symmetries $B_{u}$, with the mixed $xy$ polarization, and $A_{u}$
with polarization along the $z$ direction. As the $z$-polarized
excitations occur at higher energies, we have only considered $B_{u}$
symmetry excited states in the linear optical absorption spectra.

For those BN-GQDs in which the BN ring (rings) is (are) placed symmetrically,
the point group is $C_{2v}$. The closed-shell ground state for such
systems is of irrep $A_{1}$, and because of the lack of inversion
symmetry, $x$-polarized optical excitations are dipole allowed from
the ground state to the excited states of irrep $A_{1}$. The optical
transitions to $B_{1}$-type excited states are $\text{z}$-polarized
(out of plane), while those to $B_{2}$-type exited states are $y$
polarized. Again, $z$-polarized optical transitions occur at much
higher energies, so we have not probed $B_{1}$-type excited states
in this work.

$C_{s}$ is the point-group symmetry of BN-GQDs in which the BN ring
(rings) is (are) implanted in an asymmetric manner, with the mirror
plane in the plane of the BN-GQD. This group has irreps $A^{'}$ and
$A^{''}$ , with the closed-shell ground state being of $A^{'}$ symmetry.
The dipole transitions from the ground state to the excited states
of the same $A^{'}$ symmetry have mixed $xy$ polarization in the
plane of the BN-GQDs, while those to the excited states of $A^{''}$
symmetry are $z$-polarized. Similar to the previous cases, the $z$-polarized
transitions occur at much higher energies, and, therefore, are not
studied in this work. 

\subsubsection{Pristine GQD (dibenzo{[}bc,kl{]}coronene)}

In Fig. \ref{fig:Optical-absorption-spectrum_pristine}, we present
the linear optical absorption spectrum of the pristine GQD (C$_{30}$H$_{14}$),
in which we indicate the polarization direction of the excited state
contributing to each peak as a subscript of the peak label. We note
the following trends: (a) There are four major peaks in the energy
region (0.0--5.0 eV) explored in our calculations. (b) Peak I of
the spectrum corresponding to the optical gap is long-axis polarized
($B_{2u}$ irrep), and is located at 2.31 eV. The many-body wave function
of the excited state contributing to this peak is dominated by HOMO
($H$) to LUMO ($L$) single excitation $|H\rightarrow L\rangle$.
(c) Peak II of the spectrum at 3.46 eV is weaker in intensity compared
to I, and is $x$-polarized ($B_{3u})$, with the excited state wave
function dominated by the single excitation $|H-2\rightarrow L\rangle$.
(d) The third peak at 4.08 eV has intensity similar to that of peak
II, however, it is $y$-polarized. Its excited state wave function
is dominated by the singly-excited configuration $|H-1\rightarrow L+1\rangle$.
(e) Peak IV located near 4.91 eV is the most intense (MI) peak of
the spectrum, and is again $y$-polarized. It is due to two closely
spaced excited states both of whose many-body wave functions are dominated
by the single excitation $|H-2\rightarrow L+2\rangle$. We note that
the broad intensity profile, peak locations, and polarizations computed
in this work are in excellent agreement with results of our earlier
calculations on the same molecule (dibenzo{[}bc,kl{]}coronene), performed
using the Pariser-Parr-Pople (PPP) model Hamiltonian \citep{PhysRevB.92.205404}.
Detailed information about the excited states contributing to the
peaks are presented in Table S3 within the SM \citep{supplemental}.
\begin{center}
\begin{figure}[!ht]
\begin{centering}
\includegraphics[scale=0.5]{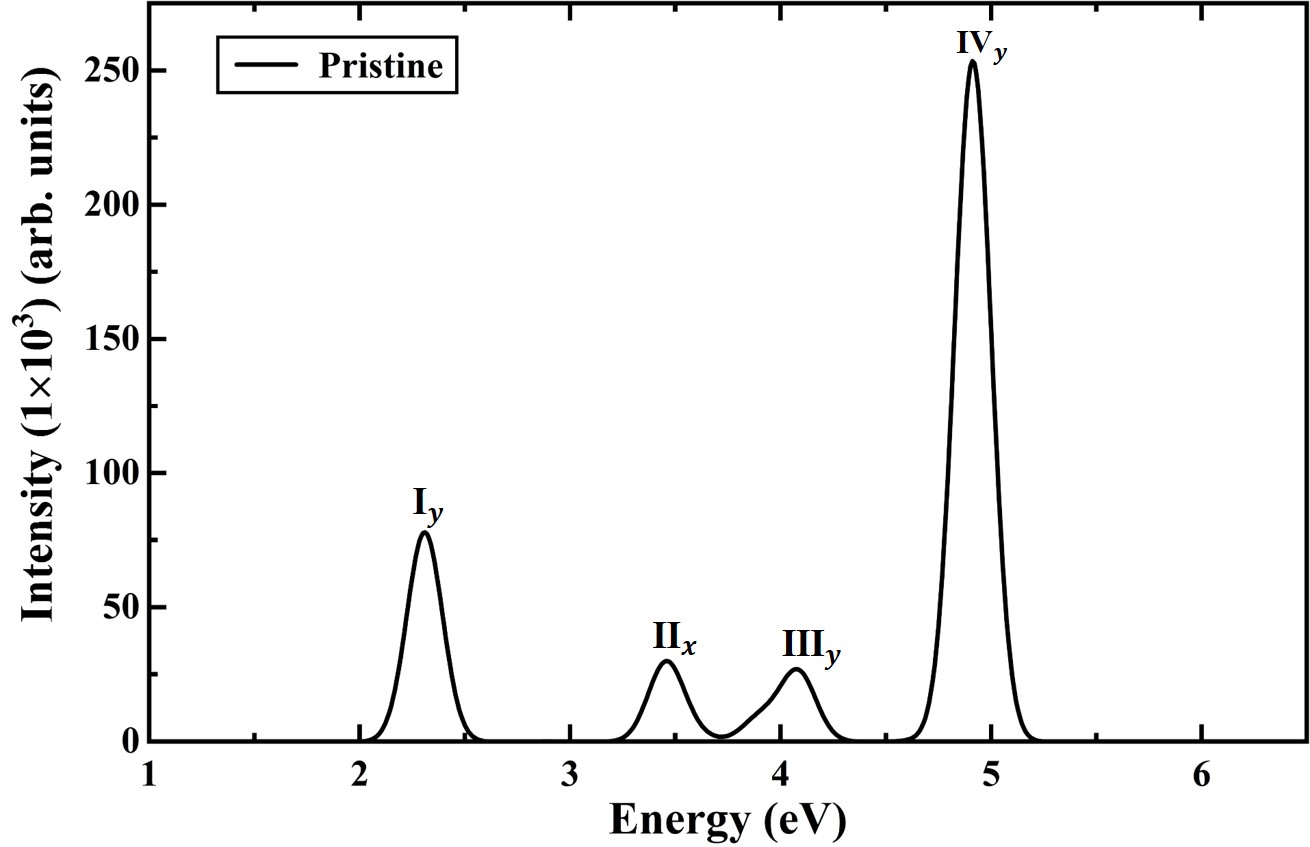}
\par\end{centering}
\caption{Optical absorption spectrum of pristine diamond shaped GQD, computed
using the TDDFT approach, and B3LYP functional. The subscript of each
peak label denotes polarization direction of the photon involved.
\protect\label{fig:Optical-absorption-spectrum_pristine}}
\end{figure}
\par\end{center}
\FloatBarrier

Next, we present and discuss the optical absorption spectra of various
BN-GQDs to understand the influence of BN-ring doping on the optical
response of the GQD. 

\subsubsection{Doping with single BN ring }

In this category four different structures (models 1--4) are possible
based on the BN-ring positions as shown in figure \ref{fig:Different-(BN)ring-doped}(a).
These model BN-GQDs are isomers of each other with common chemical
formula C\textsubscript{24}B\textsubscript{3}N\textsubscript{3}H\textsubscript{14}
of which the first two BN-GQDs (model 1, model 2) belong to the $C_{2v}$
point group, while the other two (model 3, model 4) are of the $C_{s}$
symmetry. The comparison between UV-Vis absorption spectra of these
molecules with that of the pristine GQD, obtained using the TDDFT
approach employing B3LYP functional is presented in Fig. S4 within
the SM \citep{supplemental}, while in Fig. \ref{fig:peaklabel_1_4}
we present the absorption spectra of each BN-GQD, separately. Next,
we briefly discuss the salient features of the absorption spectra
of each of the four models. The detailed information relevant to the
excited states of these BN-GQDs is presented in Tables S4--S7 within
the SM \citep{supplemental}.

\subsubsection*{Model 1: }

In this model, the BN ring is placed right at the center of the GQD
{[}see Fig. \ref{fig:Different-(BN)ring-doped}(a){]}, thereby reducing
its symmetry to $C_{2v}$. The absorption spectrum of this BN-GQD
is presented in Fig. \ref{fig:peaklabel_1_4}(a), from which it is
obvious that it has changed quite significantly as compared to that
of the parent GQD. We note the following trends based on this figure
and the information presented in Table S4 within the SM \citep{supplemental}.
The location of peak I (the optical gap) gets blueshifted by about
0.2 eV compared to the pristine GQD. The MI peak is now peak II that
is still $y$ polarized similar to the parent GQD; however, it is
comparatively redshifted by 1.4 eV. The wave function of this peak
is dominated by the single excitation $|H-1\rightarrow L+1\rangle$,
similar to peak III of the parent GQD. The first intense $x$-polarized
peak is now peak V located at 4.56 eV which is about 1.1 eV blueshifted
as compared to the corresponding $x$-polarized peak (II) of the pristine
GQD. The single excitation $|H-1\rightarrow L+3\rangle$ makes dominant
contribution to this peak of the BN-GQD, while peak II of the pristine
GQD is mainly from the excitation $\arrowvert H-2\rightarrow L\rangle$. 

\subsubsection*{Model 2: }

With the BN ring placed at the top corner of the GQD {[}see Fig. \ref{fig:Different-(BN)ring-doped}(a){]},
this BN-GQD is also $C_{2v}$ symmetric. As compared to the absorption
spectrum of the pristine GQD, model 2 exhibits a few extra peaks {[}see
Fig. \ref{fig:peaklabel_1_4}(b){]}. As far as the first peak (peak
I) is concerned, it is still $y$-polarized, but about 0.2 eV blueshifted
as compared to the parent GQD, similar to model 1. Peak VI is now
the MI peak which, however, has the same polarization ($y$) as the
MI peak of the pristine GQD, and it is also blueshifted by a very
small amount (0.05 eV) compared to the pristine one. Furthermore,
similar to the parent GQD, the wave function of the MI peak in model
2 is dominated by the excitation $\arrowvert H-2\rightarrow L+2\rangle$.
Compared to the pristine GQD which has a single $x$-polarized peak
(II), in this model we obtain two significant $x$-polarized peaks
(II and III), the first of which is redshifted compared to parent
molecule while the second one (III) is blueshifted. Therefore, we
conclude that the peaks II and III are a consequence of the splitting
of peak II of the parent GQD because of the symmetry breaking. Along
with x polarized transition, peak III also has contribution from a
closely spaced y polarized transition. Because of the change in symmetry,
different configurations contribute to these $x$-polarized excited
states in model 2 (see Table S5 within the SM \citep{supplemental}),
as compared to the one in the pristine GQD.

\subsubsection*{Model 3:}

In this model, the BN ring is placed at the right corner {[}see Fig.
\ref{fig:Different-(BN)ring-doped}(a){]} of the GQD; therefore, the
resultant symmetry $C_{s}$ is even lower than $C_{2v}$, leading
to the mixed $xy$ polarization of the transition dipole moments.
When we compare the absorption spectrum of this model {[}see Fig.
\ref{fig:peaklabel_1_4}(c){]}, we note a general redshift of the
spectral features as compared to the pristine GQD. The location of
the first absorption peak is at 1.71 eV, which amounts to a redshift
of 0.6 eV as compared to the parent molecule, although the wave functions
contributing to the peak of both the molecules are still dominated
by the $\arrowvert H\rightarrow L\rangle$ transition. The MI peak
(peak VII) of model 3 is about 0.3 eV redshifted as compared to that
of the pristine GQD, while its wave function is dominated by the excitation
$\arrowvert H-1\rightarrow L+3\rangle.$ In addition to a couple of
shoulders, there are three more (II, III, V) well-resolved peaks in
the spectrum within the energy range 2.3--3.9 eV, owing to various
other orbital excitations whose details are presented in Table S6
within the SM \citep{supplemental}.

\subsubsection*{Model 4:}

\begin{figure}[!ht]
\begin{centering}
\includegraphics[width=14cm,height=10cm]{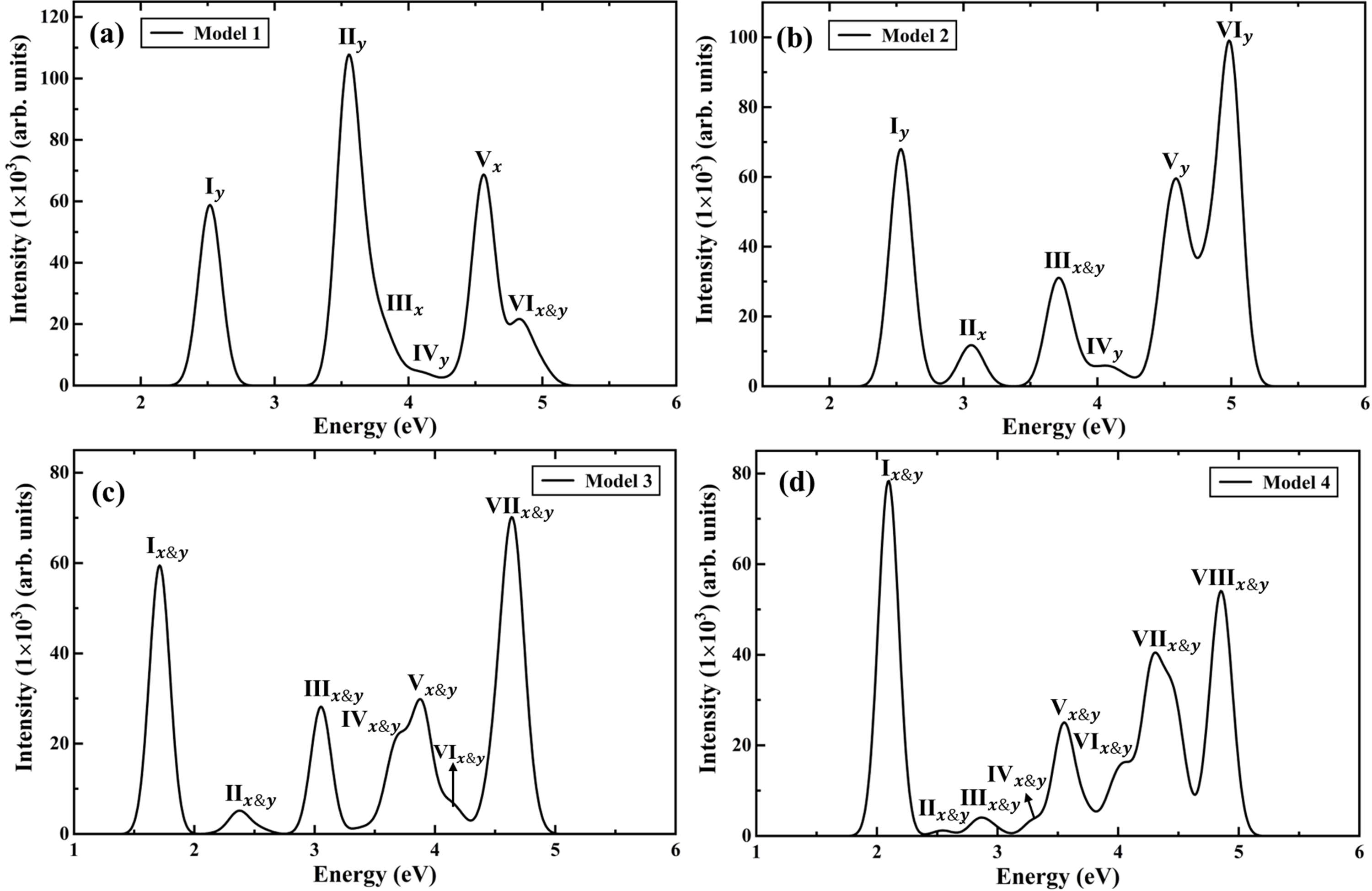}
\par\end{centering}
\caption{Linear optical absorption spectra for (a) Model 1, (b) Model 2, (c)
Model 3, and (d) Model 4, computed at the TDDFT-B3LYP level of theory,
assuming a uniform line width of 0.1 eV.\protect\label{fig:peaklabel_1_4}}
\end{figure}

The BN-GQD of model 4 is obtained by placing the BN ring at one of
the edges, which we chose to be the left-lower edge {[}Fig. \ref{fig:Different-(BN)ring-doped}(a){]}.
The symmetry of this BN-GQD is also $C_{s}$; however, its absorption
spectrum exhibits the most pronounced redistribution of the oscillator
strength as compared to the pristine GQD, with the first peak becoming
the MI peak. This peak (I) is owing to a state located at 2.09 eV
which is about 0.2 eV redshifted compared to the optical gap of the
pristine GQD, with the wave function dominated by the $\arrowvert H\rightarrow L\rangle$
transition. Because the wave functions of peak I in model 4 is identical
to that in the pristine GQD, this implies that it became the MI peak
by redistribution of the oscillator strength and not due to the redshift
of MI peak of the parent GQD. The other intense peaks {[}Fig. \ref{fig:peaklabel_1_4}(d){]}
in the spectrum of this model are V, VII, and VIII located at 3.56
eV, 4.28 eV, and 4.85 eV, owing to the excited states described by
the transitions $\arrowvert H-2\rightarrow L\rangle$, $\arrowvert H-1\rightarrow L+2\rangle$,
and $\arrowvert H-3\rightarrow L+1\rangle$, respectively. The detailed
information about various excited states in the spectrum of model
4 is presented in Table S7 within the SM \citep{supplemental}.

\subsubsection{Doping with two fused BN rings }

In Fig. \ref{fig:Different-(BN)ring-doped}(b), we depict four models
(models 5--8) of BN-GQDs obtained by doping the parent GQD with two
fused BN rings which are isomers with chemical formula C\textsubscript{20}B\textsubscript{5}N\textsubscript{5}H\textsubscript{14}.
From the figure it is obvious that only model 7 has the $C_{2v}$
symmetry, while the rest of them belong to the $C_{s}$ point group.
In Fig. S5 within the SM \citep{supplemental}, we compare the absorption
spectra of these molecules with that of the pristine GQD, obtained
using the TDDFT approach, and the B3LYP functional, while in Fig.
\ref{fig:peaklabel_5_8} we present the spectra of each BN-GQD, separately.
The detailed information about the excited states contributing to
the spectra of this class of BN-GQDs can be found in Tables S8--S11
within the SM \citep{supplemental}. Next, we briefly discuss the
main characteristics of the absorption spectra of each of the four
models. 

\subsubsection*{Model 5:}

As compared to the pristine GQD, the first peak of the spectrum of
this model at 1.77 eV {[}Fig. \ref{fig:peaklabel_5_8}(a){]}, exhibits
a red shift of about 0.5 eV. Additionally, this peak is also the MI
peak of this model, even though the excited state contributing to
it is still dominated by $\arrowvert H\rightarrow L\rangle$ excitation.
This suggests a major redistribution of the oscillator strength similar
to model 4. Additionally, this BN-GQD exhibits six more well-resolved
peaks in the energy range 2.40--4.75 eV. From the peak locations
of this BN-GQD as listed in Table S8 within the SM \citep{supplemental},
it is obvious that as compared to the parent GQD, the entire spectrum
is red shifted. The second most intense peak {[}peak VIII of Fig.
\ref{fig:peaklabel_5_8}(a){]} of this model is located at 4.75 eV
and is owing to two closely spaced excited states whose wave functions
are dominated by excitations $\arrowvert H-6\rightarrow L\rangle$
and $\arrowvert H-1\rightarrow L+4\rangle$.

\subsubsection*{Model 6:}

As compared to the pristine GQD, in this model the first peak {[}peak
I of Fig. \ref{fig:peaklabel_5_8}(b){]} undergoes a slight blue shift
of 0.1 eV, however, the MI peak (peak IV) gets red shifted by about
1.1 eV to the position 3.78 eV, and is dominated by the excitation
$\arrowvert H-1\rightarrow L+1\rangle$ (see Table S9 within the SM
\citep{supplemental}). We note that for the pristine GQD, peak III
located near 4.1 eV is also dominated by the $\arrowvert H-1\rightarrow L+1\rangle$
excitation, even though it is not the MI peak. This indicates redistribution
of the oscillator strength in this BN-GQD caused by doping.

\subsubsection*{Model 7:}

Even though this model is $C_{2v}$ symmetric, as compared to the
pristine GQD, it exhibits a significant redshift of about 1.1 eV,
from 2.31 eV to 1.17 eV, in the location of the first peak, amounting
to about 50\% reduction. Despite this, the $y$-polarization {[}Fig.
\ref{fig:peaklabel_5_8}(c){]} of the peak remains unchanged as also
the nature of the excited state with the dominant $\arrowvert H\rightarrow L\rangle$
excitation (see Table S10 within the SM \citep{supplemental}). The
MI peak (peak VII) in this model is also $y$-polarized similar to
the MI peak of the pristine GQD, although it occurs at 4.32 eV, which
is comparatively 0.6 eV red shifted. Additionally, there are several
well-resolved peaks in the energy region between the optical gap and
the MI peak.

\subsubsection*{Model 8:}

This $C_{s}$-symmetric model is obtained by replacing the two carbon
hexagons in the central row of the parent GQD by two fused BN rings,
leading to a very prominent red shift in the absorption spectrum {[}see
Fig. \ref{fig:peaklabel_5_8}(d){]}. The first peak (peak I) of this
BN-GQD is located at 1.23 eV which is also about 1.1 eV lower than
the first peak of the parent GQD at 2.31 eV. However, the relative
intensity of peak I, as compared to the first peaks of other three
fused double-ring doped models, is significantly lower. Peak III located
at 2.93 eV is the MI peak of this model, and as compared to the pristine
GQD it is red shifted by almost about 2 eV. The transition contributing
most to this peak is $|H-1\rightarrow L+1\rangle$, which is similar
to peak III of the pristine GQD located at 4.08 eV. Additionally,
there are several smaller peaks in the computed spectrum of this GQD
that are quite well resolved. The detailed information about various
excited states in the spectrum of this model is presented in Table
S11 within the SM \citep{supplemental}.
\begin{center}
\begin{figure}[!ht]
\begin{centering}
\includegraphics[scale=1.14]{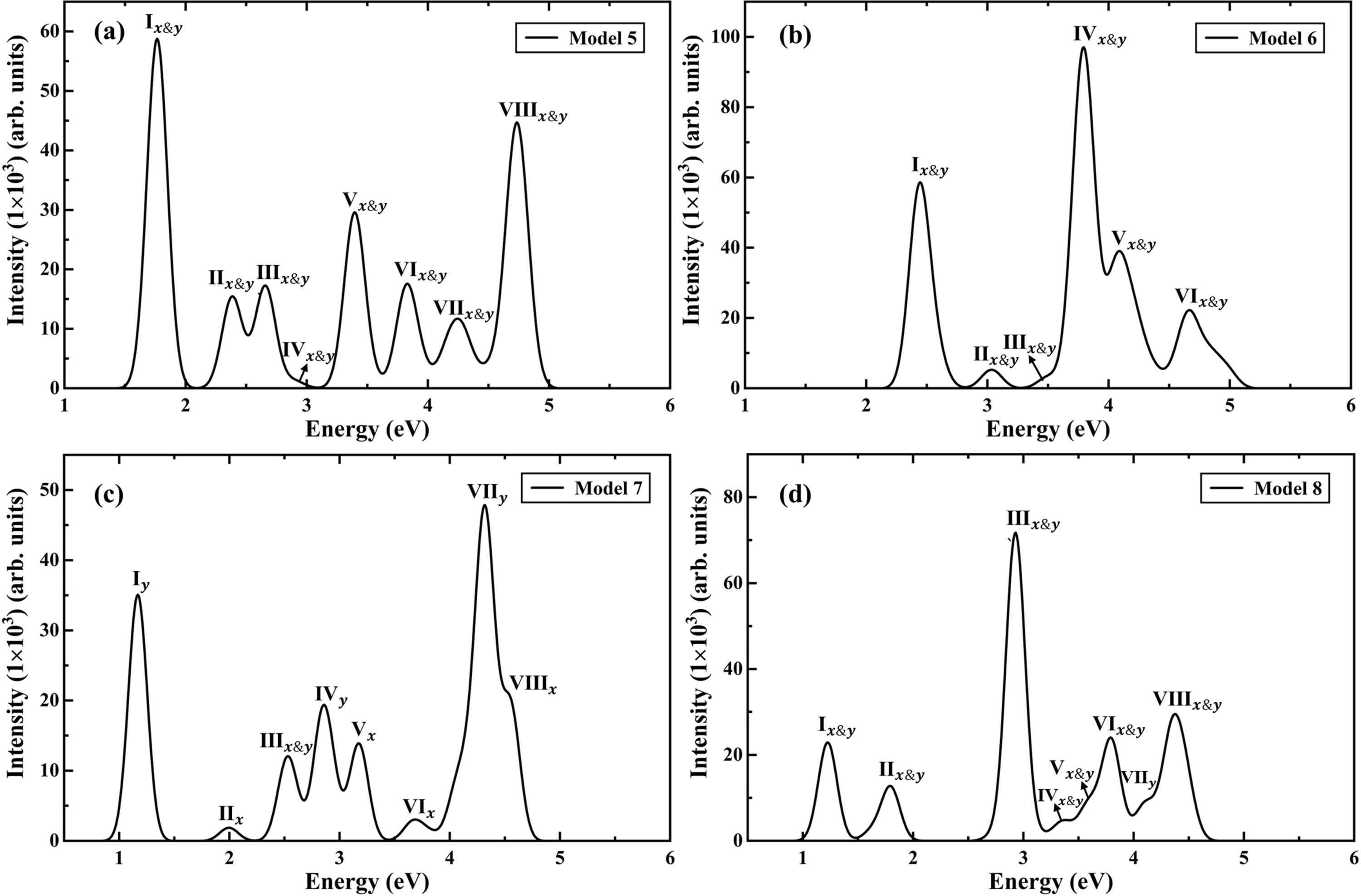}
\par\end{centering}
\caption{Linear optical absorption spectra for (a) Model 5, (b) Model 6, (c)
Model 7, and (d) Model 8, computed at the TDDFT-B3LYP level of theory,
assuming a uniform line width of 0.1 eV.\protect\label{fig:peaklabel_5_8}}

\end{figure}
\par\end{center}

\subsubsection{Doping with two separated BN rings }

BN-GQDs of this category are obtained by replacing two separated carbon
hexagons in the pristine GQD by two B$_{3}$N$_{3}$ hexagons as shown
in Fig. \ref{fig:Different-(BN)ring-doped}(c). As a result, all the
BN-GQDs in this class are isomeric with the chemical formula C\textsubscript{18}B\textsubscript{6}N\textsubscript{6}H\textsubscript{14}.
As the two rings are separated, we have more freedom in this case
as compared to the previous one because we can orient the two hexagons
in a parallel ($\uparrow-\uparrow$) or antiparallel ($\uparrow-\downarrow$)
manner. We have considered three possible arrangements of the separated
BN rings; however, combined with their relative orientations, we get
twice the number, i.e., six BN-GQDs from them. As is shown in Table
\ref{tab:Optical-gap-values}, these six models have symmetries ranging
from high ($D_{2h}$) to low ($C_{s}$). The UV-Vis spectra for all
the models of this class are compared to that of the pristine GQD
in Fig. S6 within the SM \citep{supplemental}, while for the individual
BN-GQDs they are presented in Figs. \ref{fig:peaklabel_9_11}(a)--(f).
Furthermore, in Tables S12--S17 within the SM \citep{supplemental}
all the details regarding the excited states contributing to the spectra
of each BN-GQD of this class are presented. Next, we briefly discuss
the main features of their absorption spectra.

\subsubsection*{Model 9:}

This model is obtained by substituting BN rings on the top and bottom
corners of the pristine GQD, and when we choose the parallel orientation
the point group of the BN-GQD is $C_{2v}$, while for the antiparallel
orientation it is $D_{2h}$. We note that for both the orientations
the first peak is $y$ polarized {[}see Figs. \ref{fig:peaklabel_9_11}(a)
and (d){]} and blue shifted with respect to the first peak of the
parent GQD by more than 0.4 eV. For the parallel orientation {[}Fig.
\ref{fig:peaklabel_9_11}(a){]}, there are two almost equal intensity
$y$-polarized peaks, III (4.0 eV) and V (4.8 eV) which qualify as
the MI peaks. We note that as compared to the MI peak of the pristine
GQD at 4.9 eV, peak III is red shifted, while peak V is almost unchanged.
In this case also, as compared to the parent GQD, there is a significant
transfer of oscillator strength from peak IV to peak III, because
both the polarization and orbital excitations contributing to the
peaks III ($\arrowvert H-1\rightarrow L+1\rangle$) and V ($\arrowvert H-2\rightarrow L+2\rangle$)
of this BN-GQD are the same as for the peaks III and IV of pristine
GQD. For the antiparallel orientation {[}Fig. \ref{fig:peaklabel_9_11}(d){]},
there is a single $y$-polarized MI peak (II) located at 4.06 eV which
is about 0.8 eV red shifted with respect to that in the pristine GQD.
The detailed information about the excited states for the two BN-GQDs
of this class is presented in Tables S12 and S13 within the SM \citep{supplemental}.

\subsubsection*{Model 10:}

To obtain this model, one substitutes the two separated BN rings in
the middle row of the parent GQD, and for the parallel orientation
the point group of the BN-GQD is $C_{2v}$, while for the antiparallel
orientation it is $C_{2h}$. The spectra of both the BN-GQDs of this
class are significantly red shifted as compared to the parent GQD,
as it is obvious from Figs. \ref{fig:peaklabel_9_11}(b) and (e).
Peak I for the parallel and antiparallel orientations occur at 1.27
eV and 1.46 eV, respectively, both of which are significantly red
shifted compared to the corresponding peak in the pristine GQD at
2.31 eV. As far as the MI peaks are concerned, for the parallel orientation
two peaks I ($y$ polarized) and IV ($x$ polarized at 2.86 eV) will
qualify; however, for the antiparallel orientation peak I is also
the MI peak. Additionally, in both the spectra there are several well-resolved
peaks in the energy range 2--4.5 eV. Wave functions for the first
peaks of both the BN-GQDs are dominated by the $|H\rightarrow L\rangle$
excitation, while the $x$-polarized peak IV for the parallel orientation
derives main contribution from the $\arrowvert H-2\rightarrow L\rangle$
transition. The detailed information about the excited states for
the two BN-GQDs of this class is presented in Tables S14 and S15 within
the SM \citep{supplemental}.

\subsubsection*{Model 11:}

\begin{figure}[!ht]
\centering{}\includegraphics[scale=1.33]{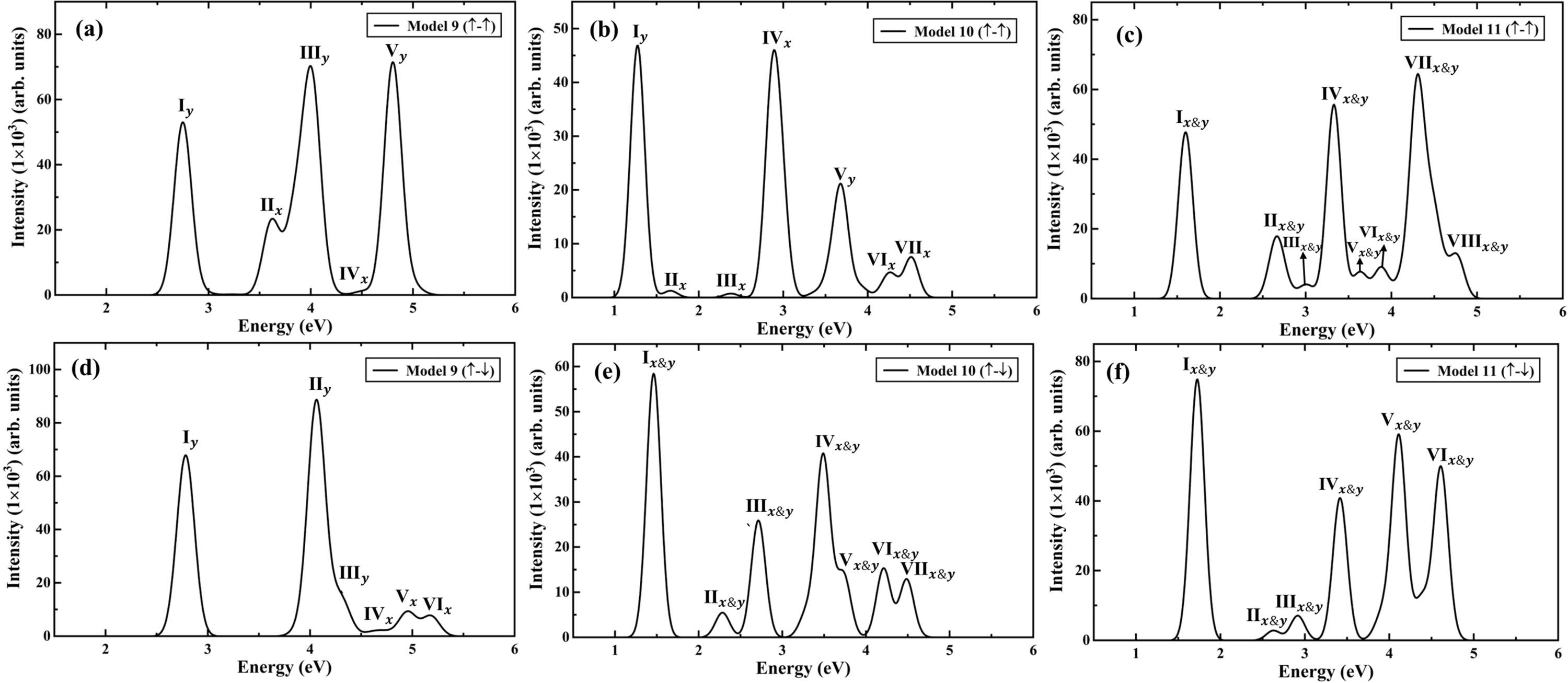}\caption{Linear optical absorption spectra for (a) Model 9 ($\uparrow-\uparrow$),
(b) Model 10 ($\uparrow-\uparrow$), (c) Model 11 ($\uparrow-\uparrow$),
(d) Model 9 ($\uparrow-\downarrow$), (e) Model 10 ($\uparrow-\downarrow$),
and (f) Model 11 ($\uparrow-\downarrow$), computed at the TDDFT-B3LYP
level of theory, assuming a uniform line width of 0.1 eV. \protect\label{fig:peaklabel_9_11}}
\end{figure}

In this model, the BN rings are placed along a diagonal of the GQD,
which for the parallel orientation corresponds to the $C_{s}$ point
group, while for the antiparallel orientation the point group is $C_{2h}$.
For the parallel and anti-parallel orientations, the computed spectra
are presented in Figs. \ref{fig:peaklabel_9_11}(c) and (f), respectively,
while the detailed information about the excited states contributing
to the spectra is given in Tables S16 and S17 within the SM \citep{supplemental}.
The location of peak I of the parallel orientation is at 1.59 eV,
while that for the antiparallel case is 1.73 eV, both of which are
quite red shifted compared to peak I of the pristine GQD. For the
parallel orientation, in addition to peak I, peaks IV (3.33 eV) and
VII (4.29 eV) also carry large intensities, although peak VII is the
MI peak. For the antiparallel orientation, peak I is the MI peak,
while peaks IV (3.42 eV), V (4.11 eV), and VI (4.61 eV) also have
significant intensities. Thus, broadly speaking, most of the features
of these BN-GQDs are red shifted as compared to the parent GQD, although
not to the extent as in model 10.

\subsubsection{Summary of the absorption spectra}

We note the following trends in the optical gaps and the absorption
spectra of the doped structures, as compared to the pristine GQD:
(a) Out of the fourteen doped structures considered, the optical gaps
of five models increase as compared to the pristine GQD, while for
nine cases the gaps decrease. (b) The point-group symmetry of the
doped structure is not a good descriptor of the shifts in the optical
gaps or in the intensity profiles of the absorption spectra. For example,
model 9 ($\uparrow-\uparrow$) with $C_{2v}$ symmetry exhibits a
blue shift in the optical gap, while model 10 ($\uparrow-\uparrow$)
also with $C_{2v}$ symmetry undergoes a significant red shift. (c)
As discussed previously, the redistribution of the charge associated
with the HOMO/LUMO of the doped structures, as compared to those of
the pristine GQD, appears to describe the changes in the optical gaps
to a reasonable extent. (d) The locations of the dopant (BN)\textsubscript{3}
rings have a strong impact on the nature of the corresponding absorption
spectra. For example, models 1 and 2 have the same point group symmetry
($C_{2v}$), and almost identical optical gaps, but their absorption
spectra are quite different from each other because of the different
locations of the dopant rings. (e) For the cases of doping with two
separated (BN)\textsubscript{3} rings (models 9, 10, and 11), for
a given model, the relative orientations of the rings ($\uparrow-\uparrow$
vs $\uparrow-\downarrow$) impact their absorption spectra significantly
(see Fig. \ref{fig:peaklabel_9_11}). Thus, we conclude that the different
doping models have significantly different electronic structures as
compared to each other, leading to distinct optical absorption spectra.

\section{Conclusions}

\label{sec:Conclusions}

In this work we considered a highly symmetric diamond-shaped GQD with
the $D_{2h}$ point group symmetry, and replaced up to two carbon
hexagons in it with B$_{3}$N$_{3}$ hexagonal rings at various places
to obtain different BN-ring doped structures. Depending on the placement
of BN rings, the resultant structures have point groups ranging from
the high-symmetry $D_{2h}$ to the lowest symmetry $C_{s}$. All the
BN-ring doped GQDs considered in this study are dynamically and energetically
stable. We found that even after BN ring doping, the resultant structures
continue to be planar, leading to $\sigma-\pi$ separation. The orbital
analysis and DOS calculations both reveal that the occupied and the
unoccupied orbitals close to the Fermi level continue to be of $\pi$
and $\pi^{*}$ type, respectively, just as in case of the parent GQD.
Subsequently, we computed their linear UV-Vis absorption spectra,
and found that they are dominated by $\pi$--$\pi^{*}$ transitions.
Furthermore, we found that by choosing suitable doping configurations,
one can tune the optical gaps of the BN-GQDs considered in this work
in a wide frequency range, spanning infrared to visible. This property
makes the considered doping strategy, and the resultant BN-GQDs potentially
quite useful in optoelectronic applications. Therefore, we hope that
our results will motivate the experimentalists to synthesize these
nanostructures and study their optical properties. 

Because the low-lying optical excitations in these structures are
of $\pi$--$\pi^{*}$ type, we plan to parametrize the Pariser-Parr-Pople
(PPP) model Hamiltonian for these systems, which is normally used
for the organic $\pi$-conjugated molecules. This will significantly
reduce the number of degrees of freedom, allowing the exploration
of much larger BN-doped, or BN-based, planar systems employing higher
level quantum many-body approaches. Furthermore, a majority of the
doped structures considered in this work lacked the inversion symmetry;
therefore, they will also exhibit second-order nonlinear optical response.
Investigations along these directions are currently underway in our
group, and the results will be communicated in future publications.

\section*{data availability}

The majority of the data are given in the main article and the Supplemental
Material. The rest of the data will be available from the authors
upon reasonable request.


\begin{thebibliography}{59}%
\makeatletter
\providecommand \@ifxundefined [1]{%
 \@ifx{#1\undefined}
}%
\providecommand \@ifnum [1]{%
 \ifnum #1\expandafter \@firstoftwo
 \else \expandafter \@secondoftwo
 \fi
}%
\providecommand \@ifx [1]{%
 \ifx #1\expandafter \@firstoftwo
 \else \expandafter \@secondoftwo
 \fi
}%
\providecommand \natexlab [1]{#1}%
\providecommand \enquote  [1]{``#1''}%
\providecommand \bibnamefont  [1]{#1}%
\providecommand \bibfnamefont [1]{#1}%
\providecommand \citenamefont [1]{#1}%
\providecommand \href@noop [0]{\@secondoftwo}%
\providecommand \href [0]{\begingroup \@sanitize@url \@href}%
\providecommand \@href[1]{\@@startlink{#1}\@@href}%
\providecommand \@@href[1]{\endgroup#1\@@endlink}%
\providecommand \@sanitize@url [0]{\catcode `\\12\catcode `\$12\catcode
  `\&12\catcode `\#12\catcode `\^12\catcode `\_12\catcode `\%12\relax}%
\providecommand \@@startlink[1]{}%
\providecommand \@@endlink[0]{}%
\providecommand \url  [0]{\begingroup\@sanitize@url \@url }%
\providecommand \@url [1]{\endgroup\@href {#1}{\urlprefix }}%
\providecommand \urlprefix  [0]{URL }%
\providecommand \Eprint [0]{\href }%
\providecommand \doibase [0]{https://doi.org/}%
\providecommand \selectlanguage [0]{\@gobble}%
\providecommand \bibinfo  [0]{\@secondoftwo}%
\providecommand \bibfield  [0]{\@secondoftwo}%
\providecommand \translation [1]{[#1]}%
\providecommand \BibitemOpen [0]{}%
\providecommand \bibitemStop [0]{}%
\providecommand \bibitemNoStop [0]{.\EOS\space}%
\providecommand \EOS [0]{\spacefactor3000\relax}%
\providecommand \BibitemShut  [1]{\csname bibitem#1\endcsname}%
\let\auto@bib@innerbib\@empty
\bibitem [{\citenamefont {Ma}\ \emph {et~al.}(2015)\citenamefont {Ma},
  \citenamefont {Zhu}, \citenamefont {Wang}, \citenamefont {Huang},
  \citenamefont {Zhang}, \citenamefont {Qi}, \citenamefont {Zhang},
  \citenamefont {Zhu}, \citenamefont {Deng}, \citenamefont {Peng} \emph
  {et~al.}}]{ma2015general}%
  \BibitemOpen
  \bibfield  {author} {\bibinfo {author} {\bibfnamefont {C.-B.}\ \bibnamefont
  {Ma}}, \bibinfo {author} {\bibfnamefont {Z.-T.}\ \bibnamefont {Zhu}},
  \bibinfo {author} {\bibfnamefont {H.-X.}\ \bibnamefont {Wang}}, \bibinfo
  {author} {\bibfnamefont {X.}~\bibnamefont {Huang}}, \bibinfo {author}
  {\bibfnamefont {X.}~\bibnamefont {Zhang}}, \bibinfo {author} {\bibfnamefont
  {X.}~\bibnamefont {Qi}}, \bibinfo {author} {\bibfnamefont {H.-L.}\
  \bibnamefont {Zhang}}, \bibinfo {author} {\bibfnamefont {Y.}~\bibnamefont
  {Zhu}}, \bibinfo {author} {\bibfnamefont {X.}~\bibnamefont {Deng}}, \bibinfo
  {author} {\bibfnamefont {Y.}~\bibnamefont {Peng}}, \emph {et~al.},\
  }\bibfield  {title} {\bibinfo {title} {A general solid-state synthesis of
  chemically-doped fluorescent graphene quantum dots for bioimaging and
  optoelectronic applications},\ }\href@noop {} {\bibfield  {journal} {\bibinfo
   {journal} {Nanoscale}\ }\textbf {\bibinfo {volume} {7}},\ \bibinfo {pages}
  {10162} (\bibinfo {year} {2015})}\BibitemShut {NoStop}%
\bibitem [{\citenamefont {Zuo}\ \emph {et~al.}(2017)\citenamefont {Zuo},
  \citenamefont {Tang}, \citenamefont {Xiang}, \citenamefont {Ji},
  \citenamefont {Luo}, \citenamefont {Rog{\'e}e},\ and\ \citenamefont
  {Ping~Lau}}]{zuo2017functionalization}%
  \BibitemOpen
  \bibfield  {author} {\bibinfo {author} {\bibfnamefont {W.}~\bibnamefont
  {Zuo}}, \bibinfo {author} {\bibfnamefont {L.}~\bibnamefont {Tang}}, \bibinfo
  {author} {\bibfnamefont {J.}~\bibnamefont {Xiang}}, \bibinfo {author}
  {\bibfnamefont {R.}~\bibnamefont {Ji}}, \bibinfo {author} {\bibfnamefont
  {L.}~\bibnamefont {Luo}}, \bibinfo {author} {\bibfnamefont {L.}~\bibnamefont
  {Rog{\'e}e}},\ and\ \bibinfo {author} {\bibfnamefont {S.}~\bibnamefont
  {Ping~Lau}},\ }\bibfield  {title} {\bibinfo {title} {Functionalization of
  graphene quantum dots by fluorine: Preparation, properties, application, and
  their mechanisms},\ }\href@noop {} {\bibfield  {journal} {\bibinfo  {journal}
  {Applied Physics Letters}\ }\textbf {\bibinfo {volume} {110}} (\bibinfo
  {year} {2017})}\BibitemShut {NoStop}%
\bibitem [{\citenamefont {Yang}\ \emph {et~al.}(2018)\citenamefont {Yang},
  \citenamefont {Wu}, \citenamefont {Luo}, \citenamefont {Liu}, \citenamefont
  {Gao}, \citenamefont {Wu}, \citenamefont {Cai},\ and\ \citenamefont
  {Saavedra}}]{yang2018exploring}%
  \BibitemOpen
  \bibfield  {author} {\bibinfo {author} {\bibfnamefont {G.}~\bibnamefont
  {Yang}}, \bibinfo {author} {\bibfnamefont {C.}~\bibnamefont {Wu}}, \bibinfo
  {author} {\bibfnamefont {X.}~\bibnamefont {Luo}}, \bibinfo {author}
  {\bibfnamefont {X.}~\bibnamefont {Liu}}, \bibinfo {author} {\bibfnamefont
  {Y.}~\bibnamefont {Gao}}, \bibinfo {author} {\bibfnamefont {P.}~\bibnamefont
  {Wu}}, \bibinfo {author} {\bibfnamefont {C.}~\bibnamefont {Cai}},\ and\
  \bibinfo {author} {\bibfnamefont {S.~S.}\ \bibnamefont {Saavedra}},\
  }\bibfield  {title} {\bibinfo {title} {Exploring the emissive states of
  heteroatom-doped graphene quantum dots},\ }\href@noop {} {\bibfield
  {journal} {\bibinfo  {journal} {The Journal of Physical Chemistry C}\
  }\textbf {\bibinfo {volume} {122}},\ \bibinfo {pages} {6483} (\bibinfo {year}
  {2018})}\BibitemShut {NoStop}%
\bibitem [{\citenamefont {Novoselov}\ \emph {et~al.}(2004)\citenamefont
  {Novoselov}, \citenamefont {Geim}, \citenamefont {Morozov}, \citenamefont
  {Jiang}, \citenamefont {Zhang}, \citenamefont {Dubonos}, \citenamefont
  {Grigorieva},\ and\ \citenamefont {Firsov}}]{graphene-synthesis-2004}%
  \BibitemOpen
  \bibfield  {author} {\bibinfo {author} {\bibfnamefont {K.~S.}\ \bibnamefont
  {Novoselov}}, \bibinfo {author} {\bibfnamefont {A.~K.}\ \bibnamefont {Geim}},
  \bibinfo {author} {\bibfnamefont {S.~V.}\ \bibnamefont {Morozov}}, \bibinfo
  {author} {\bibfnamefont {D.}~\bibnamefont {Jiang}}, \bibinfo {author}
  {\bibfnamefont {Y.}~\bibnamefont {Zhang}}, \bibinfo {author} {\bibfnamefont
  {S.~V.}\ \bibnamefont {Dubonos}}, \bibinfo {author} {\bibfnamefont {I.~V.}\
  \bibnamefont {Grigorieva}},\ and\ \bibinfo {author} {\bibfnamefont {A.~A.}\
  \bibnamefont {Firsov}},\ }\bibfield  {title} {\bibinfo {title} {Electric
  field effect in atomically thin carbon films},\ }\href
  {https://doi.org/10.1126/science.1102896} {\bibfield  {journal} {\bibinfo
  {journal} {Science}\ }\textbf {\bibinfo {volume} {306}},\ \bibinfo {pages}
  {666} (\bibinfo {year} {2004})}\BibitemShut {NoStop}%
\bibitem [{\citenamefont {Galashev}\ and\ \citenamefont
  {Rakhmanova}(2014)}]{galashev2014mechanical}%
  \BibitemOpen
  \bibfield  {author} {\bibinfo {author} {\bibfnamefont {A.~E.}\ \bibnamefont
  {Galashev}}\ and\ \bibinfo {author} {\bibfnamefont {O.~R.}\ \bibnamefont
  {Rakhmanova}},\ }\bibfield  {title} {\bibinfo {title} {Mechanical and thermal
  stability of graphene and graphene-based materials},\ }\href@noop {}
  {\bibfield  {journal} {\bibinfo  {journal} {Physics-Uspekhi}\ }\textbf
  {\bibinfo {volume} {57}},\ \bibinfo {pages} {970} (\bibinfo {year}
  {2014})}\BibitemShut {NoStop}%
\bibitem [{\citenamefont {Liu}\ \emph {et~al.}(2011{\natexlab{a}})\citenamefont
  {Liu}, \citenamefont {Yin}, \citenamefont {Ulin-Avila}, \citenamefont {Geng},
  \citenamefont {Zentgraf}, \citenamefont {Ju}, \citenamefont {Wang},\ and\
  \citenamefont {Zhang}}]{liu2011graphene}%
  \BibitemOpen
  \bibfield  {author} {\bibinfo {author} {\bibfnamefont {M.}~\bibnamefont
  {Liu}}, \bibinfo {author} {\bibfnamefont {X.}~\bibnamefont {Yin}}, \bibinfo
  {author} {\bibfnamefont {E.}~\bibnamefont {Ulin-Avila}}, \bibinfo {author}
  {\bibfnamefont {B.}~\bibnamefont {Geng}}, \bibinfo {author} {\bibfnamefont
  {T.}~\bibnamefont {Zentgraf}}, \bibinfo {author} {\bibfnamefont
  {L.}~\bibnamefont {Ju}}, \bibinfo {author} {\bibfnamefont {F.}~\bibnamefont
  {Wang}},\ and\ \bibinfo {author} {\bibfnamefont {X.}~\bibnamefont {Zhang}},\
  }\bibfield  {title} {\bibinfo {title} {A graphene-based broadband optical
  modulator},\ }\href@noop {} {\bibfield  {journal} {\bibinfo  {journal}
  {Nature}\ }\textbf {\bibinfo {volume} {474}},\ \bibinfo {pages} {64}
  (\bibinfo {year} {2011}{\natexlab{a}})}\BibitemShut {NoStop}%
\bibitem [{\citenamefont {Singh}\ \emph {et~al.}(2013)\citenamefont {Singh},
  \citenamefont {Uddin}, \citenamefont {Tolson}, \citenamefont {Maire-Afeli},
  \citenamefont {Sbrockey}, \citenamefont {Tompa}, \citenamefont {Spencer},
  \citenamefont {Vogt}, \citenamefont {Sudarshan},\ and\ \citenamefont
  {Koley}}]{singh2013electrically}%
  \BibitemOpen
  \bibfield  {author} {\bibinfo {author} {\bibfnamefont {A.}~\bibnamefont
  {Singh}}, \bibinfo {author} {\bibfnamefont {M.}~\bibnamefont {Uddin}},
  \bibinfo {author} {\bibfnamefont {J.}~\bibnamefont {Tolson}}, \bibinfo
  {author} {\bibfnamefont {H.}~\bibnamefont {Maire-Afeli}}, \bibinfo {author}
  {\bibfnamefont {N.}~\bibnamefont {Sbrockey}}, \bibinfo {author}
  {\bibfnamefont {G.}~\bibnamefont {Tompa}}, \bibinfo {author} {\bibfnamefont
  {M.}~\bibnamefont {Spencer}}, \bibinfo {author} {\bibfnamefont
  {T.}~\bibnamefont {Vogt}}, \bibinfo {author} {\bibfnamefont {T.}~\bibnamefont
  {Sudarshan}},\ and\ \bibinfo {author} {\bibfnamefont {G.}~\bibnamefont
  {Koley}},\ }\bibfield  {title} {\bibinfo {title} {Electrically tunable
  molecular doping of graphene},\ }\href@noop {} {\bibfield  {journal}
  {\bibinfo  {journal} {Applied Physics Letters}\ }\textbf {\bibinfo {volume}
  {102}} (\bibinfo {year} {2013})}\BibitemShut {NoStop}%
\bibitem [{\citenamefont {Cai}\ \emph {et~al.}(2010)\citenamefont {Cai},
  \citenamefont {Ruffieux}, \citenamefont {Jaafar}, \citenamefont {Bieri},
  \citenamefont {Braun}, \citenamefont {Blankenburg}, \citenamefont {Muoth},
  \citenamefont {Seitsonen}, \citenamefont {Saleh}, \citenamefont {Feng} \emph
  {et~al.}}]{cai2010atomically}%
  \BibitemOpen
  \bibfield  {author} {\bibinfo {author} {\bibfnamefont {J.}~\bibnamefont
  {Cai}}, \bibinfo {author} {\bibfnamefont {P.}~\bibnamefont {Ruffieux}},
  \bibinfo {author} {\bibfnamefont {R.}~\bibnamefont {Jaafar}}, \bibinfo
  {author} {\bibfnamefont {M.}~\bibnamefont {Bieri}}, \bibinfo {author}
  {\bibfnamefont {T.}~\bibnamefont {Braun}}, \bibinfo {author} {\bibfnamefont
  {S.}~\bibnamefont {Blankenburg}}, \bibinfo {author} {\bibfnamefont
  {M.}~\bibnamefont {Muoth}}, \bibinfo {author} {\bibfnamefont {A.~P.}\
  \bibnamefont {Seitsonen}}, \bibinfo {author} {\bibfnamefont {M.}~\bibnamefont
  {Saleh}}, \bibinfo {author} {\bibfnamefont {X.}~\bibnamefont {Feng}}, \emph
  {et~al.},\ }\bibfield  {title} {\bibinfo {title} {Atomically precise
  bottom-up fabrication of graphene nanoribbons},\ }\href@noop {} {\bibfield
  {journal} {\bibinfo  {journal} {Nature}\ }\textbf {\bibinfo {volume} {466}},\
  \bibinfo {pages} {470} (\bibinfo {year} {2010})}\BibitemShut {NoStop}%
\bibitem [{\citenamefont {Jiao}\ \emph {et~al.}(2010)\citenamefont {Jiao},
  \citenamefont {Wang}, \citenamefont {Diankov}, \citenamefont {Wang},\ and\
  \citenamefont {Dai}}]{jiao2010facile}%
  \BibitemOpen
  \bibfield  {author} {\bibinfo {author} {\bibfnamefont {L.}~\bibnamefont
  {Jiao}}, \bibinfo {author} {\bibfnamefont {X.}~\bibnamefont {Wang}}, \bibinfo
  {author} {\bibfnamefont {G.}~\bibnamefont {Diankov}}, \bibinfo {author}
  {\bibfnamefont {H.}~\bibnamefont {Wang}},\ and\ \bibinfo {author}
  {\bibfnamefont {H.}~\bibnamefont {Dai}},\ }\bibfield  {title} {\bibinfo
  {title} {Facile synthesis of high-quality graphene nanoribbons},\ }\href@noop
  {} {\bibfield  {journal} {\bibinfo  {journal} {Nature Nanotechnology}\
  }\textbf {\bibinfo {volume} {5}},\ \bibinfo {pages} {321} (\bibinfo {year}
  {2010})}\BibitemShut {NoStop}%
\bibitem [{\citenamefont {Li}\ and\ \citenamefont
  {Yan}(2010)}]{li2010colloidal}%
  \BibitemOpen
  \bibfield  {author} {\bibinfo {author} {\bibfnamefont {L.-s.}\ \bibnamefont
  {Li}}\ and\ \bibinfo {author} {\bibfnamefont {X.}~\bibnamefont {Yan}},\
  }\bibfield  {title} {\bibinfo {title} {Colloidal graphene quantum dots},\
  }\href@noop {} {\bibfield  {journal} {\bibinfo  {journal} {The Journal of
  Physical Chemistry Letters}\ }\textbf {\bibinfo {volume} {1}},\ \bibinfo
  {pages} {2572} (\bibinfo {year} {2010})}\BibitemShut {NoStop}%
\bibitem [{\citenamefont {Huang}\ \emph {et~al.}(2011)\citenamefont {Huang},
  \citenamefont {Yin}, \citenamefont {Wu}, \citenamefont {Qi}, \citenamefont
  {He}, \citenamefont {Zhang}, \citenamefont {Yan}, \citenamefont {Boey},\ and\
  \citenamefont {Zhang}}]{huang2011graphene}%
  \BibitemOpen
  \bibfield  {author} {\bibinfo {author} {\bibfnamefont {X.}~\bibnamefont
  {Huang}}, \bibinfo {author} {\bibfnamefont {Z.}~\bibnamefont {Yin}}, \bibinfo
  {author} {\bibfnamefont {S.}~\bibnamefont {Wu}}, \bibinfo {author}
  {\bibfnamefont {X.}~\bibnamefont {Qi}}, \bibinfo {author} {\bibfnamefont
  {Q.}~\bibnamefont {He}}, \bibinfo {author} {\bibfnamefont {Q.}~\bibnamefont
  {Zhang}}, \bibinfo {author} {\bibfnamefont {Q.}~\bibnamefont {Yan}}, \bibinfo
  {author} {\bibfnamefont {F.}~\bibnamefont {Boey}},\ and\ \bibinfo {author}
  {\bibfnamefont {H.}~\bibnamefont {Zhang}},\ }\bibfield  {title} {\bibinfo
  {title} {Graphene-based materials: synthesis, characterization, properties,
  and applications},\ }\href@noop {} {\bibfield  {journal} {\bibinfo  {journal}
  {Small}\ }\textbf {\bibinfo {volume} {7}},\ \bibinfo {pages} {1876} (\bibinfo
  {year} {2011})}\BibitemShut {NoStop}%
\bibitem [{\citenamefont {Bacon}\ \emph {et~al.}(2014)\citenamefont {Bacon},
  \citenamefont {Bradley},\ and\ \citenamefont {Nann}}]{bacon2014graphene}%
  \BibitemOpen
  \bibfield  {author} {\bibinfo {author} {\bibfnamefont {M.}~\bibnamefont
  {Bacon}}, \bibinfo {author} {\bibfnamefont {S.~J.}\ \bibnamefont {Bradley}},\
  and\ \bibinfo {author} {\bibfnamefont {T.}~\bibnamefont {Nann}},\ }\bibfield
  {title} {\bibinfo {title} {Graphene quantum dots},\ }\href@noop {} {\bibfield
   {journal} {\bibinfo  {journal} {Particle \& Particle Systems
  Characterization}\ }\textbf {\bibinfo {volume} {31}},\ \bibinfo {pages} {415}
  (\bibinfo {year} {2014})}\BibitemShut {NoStop}%
\bibitem [{\citenamefont {Kim}\ \emph {et~al.}(2012)\citenamefont {Kim},
  \citenamefont {Hwang}, \citenamefont {Kim}, \citenamefont {Shin},
  \citenamefont {Shin}, \citenamefont {Kim}, \citenamefont {Yang},
  \citenamefont {Park}, \citenamefont {Hwang}, \citenamefont {Choi} \emph
  {et~al.}}]{kim2012anomalous}%
  \BibitemOpen
  \bibfield  {author} {\bibinfo {author} {\bibfnamefont {S.}~\bibnamefont
  {Kim}}, \bibinfo {author} {\bibfnamefont {S.~W.}\ \bibnamefont {Hwang}},
  \bibinfo {author} {\bibfnamefont {M.-K.}\ \bibnamefont {Kim}}, \bibinfo
  {author} {\bibfnamefont {D.~Y.}\ \bibnamefont {Shin}}, \bibinfo {author}
  {\bibfnamefont {D.~H.}\ \bibnamefont {Shin}}, \bibinfo {author}
  {\bibfnamefont {C.~O.}\ \bibnamefont {Kim}}, \bibinfo {author} {\bibfnamefont
  {S.~B.}\ \bibnamefont {Yang}}, \bibinfo {author} {\bibfnamefont {J.~H.}\
  \bibnamefont {Park}}, \bibinfo {author} {\bibfnamefont {E.}~\bibnamefont
  {Hwang}}, \bibinfo {author} {\bibfnamefont {S.-H.}\ \bibnamefont {Choi}},
  \emph {et~al.},\ }\bibfield  {title} {\bibinfo {title} {Anomalous behaviors
  of visible luminescence from graphene quantum dots: interplay between size
  and shape},\ }\href@noop {} {\bibfield  {journal} {\bibinfo  {journal} {ACS
  Nano}\ }\textbf {\bibinfo {volume} {6}},\ \bibinfo {pages} {8203} (\bibinfo
  {year} {2012})}\BibitemShut {NoStop}%
\bibitem [{\citenamefont {Geffroy}\ \emph {et~al.}(2006)\citenamefont
  {Geffroy}, \citenamefont {Le~Roy},\ and\ \citenamefont
  {Prat}}]{geffroy2006organic}%
  \BibitemOpen
  \bibfield  {author} {\bibinfo {author} {\bibfnamefont {B.}~\bibnamefont
  {Geffroy}}, \bibinfo {author} {\bibfnamefont {P.}~\bibnamefont {Le~Roy}},\
  and\ \bibinfo {author} {\bibfnamefont {C.}~\bibnamefont {Prat}},\ }\bibfield
  {title} {\bibinfo {title} {Organic light-emitting diode {(OLED)} technology:
  materials, devices and display technologies},\ }\href@noop {} {\bibfield
  {journal} {\bibinfo  {journal} {Polymer International}\ }\textbf {\bibinfo
  {volume} {55}},\ \bibinfo {pages} {572} (\bibinfo {year} {2006})}\BibitemShut
  {NoStop}%
\bibitem [{\citenamefont {Chen}\ \emph {et~al.}(2018)\citenamefont {Chen},
  \citenamefont {Lee}, \citenamefont {Lin}, \citenamefont {Chen},\ and\
  \citenamefont {Wu}}]{chen2018liquid}%
  \BibitemOpen
  \bibfield  {author} {\bibinfo {author} {\bibfnamefont {H.-W.}\ \bibnamefont
  {Chen}}, \bibinfo {author} {\bibfnamefont {J.-H.}\ \bibnamefont {Lee}},
  \bibinfo {author} {\bibfnamefont {B.-Y.}\ \bibnamefont {Lin}}, \bibinfo
  {author} {\bibfnamefont {S.}~\bibnamefont {Chen}},\ and\ \bibinfo {author}
  {\bibfnamefont {S.-T.}\ \bibnamefont {Wu}},\ }\bibfield  {title} {\bibinfo
  {title} {Liquid crystal display and organic light-emitting diode display:
  present status and future perspectives},\ }\href@noop {} {\bibfield
  {journal} {\bibinfo  {journal} {Light: Science \& Applications}\ }\textbf
  {\bibinfo {volume} {7}},\ \bibinfo {pages} {17168} (\bibinfo {year}
  {2018})}\BibitemShut {NoStop}%
\bibitem [{\citenamefont {Liu}\ \emph {et~al.}(2018)\citenamefont {Liu},
  \citenamefont {Li}, \citenamefont {Ren}, \citenamefont {Yan},\ and\
  \citenamefont {Bryce}}]{liu2018all}%
  \BibitemOpen
  \bibfield  {author} {\bibinfo {author} {\bibfnamefont {Y.}~\bibnamefont
  {Liu}}, \bibinfo {author} {\bibfnamefont {C.}~\bibnamefont {Li}}, \bibinfo
  {author} {\bibfnamefont {Z.}~\bibnamefont {Ren}}, \bibinfo {author}
  {\bibfnamefont {S.}~\bibnamefont {Yan}},\ and\ \bibinfo {author}
  {\bibfnamefont {M.~R.}\ \bibnamefont {Bryce}},\ }\bibfield  {title} {\bibinfo
  {title} {All-organic thermally activated delayed fluorescence materials for
  organic light-emitting diodes},\ }\href@noop {} {\bibfield  {journal}
  {\bibinfo  {journal} {Nature Reviews Materials}\ }\textbf {\bibinfo {volume}
  {3}},\ \bibinfo {pages} {1} (\bibinfo {year} {2018})}\BibitemShut {NoStop}%
\bibitem [{\citenamefont {Facchetti}(2011)}]{facchetti2011pi}%
  \BibitemOpen
  \bibfield  {author} {\bibinfo {author} {\bibfnamefont {A.}~\bibnamefont
  {Facchetti}},\ }\bibfield  {title} {\bibinfo {title} {$\pi$-conjugated
  polymers for organic electronics and photovoltaic cell applications},\
  }\href@noop {} {\bibfield  {journal} {\bibinfo  {journal} {Chemistry of
  Materials}\ }\textbf {\bibinfo {volume} {23}},\ \bibinfo {pages} {733}
  (\bibinfo {year} {2011})}\BibitemShut {NoStop}%
\bibitem [{\citenamefont {Zhu}\ \emph {et~al.}(2014)\citenamefont {Zhu},
  \citenamefont {Ma}, \citenamefont {Wang}, \citenamefont {Mu}, \citenamefont
  {Fan}, \citenamefont {Du}, \citenamefont {Bai}, \citenamefont {Fan},
  \citenamefont {Yan}, \citenamefont {Phillips} \emph
  {et~al.}}]{zhu2014efficiency}%
  \BibitemOpen
  \bibfield  {author} {\bibinfo {author} {\bibfnamefont {Z.}~\bibnamefont
  {Zhu}}, \bibinfo {author} {\bibfnamefont {J.}~\bibnamefont {Ma}}, \bibinfo
  {author} {\bibfnamefont {Z.}~\bibnamefont {Wang}}, \bibinfo {author}
  {\bibfnamefont {C.}~\bibnamefont {Mu}}, \bibinfo {author} {\bibfnamefont
  {Z.}~\bibnamefont {Fan}}, \bibinfo {author} {\bibfnamefont {L.}~\bibnamefont
  {Du}}, \bibinfo {author} {\bibfnamefont {Y.}~\bibnamefont {Bai}}, \bibinfo
  {author} {\bibfnamefont {L.}~\bibnamefont {Fan}}, \bibinfo {author}
  {\bibfnamefont {H.}~\bibnamefont {Yan}}, \bibinfo {author} {\bibfnamefont
  {D.~L.}\ \bibnamefont {Phillips}}, \emph {et~al.},\ }\bibfield  {title}
  {\bibinfo {title} {Efficiency enhancement of perovskite solar cells through
  fast electron extraction: the role of graphene quantum dots},\ }\href@noop {}
  {\bibfield  {journal} {\bibinfo  {journal} {Journal of the American Chemical
  Society}\ }\textbf {\bibinfo {volume} {136}},\ \bibinfo {pages} {3760}
  (\bibinfo {year} {2014})}\BibitemShut {NoStop}%
\bibitem [{\citenamefont {Wang}\ \emph {et~al.}(2012)\citenamefont {Wang},
  \citenamefont {Dong}, \citenamefont {Hu}, \citenamefont {Liu},\ and\
  \citenamefont {Zhu}}]{wang2012semiconducting}%
  \BibitemOpen
  \bibfield  {author} {\bibinfo {author} {\bibfnamefont {C.}~\bibnamefont
  {Wang}}, \bibinfo {author} {\bibfnamefont {H.}~\bibnamefont {Dong}}, \bibinfo
  {author} {\bibfnamefont {W.}~\bibnamefont {Hu}}, \bibinfo {author}
  {\bibfnamefont {Y.}~\bibnamefont {Liu}},\ and\ \bibinfo {author}
  {\bibfnamefont {D.}~\bibnamefont {Zhu}},\ }\bibfield  {title} {\bibinfo
  {title} {Semiconducting $\pi$-conjugated systems in field-effect transistors:
  a material odyssey of organic electronics},\ }\href@noop {} {\bibfield
  {journal} {\bibinfo  {journal} {Chemical Reviews}\ }\textbf {\bibinfo
  {volume} {112}},\ \bibinfo {pages} {2208} (\bibinfo {year}
  {2012})}\BibitemShut {NoStop}%
\bibitem [{\citenamefont {Sirringhaus}(2014)}]{sirringhaus201425th}%
  \BibitemOpen
  \bibfield  {author} {\bibinfo {author} {\bibfnamefont {H.}~\bibnamefont
  {Sirringhaus}},\ }\bibfield  {title} {\bibinfo {title} {Organic field-effect
  transistors: the path beyond amorphous silicon},\ }\href@noop {} {\bibfield
  {journal} {\bibinfo  {journal} {Advanced Materials}\ }\textbf {\bibinfo
  {volume} {26}},\ \bibinfo {pages} {1319} (\bibinfo {year}
  {2014})}\BibitemShut {NoStop}%
\bibitem [{\citenamefont {Lussem}\ \emph {et~al.}(2016)\citenamefont {Lussem},
  \citenamefont {Keum}, \citenamefont {Kasemann}, \citenamefont {Naab},
  \citenamefont {Bao},\ and\ \citenamefont {Leo}}]{lussem2016doped}%
  \BibitemOpen
  \bibfield  {author} {\bibinfo {author} {\bibfnamefont {B.}~\bibnamefont
  {Lussem}}, \bibinfo {author} {\bibfnamefont {C.-M.}\ \bibnamefont {Keum}},
  \bibinfo {author} {\bibfnamefont {D.}~\bibnamefont {Kasemann}}, \bibinfo
  {author} {\bibfnamefont {B.}~\bibnamefont {Naab}}, \bibinfo {author}
  {\bibfnamefont {Z.}~\bibnamefont {Bao}},\ and\ \bibinfo {author}
  {\bibfnamefont {K.}~\bibnamefont {Leo}},\ }\bibfield  {title} {\bibinfo
  {title} {Doped organic transistors},\ }\href@noop {} {\bibfield  {journal}
  {\bibinfo  {journal} {Chemical Reviews}\ }\textbf {\bibinfo {volume} {116}},\
  \bibinfo {pages} {13714} (\bibinfo {year} {2016})}\BibitemShut {NoStop}%
\bibitem [{\citenamefont {Wehling}\ \emph {et~al.}(2008)\citenamefont
  {Wehling}, \citenamefont {Novoselov}, \citenamefont {Morozov}, \citenamefont
  {Vdovin}, \citenamefont {Katsnelson}, \citenamefont {Geim},\ and\
  \citenamefont {Lichtenstein}}]{wehling2008molecular}%
  \BibitemOpen
  \bibfield  {author} {\bibinfo {author} {\bibfnamefont {T.}~\bibnamefont
  {Wehling}}, \bibinfo {author} {\bibfnamefont {K.}~\bibnamefont {Novoselov}},
  \bibinfo {author} {\bibfnamefont {S.}~\bibnamefont {Morozov}}, \bibinfo
  {author} {\bibfnamefont {E.}~\bibnamefont {Vdovin}}, \bibinfo {author}
  {\bibfnamefont {M.}~\bibnamefont {Katsnelson}}, \bibinfo {author}
  {\bibfnamefont {A.}~\bibnamefont {Geim}},\ and\ \bibinfo {author}
  {\bibfnamefont {A.}~\bibnamefont {Lichtenstein}},\ }\bibfield  {title}
  {\bibinfo {title} {Molecular doping of graphene},\ }\href@noop {} {\bibfield
  {journal} {\bibinfo  {journal} {Nano Letters}\ }\textbf {\bibinfo {volume}
  {8}},\ \bibinfo {pages} {173} (\bibinfo {year} {2008})}\BibitemShut {NoStop}%
\bibitem [{\citenamefont {Liu}\ \emph {et~al.}(2011{\natexlab{b}})\citenamefont
  {Liu}, \citenamefont {Liu},\ and\ \citenamefont {Zhu}}]{liu2011chemical}%
  \BibitemOpen
  \bibfield  {author} {\bibinfo {author} {\bibfnamefont {H.}~\bibnamefont
  {Liu}}, \bibinfo {author} {\bibfnamefont {Y.}~\bibnamefont {Liu}},\ and\
  \bibinfo {author} {\bibfnamefont {D.}~\bibnamefont {Zhu}},\ }\bibfield
  {title} {\bibinfo {title} {Chemical doping of graphene},\ }\href@noop {}
  {\bibfield  {journal} {\bibinfo  {journal} {Journal of Materials Chemistry}\
  }\textbf {\bibinfo {volume} {21}},\ \bibinfo {pages} {3335} (\bibinfo {year}
  {2011}{\natexlab{b}})}\BibitemShut {NoStop}%
\bibitem [{\citenamefont {Kawai}\ \emph {et~al.}(2015)\citenamefont {Kawai},
  \citenamefont {Saito}, \citenamefont {Osumi}, \citenamefont {Yamaguchi},
  \citenamefont {Foster}, \citenamefont {Spijker},\ and\ \citenamefont
  {Meyer}}]{kawai2015atomically}%
  \BibitemOpen
  \bibfield  {author} {\bibinfo {author} {\bibfnamefont {S.}~\bibnamefont
  {Kawai}}, \bibinfo {author} {\bibfnamefont {S.}~\bibnamefont {Saito}},
  \bibinfo {author} {\bibfnamefont {S.}~\bibnamefont {Osumi}}, \bibinfo
  {author} {\bibfnamefont {S.}~\bibnamefont {Yamaguchi}}, \bibinfo {author}
  {\bibfnamefont {A.~S.}\ \bibnamefont {Foster}}, \bibinfo {author}
  {\bibfnamefont {P.}~\bibnamefont {Spijker}},\ and\ \bibinfo {author}
  {\bibfnamefont {E.}~\bibnamefont {Meyer}},\ }\bibfield  {title} {\bibinfo
  {title} {Atomically controlled substitutional boron-doping of graphene
  nanoribbons},\ }\href@noop {} {\bibfield  {journal} {\bibinfo  {journal}
  {Nature Communications}\ }\textbf {\bibinfo {volume} {6}},\ \bibinfo {pages}
  {8098} (\bibinfo {year} {2015})}\BibitemShut {NoStop}%
\bibitem [{\citenamefont {Xu}\ \emph {et~al.}(2010)\citenamefont {Xu},
  \citenamefont {Lu}, \citenamefont {Feng},\ and\ \citenamefont
  {Lin}}]{xu2010density}%
  \BibitemOpen
  \bibfield  {author} {\bibinfo {author} {\bibfnamefont {B.}~\bibnamefont
  {Xu}}, \bibinfo {author} {\bibfnamefont {Y.}~\bibnamefont {Lu}}, \bibinfo
  {author} {\bibfnamefont {Y.}~\bibnamefont {Feng}},\ and\ \bibinfo {author}
  {\bibfnamefont {J.}~\bibnamefont {Lin}},\ }\bibfield  {title} {\bibinfo
  {title} {Density functional theory study of {BN}-doped graphene superlattice:
  role of geometrical shape and size},\ }\href@noop {} {\bibfield  {journal}
  {\bibinfo  {journal} {Journal of Applied Physics}\ }\textbf {\bibinfo
  {volume} {108}} (\bibinfo {year} {2010})}\BibitemShut {NoStop}%
\bibitem [{\citenamefont {Ajeel}\ \emph {et~al.}(2024)\citenamefont {Ajeel},
  \citenamefont {Mutier}, \citenamefont {Mohsin}, \citenamefont {Khamees},
  \citenamefont {Khudhair},\ and\ \citenamefont
  {Ahmed}}]{bn-dimer-doped-gqds-2024}%
  \BibitemOpen
  \bibfield  {author} {\bibinfo {author} {\bibfnamefont {F.~N.}\ \bibnamefont
  {Ajeel}}, \bibinfo {author} {\bibfnamefont {M.~N.}\ \bibnamefont {Mutier}},
  \bibinfo {author} {\bibfnamefont {K.~H.}\ \bibnamefont {Mohsin}}, \bibinfo
  {author} {\bibfnamefont {S.~K.}\ \bibnamefont {Khamees}}, \bibinfo {author}
  {\bibfnamefont {A.~M.}\ \bibnamefont {Khudhair}},\ and\ \bibinfo {author}
  {\bibfnamefont {A.~B.}\ \bibnamefont {Ahmed}},\ }\bibfield  {title} {\bibinfo
  {title} {Theoretical study on electronic properties of {BN} dimers doped
  graphene quantum dots},\ }\href {https://doi.org/10.1007/s12668-024-01422-z}
  {\bibfield  {journal} {\bibinfo  {journal} {BioNanoScience}\ }\textbf
  {\bibinfo {volume} {14}},\ \bibinfo {pages} {1110} (\bibinfo {year}
  {2024})}\BibitemShut {NoStop}%
\bibitem [{\citenamefont {Fei}\ \emph {et~al.}(2014)\citenamefont {Fei},
  \citenamefont {Ye}, \citenamefont {Ye}, \citenamefont {Gong}, \citenamefont
  {Peng}, \citenamefont {Fan}, \citenamefont {Samuel}, \citenamefont {Ajayan},\
  and\ \citenamefont {Tour}}]{fei2014boron}%
  \BibitemOpen
  \bibfield  {author} {\bibinfo {author} {\bibfnamefont {H.}~\bibnamefont
  {Fei}}, \bibinfo {author} {\bibfnamefont {R.}~\bibnamefont {Ye}}, \bibinfo
  {author} {\bibfnamefont {G.}~\bibnamefont {Ye}}, \bibinfo {author}
  {\bibfnamefont {Y.}~\bibnamefont {Gong}}, \bibinfo {author} {\bibfnamefont
  {Z.}~\bibnamefont {Peng}}, \bibinfo {author} {\bibfnamefont {X.}~\bibnamefont
  {Fan}}, \bibinfo {author} {\bibfnamefont {E.~L.}\ \bibnamefont {Samuel}},
  \bibinfo {author} {\bibfnamefont {P.~M.}\ \bibnamefont {Ajayan}},\ and\
  \bibinfo {author} {\bibfnamefont {J.~M.}\ \bibnamefont {Tour}},\ }\bibfield
  {title} {\bibinfo {title} {Boron-and nitrogen-doped graphene quantum
  dots/graphene hybrid nanoplatelets as efficient electrocatalysts for oxygen
  reduction},\ }\href@noop {} {\bibfield  {journal} {\bibinfo  {journal} {ACS
  Nano}\ }\textbf {\bibinfo {volume} {8}},\ \bibinfo {pages} {10837} (\bibinfo
  {year} {2014})}\BibitemShut {NoStop}%
\bibitem [{\citenamefont {Li}\ \emph {et~al.}(2017)\citenamefont {Li},
  \citenamefont {Yuan}, \citenamefont {Liu}, \citenamefont {Liu}, \citenamefont
  {Gao}, \citenamefont {Li}, \citenamefont {Li},\ and\ \citenamefont
  {Huang}}]{bn-codoped-gqds-for-bioimaging-synthesis-2017}%
  \BibitemOpen
  \bibfield  {author} {\bibinfo {author} {\bibfnamefont {R.~S.}\ \bibnamefont
  {Li}}, \bibinfo {author} {\bibfnamefont {B.}~\bibnamefont {Yuan}}, \bibinfo
  {author} {\bibfnamefont {J.~H.}\ \bibnamefont {Liu}}, \bibinfo {author}
  {\bibfnamefont {M.~L.}\ \bibnamefont {Liu}}, \bibinfo {author} {\bibfnamefont
  {P.~F.}\ \bibnamefont {Gao}}, \bibinfo {author} {\bibfnamefont {Y.~F.}\
  \bibnamefont {Li}}, \bibinfo {author} {\bibfnamefont {M.}~\bibnamefont
  {Li}},\ and\ \bibinfo {author} {\bibfnamefont {C.~Z.}\ \bibnamefont
  {Huang}},\ }\bibfield  {title} {\bibinfo {title} {Boron and nitrogen co-doped
  single-layered graphene quantum dots: a high-affinity platform for
  visualizing the dynamic invasion of {HIV DNA} into living cells through
  fluorescence resonance energy transfer},\ }\href
  {https://doi.org/10.1039/C7TB02356A} {\bibfield  {journal} {\bibinfo
  {journal} {J. Mater. Chem. B}\ }\textbf {\bibinfo {volume} {5}},\ \bibinfo
  {pages} {8719} (\bibinfo {year} {2017})}\BibitemShut {NoStop}%
\bibitem [{\citenamefont {Yang}\ \emph {et~al.}(2019)\citenamefont {Yang},
  \citenamefont {Su}, \citenamefont {Guo}, \citenamefont {Yao},\ and\
  \citenamefont {Yuan}}]{bn-doped-gqds-for-sensors-synthesis-2019}%
  \BibitemOpen
  \bibfield  {author} {\bibinfo {author} {\bibfnamefont {P.}~\bibnamefont
  {Yang}}, \bibinfo {author} {\bibfnamefont {J.}~\bibnamefont {Su}}, \bibinfo
  {author} {\bibfnamefont {R.}~\bibnamefont {Guo}}, \bibinfo {author}
  {\bibfnamefont {F.}~\bibnamefont {Yao}},\ and\ \bibinfo {author}
  {\bibfnamefont {C.}~\bibnamefont {Yuan}},\ }\bibfield  {title} {\bibinfo
  {title} {B{,N}-co-doped graphene quantum dots as fluorescence sensor for
  detection of {Hg$^{2+}$} and {F$^{-}$} ions},\ }\href
  {https://doi.org/10.1039/C9AY00249A} {\bibfield  {journal} {\bibinfo
  {journal} {Anal. Methods}\ }\textbf {\bibinfo {volume} {11}},\ \bibinfo
  {pages} {1879} (\bibinfo {year} {2019})}\BibitemShut {NoStop}%
\bibitem [{\citenamefont {Budak}\ and\ \citenamefont
  {{\"U}nl{\"u}}(2021)}]{BUDAK-bn-doped-dots-synthesis-2021}%
  \BibitemOpen
  \bibfield  {author} {\bibinfo {author} {\bibfnamefont {E.}~\bibnamefont
  {Budak}}\ and\ \bibinfo {author} {\bibfnamefont {C.}~\bibnamefont
  {{\"U}nl{\"u}}},\ }\bibfield  {title} {\bibinfo {title} {Boron regulated dual
  emission in {B}, {N} doped graphene quantum dots},\ }\href
  {https://doi.org/https://doi.org/10.1016/j.optmat.2020.110577} {\bibfield
  {journal} {\bibinfo  {journal} {Optical Materials}\ }\textbf {\bibinfo
  {volume} {111}},\ \bibinfo {pages} {110577} (\bibinfo {year}
  {2021})}\BibitemShut {NoStop}%
\bibitem [{\citenamefont {Chen}\ \emph {et~al.}(2022)\citenamefont {Chen},
  \citenamefont {Tan},\ and\ \citenamefont
  {Yang}}]{multi-bn-doped-pi-conjugated-systems-review-2022}%
  \BibitemOpen
  \bibfield  {author} {\bibinfo {author} {\bibfnamefont {X.}~\bibnamefont
  {Chen}}, \bibinfo {author} {\bibfnamefont {D.}~\bibnamefont {Tan}},\ and\
  \bibinfo {author} {\bibfnamefont {D.-T.}\ \bibnamefont {Yang}},\ }\bibfield
  {title} {\bibinfo {title} {Multiple-boron-nitrogen (multi-{BN}) doped
  $\pi$-conjugated systems for optoelectronics},\ }\href
  {https://doi.org/10.1039/D2TC01106A} {\bibfield  {journal} {\bibinfo
  {journal} {J. Mater. Chem. C}\ }\textbf {\bibinfo {volume} {10}},\ \bibinfo
  {pages} {13499} (\bibinfo {year} {2022})}\BibitemShut {NoStop}%
\bibitem [{\citenamefont {S{\'{a}}nchez-S{\'{a}}nchez}\ \emph
  {et~al.}(2015)\citenamefont {S{\'{a}}nchez-S{\'{a}}nchez}, \citenamefont
  {Br{\"u}ller}, \citenamefont {Sachdev}, \citenamefont {M{\"u}llen},
  \citenamefont {Krieg}, \citenamefont {Bettinger}, \citenamefont
  {Nicola{\"i}}, \citenamefont {Meunier}, \citenamefont {Talirz}, \citenamefont
  {Fasel},\ and\ \citenamefont {Ruffieux}}]{doi:10.1021/acsnano.5b03895}%
  \BibitemOpen
  \bibfield  {author} {\bibinfo {author} {\bibfnamefont {C.}~\bibnamefont
  {S{\'{a}}nchez-S{\'{a}}nchez}}, \bibinfo {author} {\bibfnamefont
  {S.}~\bibnamefont {Br{\"u}ller}}, \bibinfo {author} {\bibfnamefont
  {H.}~\bibnamefont {Sachdev}}, \bibinfo {author} {\bibfnamefont
  {K.}~\bibnamefont {M{\"u}llen}}, \bibinfo {author} {\bibfnamefont
  {M.}~\bibnamefont {Krieg}}, \bibinfo {author} {\bibfnamefont {H.~F.}\
  \bibnamefont {Bettinger}}, \bibinfo {author} {\bibfnamefont {A.}~\bibnamefont
  {Nicola{\"i}}}, \bibinfo {author} {\bibfnamefont {V.}~\bibnamefont
  {Meunier}}, \bibinfo {author} {\bibfnamefont {L.}~\bibnamefont {Talirz}},
  \bibinfo {author} {\bibfnamefont {R.}~\bibnamefont {Fasel}},\ and\ \bibinfo
  {author} {\bibfnamefont {P.}~\bibnamefont {Ruffieux}},\ }\bibfield  {title}
  {\bibinfo {title} {On-surface synthesis of {BN}-substituted heteroaromatic
  networks},\ }\href {https://doi.org/10.1021/acsnano.5b03895} {\bibfield
  {journal} {\bibinfo  {journal} {ACS Nano}\ }\textbf {\bibinfo {volume} {9}},\
  \bibinfo {pages} {9228} (\bibinfo {year} {2015})}\BibitemShut {NoStop}%
\bibitem [{\citenamefont {Chen}\ \emph {et~al.}(2019)\citenamefont {Chen},
  \citenamefont {Guo}, \citenamefont {Zhu}, \citenamefont {Wang}, \citenamefont
  {Zhang},\ and\ \citenamefont {Qi}}]{doi:10.1021/acsami.9b10582}%
  \BibitemOpen
  \bibfield  {author} {\bibinfo {author} {\bibfnamefont {C.}~\bibnamefont
  {Chen}}, \bibinfo {author} {\bibfnamefont {K.}~\bibnamefont {Guo}}, \bibinfo
  {author} {\bibfnamefont {Y.}~\bibnamefont {Zhu}}, \bibinfo {author}
  {\bibfnamefont {F.}~\bibnamefont {Wang}}, \bibinfo {author} {\bibfnamefont
  {W.}~\bibnamefont {Zhang}},\ and\ \bibinfo {author} {\bibfnamefont
  {H.}~\bibnamefont {Qi}},\ }\bibfield  {title} {\bibinfo {title} {Construction
  of layered {B3N3}-doped graphene sheets from an acetylenic compound
  containing {B3N3} by a semisynthetic strategy},\ }\href
  {https://doi.org/10.1021/acsami.9b10582} {\bibfield  {journal} {\bibinfo
  {journal} {ACS Applied Materials \& Interfaces}\ }\textbf {\bibinfo {volume}
  {11}},\ \bibinfo {pages} {33245} (\bibinfo {year} {2019})}\BibitemShut
  {NoStop}%
\bibitem [{\citenamefont {Herrera-Reinoza}\ \emph {et~al.}(2021)\citenamefont
  {Herrera-Reinoza}, \citenamefont {dos Santos}, \citenamefont {de~Lima},
  \citenamefont {Landers},\ and\ \citenamefont
  {de~Siervo}}]{herrera2021atomically}%
  \BibitemOpen
  \bibfield  {author} {\bibinfo {author} {\bibfnamefont {N.}~\bibnamefont
  {Herrera-Reinoza}}, \bibinfo {author} {\bibfnamefont {A.~C.}\ \bibnamefont
  {dos Santos}}, \bibinfo {author} {\bibfnamefont {L.~H.}\ \bibnamefont
  {de~Lima}}, \bibinfo {author} {\bibfnamefont {R.}~\bibnamefont {Landers}},\
  and\ \bibinfo {author} {\bibfnamefont {A.}~\bibnamefont {de~Siervo}},\
  }\bibfield  {title} {\bibinfo {title} {Atomically precise bottom-up synthesis
  of {h-BNC}: graphene doped with {h-BN} nanoclusters},\ }\href@noop {}
  {\bibfield  {journal} {\bibinfo  {journal} {Chemistry of Materials}\ }\textbf
  {\bibinfo {volume} {33}},\ \bibinfo {pages} {2871} (\bibinfo {year}
  {2021})}\BibitemShut {NoStop}%
\bibitem [{\citenamefont {Caputo}\ \emph {et~al.}(2022)\citenamefont {Caputo},
  \citenamefont {Nguyen},\ and\ \citenamefont
  {Charlier}}]{PhysRevMaterials.6.114001}%
  \BibitemOpen
  \bibfield  {author} {\bibinfo {author} {\bibfnamefont {L.}~\bibnamefont
  {Caputo}}, \bibinfo {author} {\bibfnamefont {V.-H.}\ \bibnamefont {Nguyen}},\
  and\ \bibinfo {author} {\bibfnamefont {J.-C.}\ \bibnamefont {Charlier}},\
  }\bibfield  {title} {\bibinfo {title} {First-principles study of the
  structural and electronic properties of {BN}-ring doped graphene},\ }\href
  {https://doi.org/10.1103/PhysRevMaterials.6.114001} {\bibfield  {journal}
  {\bibinfo  {journal} {Phys. Rev. Mater.}\ }\textbf {\bibinfo {volume} {6}},\
  \bibinfo {pages} {114001} (\bibinfo {year} {2022})}\BibitemShut {NoStop}%
\bibitem [{\citenamefont {Nguyen}\ \emph {et~al.}(2021)\citenamefont {Nguyen},
  \citenamefont {Hoang}, \citenamefont {Phuc}, \citenamefont {Sin},\ and\
  \citenamefont {Nguyen}}]{doi:10.1021/acs.jpclett.1c01284}%
  \BibitemOpen
  \bibfield  {author} {\bibinfo {author} {\bibfnamefont {C.}~\bibnamefont
  {Nguyen}}, \bibinfo {author} {\bibfnamefont {N.~V.}\ \bibnamefont {Hoang}},
  \bibinfo {author} {\bibfnamefont {H.~V.}\ \bibnamefont {Phuc}}, \bibinfo
  {author} {\bibfnamefont {A.~Y.}\ \bibnamefont {Sin}},\ and\ \bibinfo {author}
  {\bibfnamefont {C.~V.}\ \bibnamefont {Nguyen}},\ }\bibfield  {title}
  {\bibinfo {title} {Two-dimensional boron phosphide/{MoGe2N4} van der {Waals}
  heterostructure: A promising tunable optoelectronic material},\ }\href@noop
  {} {\bibfield  {journal} {\bibinfo  {journal} {The Journal of Physical
  Chemistry Letters}\ }\textbf {\bibinfo {volume} {12}},\ \bibinfo {pages}
  {5076} (\bibinfo {year} {2021})}\BibitemShut {NoStop}%
\bibitem [{\citenamefont {Nguyen}\ \emph {et~al.}(2025)\citenamefont {Nguyen},
  \citenamefont {Truong}, \citenamefont {Phuc}, \citenamefont {Nguyen},
  \citenamefont {Hiep},\ and\ \citenamefont
  {Hieu}}]{doi:10.1021/acs.nanolett.5c02560}%
  \BibitemOpen
  \bibfield  {author} {\bibinfo {author} {\bibfnamefont {C.~V.}\ \bibnamefont
  {Nguyen}}, \bibinfo {author} {\bibfnamefont {P.~T.}\ \bibnamefont {Truong}},
  \bibinfo {author} {\bibfnamefont {H.~V.}\ \bibnamefont {Phuc}}, \bibinfo
  {author} {\bibfnamefont {C.~Q.}\ \bibnamefont {Nguyen}}, \bibinfo {author}
  {\bibfnamefont {N.~T.}\ \bibnamefont {Hiep}},\ and\ \bibinfo {author}
  {\bibfnamefont {N.~N.}\ \bibnamefont {Hieu}},\ }\bibfield  {title} {\bibinfo
  {title} {Rationally designed versatile heterostructures consisted of
  two-dimensional {Goldene and MXene Sc2CF2}},\ }\href@noop {} {\bibfield
  {journal} {\bibinfo  {journal} {Nano Letters}\ }\textbf {\bibinfo {volume}
  {25}},\ \bibinfo {pages} {10673} (\bibinfo {year} {2025})}\BibitemShut
  {NoStop}%
\bibitem [{\citenamefont {Truong}\ \emph {et~al.}(2025)\citenamefont {Truong},
  \citenamefont {Hieu}, \citenamefont {Nguyen}, \citenamefont {Nguyen},
  \citenamefont {Linh}, \citenamefont {Phuc},\ and\ \citenamefont
  {Nguyen}}]{linh2025electronic}%
  \BibitemOpen
  \bibfield  {author} {\bibinfo {author} {\bibfnamefont {P.~T.}\ \bibnamefont
  {Truong}}, \bibinfo {author} {\bibfnamefont {N.~N.}\ \bibnamefont {Hieu}},
  \bibinfo {author} {\bibfnamefont {H.~V.}\ \bibnamefont {Nguyen}}, \bibinfo
  {author} {\bibfnamefont {C.~Q.}\ \bibnamefont {Nguyen}}, \bibinfo {author}
  {\bibfnamefont {T.~P.~T.}\ \bibnamefont {Linh}}, \bibinfo {author}
  {\bibfnamefont {H.~V.}\ \bibnamefont {Phuc}},\ and\ \bibinfo {author}
  {\bibfnamefont {C.~V.}\ \bibnamefont {Nguyen}},\ }\bibfield  {title}
  {\bibinfo {title} {Electronic properties and interfacial engineering of
  metal-semiconductor {1T-, 2H-Ta2B MBene/Janus MoSSe} heterostructures},\
  }\href@noop {} {\bibfield  {journal} {\bibinfo  {journal} {Nanoscale}\
  }\textbf {\bibinfo {volume} {17}},\ \bibinfo {pages} {17182} (\bibinfo {year}
  {2025})}\BibitemShut {NoStop}%
\bibitem [{\citenamefont {Frisch}\ \emph {et~al.}(2016)\citenamefont {Frisch},
  \citenamefont {Trucks}, \citenamefont {Schlegel}, \citenamefont {Scuseria},
  \citenamefont {Robb}, \citenamefont {Cheeseman}, \citenamefont {Scalmani},
  \citenamefont {Barone}, \citenamefont {Petersson} \emph {et~al.}}]{g16}%
  \BibitemOpen
  \bibfield  {author} {\bibinfo {author} {\bibfnamefont {M.~J.}\ \bibnamefont
  {Frisch}}, \bibinfo {author} {\bibfnamefont {G.~W.}\ \bibnamefont {Trucks}},
  \bibinfo {author} {\bibfnamefont {H.~B.}\ \bibnamefont {Schlegel}}, \bibinfo
  {author} {\bibfnamefont {G.~E.}\ \bibnamefont {Scuseria}}, \bibinfo {author}
  {\bibfnamefont {M.~A.}\ \bibnamefont {Robb}}, \bibinfo {author}
  {\bibfnamefont {J.~R.}\ \bibnamefont {Cheeseman}}, \bibinfo {author}
  {\bibfnamefont {G.}~\bibnamefont {Scalmani}}, \bibinfo {author}
  {\bibfnamefont {V.}~\bibnamefont {Barone}}, \bibinfo {author} {\bibfnamefont
  {G.~A.}\ \bibnamefont {Petersson}}, \emph {et~al.},\ }\href@noop {} {\bibinfo
  {title} {{Gaussian 16, Revision C.01}}} (\bibinfo {year} {2016}),\ \bibinfo
  {note} {{Gaussian, Inc., Wallingford CT}}\BibitemShut {NoStop}%
\bibitem [{\citenamefont {Lee}\ \emph {et~al.}(1988)\citenamefont {Lee},
  \citenamefont {Yang},\ and\ \citenamefont {Parr}}]{lee1988development}%
  \BibitemOpen
  \bibfield  {author} {\bibinfo {author} {\bibfnamefont {C.}~\bibnamefont
  {Lee}}, \bibinfo {author} {\bibfnamefont {W.}~\bibnamefont {Yang}},\ and\
  \bibinfo {author} {\bibfnamefont {R.~G.}\ \bibnamefont {Parr}},\ }\bibfield
  {title} {\bibinfo {title} {{Development of the Colle-Salvetti
  correlation-energy formula into a functional of the electron density}},\
  }\href@noop {} {\bibfield  {journal} {\bibinfo  {journal} {Physical review
  B}\ }\textbf {\bibinfo {volume} {37}},\ \bibinfo {pages} {785} (\bibinfo
  {year} {1988})}\BibitemShut {NoStop}%
\bibitem [{\citenamefont {Becke}(1992)}]{becke1992density}%
  \BibitemOpen
  \bibfield  {author} {\bibinfo {author} {\bibfnamefont {A.~D.}\ \bibnamefont
  {Becke}},\ }\bibfield  {title} {\bibinfo {title} {{Density-functional
  thermochemistry. I. The effect of the exchange-only gradient correction}},\
  }\href@noop {} {\bibfield  {journal} {\bibinfo  {journal} {The Journal of
  Chemical Physics}\ }\textbf {\bibinfo {volume} {96}},\ \bibinfo {pages}
  {2155} (\bibinfo {year} {1992})}\BibitemShut {NoStop}%
\bibitem [{\citenamefont {Becke}(1993)}]{becke1993density}%
  \BibitemOpen
  \bibfield  {author} {\bibinfo {author} {\bibfnamefont {A.}~\bibnamefont
  {Becke}},\ }\href@noop {} {\bibinfo {title} {{Density-Functional
  Thermochemistry. III. The Role of Exact Exchange. J. Chem. Phys., 98:
  5648-5652}}} (\bibinfo {year} {1993})\BibitemShut {NoStop}%
\bibitem [{\citenamefont {Runge}\ and\ \citenamefont
  {Gross}(1984)}]{runge-gross-tddft-1984}%
  \BibitemOpen
  \bibfield  {author} {\bibinfo {author} {\bibfnamefont {E.}~\bibnamefont
  {Runge}}\ and\ \bibinfo {author} {\bibfnamefont {E.~K.~U.}\ \bibnamefont
  {Gross}},\ }\bibfield  {title} {\bibinfo {title} {Density-functional theory
  for time-dependent systems},\ }\href
  {https://doi.org/10.1103/PhysRevLett.52.997} {\bibfield  {journal} {\bibinfo
  {journal} {Phys. Rev. Lett.}\ }\textbf {\bibinfo {volume} {52}},\ \bibinfo
  {pages} {997} (\bibinfo {year} {1984})}\BibitemShut {NoStop}%
\bibitem [{\citenamefont {Marques}\ and\ \citenamefont
  {Gross}(2004)}]{marques2004time}%
  \BibitemOpen
  \bibfield  {author} {\bibinfo {author} {\bibfnamefont {M.}~\bibnamefont
  {Marques}}\ and\ \bibinfo {author} {\bibfnamefont {E.}~\bibnamefont
  {Gross}},\ }\bibfield  {title} {\bibinfo {title} {Time-dependent density
  functional theory},\ }\href@noop {} {\bibfield  {journal} {\bibinfo
  {journal} {Annual Review of Physical Chemistry}\ }\textbf {\bibinfo {volume}
  {55}},\ \bibinfo {pages} {427} (\bibinfo {year} {2004})}\BibitemShut
  {NoStop}%
\bibitem [{\citenamefont {Hehre}\ \emph {et~al.}(1972)\citenamefont {Hehre},
  \citenamefont {Ditchfield},\ and\ \citenamefont {Pople}}]{hehre1972self}%
  \BibitemOpen
  \bibfield  {author} {\bibinfo {author} {\bibfnamefont {W.~J.}\ \bibnamefont
  {Hehre}}, \bibinfo {author} {\bibfnamefont {R.}~\bibnamefont {Ditchfield}},\
  and\ \bibinfo {author} {\bibfnamefont {J.~A.}\ \bibnamefont {Pople}},\
  }\bibfield  {title} {\bibinfo {title} {{Self-consistent molecular orbital
  methods. XII. Further extensions of Gaussian-type basis sets for use in
  molecular orbital studies of organic molecules}},\ }\href@noop {} {\bibfield
  {journal} {\bibinfo  {journal} {The Journal of Chemical Physics}\ }\textbf
  {\bibinfo {volume} {56}},\ \bibinfo {pages} {2257} (\bibinfo {year}
  {1972})}\BibitemShut {NoStop}%
\bibitem [{\citenamefont {Hariharan}\ and\ \citenamefont
  {Pople}(1973)}]{hariharan1973influence}%
  \BibitemOpen
  \bibfield  {author} {\bibinfo {author} {\bibfnamefont {P.~C.}\ \bibnamefont
  {Hariharan}}\ and\ \bibinfo {author} {\bibfnamefont {J.~A.}\ \bibnamefont
  {Pople}},\ }\bibfield  {title} {\bibinfo {title} {The influence of
  polarization functions on molecular orbital hydrogenation energies},\
  }\href@noop {} {\bibfield  {journal} {\bibinfo  {journal} {Theoretica Chimica
  Acta}\ }\textbf {\bibinfo {volume} {28}},\ \bibinfo {pages} {213} (\bibinfo
  {year} {1973})}\BibitemShut {NoStop}%
\bibitem [{\citenamefont {Kohn}\ and\ \citenamefont
  {Sham}(1965)}]{kohn1965self}%
  \BibitemOpen
  \bibfield  {author} {\bibinfo {author} {\bibfnamefont {W.}~\bibnamefont
  {Kohn}}\ and\ \bibinfo {author} {\bibfnamefont {L.~J.}\ \bibnamefont
  {Sham}},\ }\bibfield  {title} {\bibinfo {title} {Self-consistent equations
  including exchange and correlation effects},\ }\href@noop {} {\bibfield
  {journal} {\bibinfo  {journal} {Physical Review}\ }\textbf {\bibinfo {volume}
  {140}},\ \bibinfo {pages} {A1133} (\bibinfo {year} {1965})}\BibitemShut
  {NoStop}%
\bibitem [{\citenamefont {Bauernschmitt}\ and\ \citenamefont
  {Ahlrichs}(1996)}]{BAUERNSCHMITT1996454}%
  \BibitemOpen
  \bibfield  {author} {\bibinfo {author} {\bibfnamefont {R.}~\bibnamefont
  {Bauernschmitt}}\ and\ \bibinfo {author} {\bibfnamefont {R.}~\bibnamefont
  {Ahlrichs}},\ }\bibfield  {title} {\bibinfo {title} {Treatment of electronic
  excitations within the adiabatic approximation of time dependent density
  functional theory},\ }\href@noop {} {\bibfield  {journal} {\bibinfo
  {journal} {Chemical Physics Letters}\ }\textbf {\bibinfo {volume} {256}},\
  \bibinfo {pages} {454} (\bibinfo {year} {1996})}\BibitemShut {NoStop}%
\bibitem [{\citenamefont {Scalmani}\ \emph {et~al.}(2006)\citenamefont
  {Scalmani}, \citenamefont {Frisch}, \citenamefont {Mennucci}, \citenamefont
  {Tomasi}, \citenamefont {Cammi},\ and\ \citenamefont
  {Barone}}]{scalmani2006geometries}%
  \BibitemOpen
  \bibfield  {author} {\bibinfo {author} {\bibfnamefont {G.}~\bibnamefont
  {Scalmani}}, \bibinfo {author} {\bibfnamefont {M.~J.}\ \bibnamefont
  {Frisch}}, \bibinfo {author} {\bibfnamefont {B.}~\bibnamefont {Mennucci}},
  \bibinfo {author} {\bibfnamefont {J.}~\bibnamefont {Tomasi}}, \bibinfo
  {author} {\bibfnamefont {R.}~\bibnamefont {Cammi}},\ and\ \bibinfo {author}
  {\bibfnamefont {V.}~\bibnamefont {Barone}},\ }\bibfield  {title} {\bibinfo
  {title} {Geometries and properties of excited states in the gas phase and in
  solution: Theory and application of a time-dependent density functional
  theory polarizable continuum model},\ }\href@noop {} {\bibfield  {journal}
  {\bibinfo  {journal} {The Journal of Chemical Physics}\ }\textbf {\bibinfo
  {volume} {124}} (\bibinfo {year} {2006})}\BibitemShut {NoStop}%
\bibitem [{\citenamefont {Furche}\ and\ \citenamefont
  {Ahlrichs}(2002)}]{furche2002adiabatic}%
  \BibitemOpen
  \bibfield  {author} {\bibinfo {author} {\bibfnamefont {F.}~\bibnamefont
  {Furche}}\ and\ \bibinfo {author} {\bibfnamefont {R.}~\bibnamefont
  {Ahlrichs}},\ }\bibfield  {title} {\bibinfo {title} {Adiabatic time-dependent
  density functional methods for excited state properties},\ }\href@noop {}
  {\bibfield  {journal} {\bibinfo  {journal} {The Journal of Chemical Physics}\
  }\textbf {\bibinfo {volume} {117}},\ \bibinfo {pages} {7433} (\bibinfo {year}
  {2002})}\BibitemShut {NoStop}%
\bibitem [{VAN(1999)}]{VANCAILLIE1999249}%
  \BibitemOpen
  \bibfield  {title} {\bibinfo {title} {Geometric derivatives of excitation
  energies using {SCF} and {DFT}},\ }\href@noop {} {\bibfield  {journal}
  {\bibinfo  {journal} {Chemical Physics Letters}\ }\textbf {\bibinfo {volume}
  {308}},\ \bibinfo {pages} {249} (\bibinfo {year} {1999})}\BibitemShut
  {NoStop}%
\bibitem [{VAN(2000)}]{VANCAILLIE2000159}%
  \BibitemOpen
  \bibfield  {title} {\bibinfo {title} {Geometric derivatives of density
  functional theory excitation energies using gradient-corrected functionals},\
  }\href@noop {} {\bibfield  {journal} {\bibinfo  {journal} {Chemical Physics
  Letters}\ }\textbf {\bibinfo {volume} {317}},\ \bibinfo {pages} {159}
  (\bibinfo {year} {2000})}\BibitemShut {NoStop}%
\bibitem [{\citenamefont {Stratmann}\ \emph {et~al.}(1998)\citenamefont
  {Stratmann}, \citenamefont {Scuseria},\ and\ \citenamefont
  {Frisch}}]{stratmann1998efficient}%
  \BibitemOpen
  \bibfield  {author} {\bibinfo {author} {\bibfnamefont {R.~E.}\ \bibnamefont
  {Stratmann}}, \bibinfo {author} {\bibfnamefont {G.~E.}\ \bibnamefont
  {Scuseria}},\ and\ \bibinfo {author} {\bibfnamefont {M.~J.}\ \bibnamefont
  {Frisch}},\ }\bibfield  {title} {\bibinfo {title} {An efficient
  implementation of time-dependent density-functional theory for the
  calculation of excitation energies of large molecules},\ }\href@noop {}
  {\bibfield  {journal} {\bibinfo  {journal} {The Journal of Chemical Physics}\
  }\textbf {\bibinfo {volume} {109}},\ \bibinfo {pages} {8218} (\bibinfo {year}
  {1998})}\BibitemShut {NoStop}%
\bibitem [{\citenamefont {Casida}\ \emph {et~al.}(1998)\citenamefont {Casida},
  \citenamefont {Jamorski}, \citenamefont {Casida},\ and\ \citenamefont
  {Salahub}}]{casida1998molecular}%
  \BibitemOpen
  \bibfield  {author} {\bibinfo {author} {\bibfnamefont {M.~E.}\ \bibnamefont
  {Casida}}, \bibinfo {author} {\bibfnamefont {C.}~\bibnamefont {Jamorski}},
  \bibinfo {author} {\bibfnamefont {K.~C.}\ \bibnamefont {Casida}},\ and\
  \bibinfo {author} {\bibfnamefont {D.~R.}\ \bibnamefont {Salahub}},\
  }\bibfield  {title} {\bibinfo {title} {Molecular excitation energies to
  high-lying bound states from time-dependent density-functional response
  theory: Characterization and correction of the time-dependent local density
  approximation ionization threshold},\ }\href@noop {} {\bibfield  {journal}
  {\bibinfo  {journal} {The Journal of Chemical Physics}\ }\textbf {\bibinfo
  {volume} {108}},\ \bibinfo {pages} {4439} (\bibinfo {year}
  {1998})}\BibitemShut {NoStop}%
\bibitem [{\citenamefont {Das}\ and\ \citenamefont
  {Shukla}(2024)}]{samayita-jpca}%
  \BibitemOpen
  \bibfield  {author} {\bibinfo {author} {\bibfnamefont {S.}~\bibnamefont
  {Das}}\ and\ \bibinfo {author} {\bibfnamefont {A.}~\bibnamefont {Shukla}},\
  }\bibfield  {title} {\bibinfo {title} {{DFT} and model hamiltonian study of
  optoelectronic properties of some low-symmetry graphene quantum dots},\
  }\href {https://doi.org/10.1021/acs.jpca.4c03439} {\bibfield  {journal}
  {\bibinfo  {journal} {The Journal of Physical Chemistry A}\ }\textbf
  {\bibinfo {volume} {128}},\ \bibinfo {pages} {10099} (\bibinfo {year}
  {2024})}\BibitemShut {NoStop}%
\bibitem [{sup()}]{supplemental}%
  \BibitemOpen
  \href@noop {} {}\bibinfo {note} {See Supplemental Material for optimized
  geometry parameters, information about the excited states contributing to the
  peaks in the calculated absorption spectra, orbital projected DOS plots,
  comparison of absorption spectra computed using the B3LYP and HSE06
  functionals, UV-Vis absorption spectra computed using different number of
  excited states and UV-Vis spectra computed for various doping models compared
  to that of pristine GQD}\BibitemShut {NoStop}%
\bibitem [{\citenamefont {Lu}\ and\ \citenamefont
  {Chen}(2012)}]{lu2012multiwfn}%
  \BibitemOpen
  \bibfield  {author} {\bibinfo {author} {\bibfnamefont {T.}~\bibnamefont
  {Lu}}\ and\ \bibinfo {author} {\bibfnamefont {F.}~\bibnamefont {Chen}},\
  }\bibfield  {title} {\bibinfo {title} {Multiwfn: A multifunctional
  wavefunction analyzer},\ }\href@noop {} {\bibfield  {journal} {\bibinfo
  {journal} {Journal of Computational Chemistry}\ }\textbf {\bibinfo {volume}
  {33}},\ \bibinfo {pages} {580} (\bibinfo {year} {2012})}\BibitemShut
  {NoStop}%
\bibitem [{\citenamefont {Clar}\ and\ \citenamefont
  {Schmidt}(1978)}]{CLAR19783219}%
  \BibitemOpen
  \bibfield  {author} {\bibinfo {author} {\bibfnamefont {E.}~\bibnamefont
  {Clar}}\ and\ \bibinfo {author} {\bibfnamefont {W.}~\bibnamefont {Schmidt}},\
  }\bibfield  {title} {\bibinfo {title} {Correlations between photoelectron and
  ultraviolet absorption spectra of polycyclic hydrocarbons: The terrylene and
  peropyrene series},\ }\href
  {https://doi.org/https://doi.org/10.1016/0040-4020(78)87020-3} {\bibfield
  {journal} {\bibinfo  {journal} {Tetrahedron}\ }\textbf {\bibinfo {volume}
  {34}},\ \bibinfo {pages} {3219} (\bibinfo {year} {1978})}\BibitemShut
  {NoStop}%
\bibitem [{\citenamefont {Basak}\ \emph {et~al.}(2015)\citenamefont {Basak},
  \citenamefont {Chakraborty},\ and\ \citenamefont
  {Shukla}}]{PhysRevB.92.205404}%
  \BibitemOpen
  \bibfield  {author} {\bibinfo {author} {\bibfnamefont {T.}~\bibnamefont
  {Basak}}, \bibinfo {author} {\bibfnamefont {H.}~\bibnamefont {Chakraborty}},\
  and\ \bibinfo {author} {\bibfnamefont {A.}~\bibnamefont {Shukla}},\
  }\bibfield  {title} {\bibinfo {title} {Theory of linear optical absorption in
  diamond-shaped graphene quantum dots},\ }\href
  {https://doi.org/10.1103/PhysRevB.92.205404} {\bibfield  {journal} {\bibinfo
  {journal} {Phys. Rev. B}\ }\textbf {\bibinfo {volume} {92}},\ \bibinfo
  {pages} {205404} (\bibinfo {year} {2015})}\BibitemShut {NoStop}%
\end{thebibliography}
\end{document}


\title{Supplemental Material for ``Tuning the optoelectronic properties
of graphene quantum dots by BN-ring doping: A density functional theory
study''}
\author{Samayita Das}
\email{samayitadas5@gmail.com}

\address{Department of Physics, Indian Institute of Technology Bombay, Powai,
Mumbai 400076, India}
\author{Alok Shukla}
\email{shukla@iitb.ac.in}

\address{Department of Physics, Indian Institute of Technology Bombay, Powai,
Mumbai 400076, India}

\maketitle
In this Supplemental Material (SM), Tables S1 and S2 represent the
optimized bond lengths and bond angles, respectively, showing the
minima and maxima for all the doped structures considered.

Next, Tables S3$-$S17 contain the information regarding the excited
states contributing to important peaks in the calculated absorption
spectra of the pristine and the BN-GQDs considered in this work. This
includes peak positions E, symmetries of the excited states contributing
to various peaks, oscillator strengths $f$, polarization direction
of the absorbed photon indicated by the subscripts in the 'Peak' column,
and the many-body wave functions of the excited states. Here, $f$
(= $\frac{2\times m_{e}}{3\times\hbar^{2}}(E)\sum_{j=x,y,z}|\langle e|O_{j}|r\rangle|^{2}$)
indicates the oscillator strength for a particular transition where
$E$, $|e\rangle$, $|r\rangle$, and $O_{j}$ are representing the
peak energy, excited states of the corresponding peak, reference state,
and the Cartesian components of the electric dipole operator, respectively.
In the last column, the numbers inside the parentheses represent the
contributing coefficients of the wave functions for each corresponding
configuration.

This SM also contains several figures depicting: the total and orbital
projected density of states (Fig. \ref{fig:orb-pdos}), UV-Vis absorption
spectra computed using the B3LYP and HSE06 functionals (Fig. \ref{fig:Comparison-of-TDDFTfunctional}),
UV-Vis absorption spectra computed using different numbers of excited
states (Fig. \ref{fig:(a)-pristine-GQDdifferentstates=000020}), and
UV-Vis absorption spectra computed for various doping models compared
to that of the pristine GQD (Figs. \ref{fig:Variation-of-UV-Vis(1-4)}$-$\ref{fig:Variation-of-UV-Vis(9-11)}).

\subsection{Optimized Geometry Parameters}

\begin{table}[H]
\caption{Optimized bond lengths for the fourteen doped structures. \protect\label{tab:optimized_parameters}}

\centering{}%
\begin{tabular}{|cccccccc|}
\hline 
BN-ring doped & $\text{(C \ensuremath{-} C)\textrm{\ensuremath{\mathring{A}}}}$ & $\text{(B \ensuremath{-} N\textrm{)\ensuremath{\mathring{A}}}}$ & $\text{(C \ensuremath{-} B\textrm{)\ensuremath{\mathring{A}}}}$ & $\text{(C \ensuremath{-} N\textrm{)\ensuremath{\mathring{A}}}}$ & $\text{(C \ensuremath{-} H\textrm{)\ensuremath{\mathring{A}}}}$ & $\text{(N \ensuremath{-} H\textrm{)\ensuremath{\mathring{A}}}}$ & $\text{(B \ensuremath{-} H\textrm{)\ensuremath{\mathring{A}}}}$\tabularnewline
structures & (Min - Max) & (Min - Max) & (Min - Max) & (Min - Max) & (Min - Max) & (Min - Max) & (Min - Max)\tabularnewline
\hline 
\hline 
Model 1 & $1.399-1.440$ & $1.437-1.450$ & $1.497-1.522$ & $1.396-1.410$ & $1.085-1.091$ & $-$ & $-$\tabularnewline
\hline 
Model 2 & $1.379-1.462$ & $1.409-1.426$ & $1.517$ & $1.396$ & $1.086-1.090$ & $1.011$ & $1.194$\tabularnewline
\hline 
Model 3 & $1.357-1.445$ & $1.399-1.470$ & $1.484-1.499$ & $1.357-1.401$ & $1.086-1.089$ & $1.012$ & $1.193$\tabularnewline
\hline 
Model 4 & $1.375-1.447$ & $1.425-1.457$ & $1.491-1.516$ & $1.384-1.410$ & $1.085-1.090$ & $-$ & $1.192$\tabularnewline
\hline 
Model 5 & $1.354-1.445$ & $1.404-1.476$ & $1.491-1.516$ & $1.383-1.399$ & $1.085-1.090$ & $1.011$ & $1.191-1.193$\tabularnewline
\hline 
Model 6 & $1.353-1.427$ & $1.430-1.450$ & $1.499-1.532$ & $1.385-1.410$ & $1.084-1.093$ & $-$ & $1.192$\tabularnewline
\hline 
Model 7 & $1.357-1.436$ & $1.405-1.451$ & $1.496-1.497$ & $1.369-1.411$ & $1.083-1.093$ & $-$ & $1.192-1.193$\tabularnewline
\hline 
Model 8 & $1.363-1.442$ & $1.396-1.483$ & $1.494-1.525$ & $1.345-1.411$ & $1.085-1.090$ & $1.013$ & $1.193$\tabularnewline
\hline 
Model 9 ($\uparrow-\uparrow$) & $1.354-1.457$ & $1.426-1.471$ & $1.499-1.519$ & $1.391$ & $1.086-1.090$ & $1.011$ & $1.192-1.195$\tabularnewline
\hline 
Model 9 ($\uparrow-\downarrow$) & $1.345-1.467$ & $1.425-1.463$ & $1.515-1.516$ & $1.404$ & $1.087-1.089$ & $1.011$ & $1.194$\tabularnewline
\hline 
Model 10 ($\uparrow-\uparrow$) & $1.390-1.460$ & $1.399-1.472$ & $1.480-1.482$ & $1.361-1.408$ & $1.086-1.089$ & $1.012$ & $1.193$\tabularnewline
\hline 
Model 10 ($\uparrow-\downarrow$) & $1.392-1.456$ & $1.399-1475$ & $1.491-1.524$ & $1.354-1.396$ & $1.086-1.089$ & $1.012$ & $1.193$\tabularnewline
\hline 
Model 11 ($\uparrow-\uparrow$) & $1.359-1.453$ & $1.414-1.488$ & $1.459-1.533$ & $1.371-1.414$ & $1.085-1.090$ & $1.012$ & $1.191$\tabularnewline
\hline 
Model 11 ($\uparrow-\downarrow$) & $1.357-1.455$ & $1.419-1.474$ & $1.514-1.539$ & $1.378-1399$ & $1.085-1.090$ & $1.012$ & $-$\tabularnewline
\hline 
\end{tabular}
\end{table}

\begin{table}[H]
\caption{Optimized bond angles for the fourteen doped structures. \protect\label{tab:optimized_parameters-1}}

\centering{}%
\begin{tabular}{|cccccccc|}
\hline 
{\footnotesize BN-ring doped} & {\footnotesize$\angle C-B-N$ $({^\circ})$} & {\footnotesize$\angle B-N-B$ $({^\circ})$} & {\footnotesize$\angle N-B-N$ $({^\circ})$} & {\footnotesize$\angle C-C-C$ $({^\circ})$} & {\footnotesize$\angle C-N-B$ $({^\circ})$} & {\footnotesize$\angle C-C-B$ $({^\circ})$} & {\footnotesize$\angle C-C-N$ $({^\circ})$}\tabularnewline
{\footnotesize structures} & {\footnotesize (Min - Max)} & {\footnotesize (Min - Max)} & {\footnotesize (Min - Max)} & {\footnotesize (Min - Max)} & {\footnotesize (Min - Max)} & {\footnotesize (Min - Max)} & {\footnotesize (Min - Max)}\tabularnewline
\hline 
\hline 
{\footnotesize Model 1} & {\footnotesize$120.05-120.42$} & {\footnotesize$120.43-120.44$} & {\footnotesize$119.39-119.91$} & {\footnotesize$119.79-121.80$} & {\footnotesize$118.91-120.6$} & {\footnotesize$117.04-118.78$} & {\footnotesize$119.91-120.71$}\tabularnewline
\hline 
{\footnotesize Model 2} & {\footnotesize$118$} & {\footnotesize$118.7-122$} & {\footnotesize$118.44$} & {\footnotesize$118.6-122$} & {\footnotesize$120.6$} & {\footnotesize$119.6$} & {\footnotesize$120.24$}\tabularnewline
\hline 
{\footnotesize Model 3} & {\footnotesize$118.35-120.31$} & {\footnotesize$119.42-119.68$} & {\footnotesize$117-121.8$} & {\footnotesize$119.47-121.26$} & {\footnotesize$119.07-119.17$} & {\footnotesize$119.13-119.83$} & {\footnotesize$119.64-121$}\tabularnewline
\hline 
{\footnotesize Model 4} & {\footnotesize$119.44-120.23$} & {\footnotesize$119.29-121.48$} & {\footnotesize$118-120.3$} & {\footnotesize$118.12-120.3$} & {\footnotesize$117.5-120.47$} & {\footnotesize$117.15-120$} & {\footnotesize$118.86-122$}\tabularnewline
\hline 
{\footnotesize Model 5} & {\footnotesize$118-120.48$} & {\footnotesize$119.34-121.6$} & {\footnotesize$118-121.48$} & {\footnotesize$118.74-121.6$} & {\footnotesize$119.51-120.22$} & {\footnotesize$117.47-119.50$} & {\footnotesize$118.98-120.42$}\tabularnewline
\hline 
{\footnotesize Model 6} & {\footnotesize$119.24-121.44$} & {\footnotesize$119.27-121.95$} & {\footnotesize$119-120.74$} & {\footnotesize$119-121.69$} & {\footnotesize$118.92-121.6$} & {\footnotesize$118.60-119.86$} & {\footnotesize$118.1-122$}\tabularnewline
\hline 
{\footnotesize Model 7} & {\footnotesize$120.3-120.99$} & {\footnotesize$119.85-121.38$} & {\footnotesize$118.77-119.75$} & {\footnotesize$119.42-121.34$} & {\footnotesize$117.86-118$} & {\footnotesize$119.54-120.95$} & {\footnotesize$120.52-122$}\tabularnewline
\hline 
{\footnotesize Model 8} & {\footnotesize$119.59-120.32$} & {\footnotesize$118.8-120.89$} & {\footnotesize$119.1-121.14$} & {\footnotesize$119.43-120.27$} & {\footnotesize$118.94-121.13$} & {\footnotesize$118-121.15$} & {\footnotesize$118.22-120.90$}\tabularnewline
\hline 
{\footnotesize Model 9 ($\uparrow-\uparrow$)} & {\footnotesize$118.7-119.6$} & {\footnotesize$119.27-120$} & {\footnotesize$181-120.87$} & {\footnotesize$119.5-121$} & {\footnotesize$118-120.36$} & {\footnotesize$119.5-120$} & {\footnotesize$119.96-120.01$}\tabularnewline
\hline 
{\footnotesize Model 9 ($\uparrow-\downarrow$)} & {\footnotesize$118.3$} & {\footnotesize$118.8-122$} & {\footnotesize$118.96-119.02$} & {\footnotesize$119.45-121$} & {\footnotesize$120.91-120.96$} & {\footnotesize$119.72$} & {\footnotesize$120.52$}\tabularnewline
\hline 
{\footnotesize Model 10 ($\uparrow-\uparrow$)} & {\footnotesize$118.6-119$} & {\footnotesize$119.26-122.3$} & {\footnotesize$117-122$} & {\footnotesize$119-120.7$} & {\footnotesize$118.8-120.74$} & {\footnotesize$119-121.5$} & {\footnotesize$120.1-122$}\tabularnewline
\hline 
{\footnotesize Model 10 ($\uparrow-\downarrow$)} & {\footnotesize$118-121.6$} & {\footnotesize$120.15-121.99$} & {\footnotesize$118-119.63$} & {\footnotesize$119.6-121.13$} & {\footnotesize$119.38-120.42$} & {\footnotesize$119.34-119.7$} & {\footnotesize$120.95-122$}\tabularnewline
\hline 
{\footnotesize Model 11 ($\uparrow-\uparrow$)} & {\footnotesize$118.79-120.37$} & {\footnotesize$118-122$} & {\footnotesize$118-122.03$} & {\footnotesize$119.98-121.85$} & {\footnotesize$118.4-121.68$} & {\footnotesize$118-121.65$} & {\footnotesize$119.1-122$}\tabularnewline
\hline 
{\footnotesize Model 11 ($\uparrow-\downarrow$)} & {\footnotesize$118.1-120.25$} & {\footnotesize$120.34-122$} & {\footnotesize$117-120.6$} & {\footnotesize$119-121.53$} & {\footnotesize$119.32-120.06$} & {\footnotesize$118-119.1$} & {\footnotesize$119.96-120.98$}\tabularnewline
\hline 
\end{tabular}
\end{table}

\vspace{1cm}

\subsection{Data from the TDDFT calculations of the UV-Vis absorption spectra}

The wave function analysis of the different peaks in the UV-Vis optical
absorption spectra, obtained by TDDFT calculations using B3LYP functional
for all the structures (pristine and doped), is presented in the following
tables.

\subsubsection{Pristine diamond-shaped GQD}

\begin{table}[H]
\caption{Information about the excited states contributing to the peaks in
the linear optical absorption spectrum of the pristine GQD of $D_{2h}$
symmetry {[}Fig. 6, main article{]}, computed using the first-principles
TDDFT method, and the B3LYP functional.}

\centering{}%
\begin{tabular}{ccccccccc}
\hline 
Peak & \hspace{1.7cm} & Symmetry & \hspace{1.7cm} & E (eV) & \hspace{1.7cm} & $f$ & \hspace{1.7cm} & Wave function\tabularnewline
\hline 
\hline 
I\textsubscript{$y$} &  & \textsuperscript{1}B\textsubscript{2u} &  & 2.31 &  & 0.578 &  & $\arrowvert H\rightarrow L\rangle(0.69200)$\tabularnewline
II\textsubscript{$x$} &  & \textsuperscript{1}B\textsubscript{3u} &  & 3.46 &  & 0.220 &  & $\arrowvert H-2\rightarrow L\rangle(0.53469)$\tabularnewline
III\textsubscript{$y$} &  & \textsuperscript{1}B\textsubscript{2u} &  & 4.08 &  & 0.193 &  & $\arrowvert H-1\rightarrow L+1\rangle(0.58160)$\tabularnewline
IV\textsubscript{$y$} &  & \textsuperscript{1}B\textsubscript{2u} &  & 4.90 &  & 0.968 &  & $\arrowvert H-2\rightarrow L+2\rangle(0.44856)$\tabularnewline
 &  & \textsuperscript{1}B\textsubscript{2u} &  & 4.93 &  & 0.950 &  & $\arrowvert H-2\rightarrow L+2\rangle(0.44827)$\tabularnewline
\hline 
\end{tabular}
\end{table}

\subsubsection{Doping with single BN-ring}

\begin{table}[H]
\caption{Information about the excited states contributing to the peaks in
the linear optical absorption spectrum of model 1 of $C_{2v}$ symmetry
{[}Fig. 7(a), main article{]}, computed using the first principles
TDDFT method, and the B3LYP functional.}

\centering{}%
\begin{tabular}{ccccccccc}
\hline 
Peak & \hspace{1.7cm} & Symmetry & \hspace{1.7cm} & E (eV) & \hspace{1.7cm} & $f$ & \hspace{1.7cm} & Wave function\tabularnewline
\hline 
\hline 
I\textsubscript{$y$} &  & \textsuperscript{1}B\textsubscript{2} &  & 2.52 &  & 0.427 &  & $\arrowvert H\rightarrow L\rangle(0.66395)$\tabularnewline
II\textsubscript{$y$} &  & \textsuperscript{1}B\textsubscript{2} &  & 3.55 &  & 0.748 &  & $\arrowvert H-1\rightarrow L+1\rangle(0.64281)$\tabularnewline
III\textsubscript{$x$} &  & \textsuperscript{1}A\textsubscript{1} &  & 3.83 &  & 0.099 &  & $\arrowvert H-3\rightarrow L\rangle(0.53543)$\tabularnewline
IV\textsubscript{$y$} &  & \textsuperscript{1}B\textsubscript{2} &  & 4.13 &  & 0.017 &  & $\arrowvert H-1\rightarrow L+2\rangle(0.59847)$\tabularnewline
V\textsubscript{$x$} &  & \textsuperscript{1}A\textsubscript{1} &  & 4.56 &  & 0.497 &  & $\arrowvert H-1\rightarrow L+3\rangle(0.44547)$\tabularnewline
VI\textsubscript{$xy$} &  & \textsuperscript{1}B\textsubscript{2} &  & 4.82 &  & 0.059 &  & $\arrowvert H-1\rightarrow L+4\rangle(0.59345)$\tabularnewline
 &  & \textsuperscript{1}A\textsubscript{1} &  & 4.84 &  & 0.074 &  & $\arrowvert H-1\rightarrow L+7\rangle(0.63750)$\tabularnewline
\hline 
\end{tabular}
\end{table}

\begin{table}[H]
\caption{Information about the excited states contributing to the peaks in
the linear optical absorption spectrum of model 2 of $C_{2v}$ symmetry
{[}Fig. 7(b), main article{]}, computed using the first-principles
TDDFT method, and the B3LYP functional.}

\centering{}%
\begin{tabular}{ccccccccc}
\hline 
Peak & \hspace{1.7cm} & Symmetry & \hspace{1.7cm} & E (eV) & \hspace{1.7cm} & $f$ & \hspace{1.7cm} & Wave function\tabularnewline
\hline 
\hline 
I\textsubscript{$y$} &  & \textsuperscript{1}B\textsubscript{2} &  & 2.53 &  & 0.495 &  & $\arrowvert H\rightarrow L\rangle(0.68438)$\tabularnewline
II\textsubscript{$x$} &  & \textsuperscript{1}A\textsubscript{1} &  & 3.03 &  & 0.054 &  & $\arrowvert H\rightarrow L+1\rangle(0.63418)$\tabularnewline
III\textsubscript{$xy$} &  & \textsuperscript{1}B\textsubscript{2} &  & 3.68 &  & 0.173 &  & $\arrowvert H-1\rightarrow L+1\rangle(0.65621)$\tabularnewline
 &  & \textsuperscript{1}A\textsubscript{1} &  & 3.78 &  & 0.092 &  & $\arrowvert H\rightarrow L+2\rangle(0.63005)$\tabularnewline
IV\textsubscript{$y$} &  & \textsuperscript{1}B\textsubscript{2} &  & 4.10 &  & 0.026 &  & $\arrowvert H-2\rightarrow L+1\rangle(0.34330)$\tabularnewline
V\textsubscript{$y$} &  & \textsuperscript{1}B\textsubscript{2} &  & 4.51 &  & 0.183 &  & $\arrowvert H\rightarrow L+5\rangle(0.47527)$\tabularnewline
 &  & \textsuperscript{1}B\textsubscript{2} &  & 4.61 &  & 0.327 &  & $\arrowvert H-1\rightarrow L+2\rangle(0.45181)$\tabularnewline
VI\textsubscript{$y$} &  & \textsuperscript{1}B\textsubscript{2} &  & 4.99 &  & 0.689 &  & $\arrowvert H-2\rightarrow L+2\rangle(0.56689)$\tabularnewline
\hline 
\end{tabular}
\end{table}

\begin{table}[H]
\caption{Information about the excited states contributing to the peaks in
the linear optical absorption spectrum of model 3 of $C_{s}$ symmetry
{[}Fig. 7(c), main article{]}, computed using the first-principles
TDDFT method, and the B3LYP functional.}

\centering{}%
\begin{tabular}{ccccccccc}
\hline 
Peak & \hspace{1.7cm} & Symmetry & \hspace{1.7cm} & E (eV) & \hspace{1.7cm} & $f$ & \hspace{1.7cm} & Wave function\tabularnewline
\hline 
\hline 
I\textsubscript{$xy$} &  & \textsuperscript{1}A$'$ &  & 1.71 &  & 0.441 &  & $\arrowvert H\rightarrow L\rangle(0.71610)$\tabularnewline
II\textsubscript{$xy$} &  & \textsuperscript{1}A$'$ &  & 2.37 &  & 0.037 &  & $\arrowvert H-1\rightarrow L\rangle(0.53737)$\tabularnewline
III\textsubscript{$xy$} &  & \textsuperscript{1}A$'$ &  & 3.05 &  & 0.209 &  & $\arrowvert H-2\rightarrow L\rangle(0.53447)$\tabularnewline
IV\textsubscript{$xy$} &  & \textsuperscript{1}A$'$ &  & 3.65 &  & 0.064 &  & $\arrowvert H-3\rightarrow L\rangle(0.42588)$\tabularnewline
 &  & \textsuperscript{1}A$'$ &  & 3.69 &  & 0.061 &  & $\arrowvert H-1\rightarrow L+1\rangle(0.46739)$\tabularnewline
V\textsubscript{$xy$} &  & \textsuperscript{1}A$'$ &  & 3.89 &  & 0.198 &  & $\arrowvert H-1\rightarrow L+2\rangle(0.46482)$\tabularnewline
VI\textsubscript{$xy$} &  & \textsuperscript{1}A$'$ &  & 4.08 &  & 0.036 &  & $\arrowvert H-5\rightarrow L\rangle(0.46643)$\tabularnewline
VII\textsubscript{$xy$} &  & \textsuperscript{1}A$'$ &  & 4.60 &  & 0.189 &  & $\arrowvert H-1\rightarrow L+3\rangle(0.39933)$\tabularnewline
 &  & \textsuperscript{1}A$'$ &  & 4.67 &  & 0.283 &  & $\arrowvert H-1\rightarrow L+3\rangle(0.42775)$\tabularnewline
\hline 
\end{tabular}
\end{table}

\begin{table}[H]
\caption{Information about the excited states contributing to the peaks in
the linear optical absorption spectrum of model 4 of $C_{s}$ symmetry
{[}Fig. 7(d), main article{]}, computed using the first-principles
TDDFT method, and the B3LYP functional.}

\centering{}%
\begin{tabular}{ccccccccc}
\hline 
Peak & \hspace{1.7cm} & Symmetry & \hspace{1.7cm} & E (eV) & \hspace{1.7cm} & $f$ & \hspace{1.7cm} & Wave function\tabularnewline
\hline 
\hline 
I\textsubscript{$xy$} &  & \textsuperscript{1}A$'$ &  & 2.09 &  & 0.581 &  & $\arrowvert H\rightarrow L\rangle(0.70236)$\tabularnewline
II\textsubscript{$xy$} &  & \textsuperscript{1}A$'$ &  & 2.54 &  & 0.009 &  & $\arrowvert H-1\rightarrow L\rangle(0.68658)$\tabularnewline
III\textsubscript{$xy$} &  & \textsuperscript{1}A\textsubscript{}$'$ &  & 2.84 &  & 0.026 &  & $\arrowvert H\rightarrow L+1\rangle(0.66242)$\tabularnewline
IV\textsubscript{$xy$} &  & \textsuperscript{1}A$'$ &  & 3.30 &  & 0.026 &  & $\arrowvert H\rightarrow L+3\rangle(0.50235)$\tabularnewline
V\textsubscript{$xy$} &  & \textsuperscript{1}A\textsubscript{}$'$ &  & 3.56 &  & 0.143 &  & $\arrowvert H-2\rightarrow L\rangle(0.44978)$\tabularnewline
VI\textsubscript{$xy$} &  & \textsuperscript{1}A\textsubscript{}$'$ &  & 4.07 &  & 0.085 &  & $\arrowvert H-4\rightarrow L\rangle(0.56338)$\tabularnewline
VII\textsubscript{$xy$} &  & \textsuperscript{1}A$'$ &  & 4.28 &  & 0.220 &  & $\arrowvert H-1\rightarrow L+2\rangle(0.47600)$\tabularnewline
VIII\textsubscript{$xy$} &  & \textsuperscript{1}A$'$ &  & 4.85 &  & 0.226 &  & $\arrowvert H-3\rightarrow L+1\rangle(0.50758)$\tabularnewline
\hline 
\end{tabular}
\end{table}

\subsubsection{Doping with two fused BN-rings}

\begin{table}[H]
\caption{Information about the excited states contributing to the peaks in
the linear optical absorption spectrum of model 5 of $C_{s}$ symmetry
{[}Fig. 8(a), main article{]}, computed using the first-principles
TDDFT method, and the B3LYP functional.}

\centering{}%
\begin{tabular}{ccccccccc}
\hline 
Peak & \hspace{1.7cm} & Symmetry & \hspace{1.7cm} & E (eV) & \hspace{1.7cm} & $f$ & \hspace{1.7cm} & Wave function\tabularnewline
\hline 
\hline 
I\textsubscript{$xy$} &  & \textsuperscript{1}A$'$ &  & 1.77 &  & 0.436 &  & $\arrowvert H\rightarrow L\rangle(0.69855)$\tabularnewline
II\textsubscript{$xy$} &  & \textsuperscript{1}A$'$ &  & 2.38 &  & 0.114 &  & $\arrowvert H\rightarrow L+1\rangle(0.50494)$\tabularnewline
III\textsubscript{$xy$} &  & \textsuperscript{1}A\textsubscript{}$'$ &  & 2.66 &  & 0.128 &  & $\arrowvert H-1\rightarrow L\rangle(0.48027)$\tabularnewline
IV\textsubscript{$xy$} &  & \textsuperscript{1}A\textsubscript{}$'$ &  & 2.89 &  & 0.009 &  & $\arrowvert H\rightarrow L+2\rangle(0.66386)$\tabularnewline
V\textsubscript{$xy$} &  & \textsuperscript{1}A$'$ &  & 3.39 &  & 0.216 &  & $\arrowvert H-2\rightarrow L\rangle(0.57870)$\tabularnewline
VI\textsubscript{$xy$} &  & \textsuperscript{1}A\textsubscript{}$'$ &  & 3.82 &  & 0.122 &  & $\arrowvert H-3\rightarrow L\rangle(0.41818)$\tabularnewline
VII\textsubscript{$xy$} &  & \textsuperscript{1}A$'$ &  & 4.27 &  & 0.045 &  & $\arrowvert H-5\rightarrow L\rangle(0.34716)$\tabularnewline
VIII\textsubscript{$xy$} &  & \textsuperscript{1}A\textsubscript{}$'$ &  & 4.71 &  & 0.160 &  & $\arrowvert H-6\rightarrow L\rangle(0.44332)$\tabularnewline
 &  & \textsuperscript{1}A$'$ &  & 4.76 &  & 0.157 &  & $\arrowvert H-1\rightarrow L+4\rangle(0.43532)$\tabularnewline
\hline 
\end{tabular}
\end{table}

\begin{table}[H]
\caption{Information about the excited states contributing to the peaks in
the linear optical absorption spectrum of model 6 of $C_{s}$ symmetry
{[}Fig. 8(b), main article{]}, computed using the first-principles
TDDFT method, and the B3LYP functional.}

\centering{}%
\begin{tabular}{ccccccccc}
\hline 
Peak & \hspace{1.7cm} & Symmetry & \hspace{1.7cm} & E (eV) & \hspace{1.7cm} & $f$ & \hspace{1.7cm} & Wave function\tabularnewline
\hline 
\hline 
I\textsubscript{$xy$} &  & \textsuperscript{1}A$'$ &  & 2.44 &  & 0.416 &  & $\arrowvert H\rightarrow L\rangle(0.66788)$\tabularnewline
II\textsubscript{$xy$} &  & \textsuperscript{1}A$'$ &  & 3.03 &  & 0.039 &  & $\arrowvert H\rightarrow L+1\rangle(0.68229)$\tabularnewline
III\textsubscript{$xy$} &  & \textsuperscript{1}A$'$ &  & 3.50 &  & 0.024 &  & $\arrowvert H\rightarrow L+2\rangle(0.47925)$\tabularnewline
IV\textsubscript{$xy$} &  & \textsuperscript{1}A$'$ &  & 3.78 &  & 0.571 &  & $\arrowvert H-1\rightarrow L+1\rangle(0.56118)$\tabularnewline
V\textsubscript{$xy$} &  & \textsuperscript{1}A$'$ &  & 4.07 &  & 0.230 &  & $\arrowvert H-1\rightarrow L+2\rangle(0.58977)$\tabularnewline
VI\textsubscript{$xy$} &  & \textsuperscript{1}A$'$ &  & 4.67 &  & 0.134 &  & $\arrowvert H-1\rightarrow L+5\rangle(0.58280)$\tabularnewline
\hline 
\end{tabular}
\end{table}

\begin{table}[H]
\caption{Information about the excited states contributing to the peaks in
the linear optical absorption spectrum of model 7 of $C_{2v}$ symmetry
{[}Fig. 8(c), main article{]}, computed using the first-principles
TDDFT method, and the B3LYP functional.}

\centering{}%
\begin{tabular}{ccccccccc}
\hline 
Peak & \hspace{1.7cm} & Symmetry & \hspace{1.7cm} & E (eV) & \hspace{1.7cm} & $f$ & \hspace{1.7cm} & Wave function\tabularnewline
\hline 
\hline 
I\textsubscript{$y$} &  & \textsuperscript{1}B\textsubscript{2} &  & 1.17 &  & 0.260 &  & $\arrowvert H\rightarrow L\rangle(0.75936)$\tabularnewline
II\textsubscript{$x$} &  & \textsuperscript{1}A\textsubscript{1} &  & 1.99 &  & 0.014 &  & $\arrowvert H-1\rightarrow L\rangle(0.69825)$\tabularnewline
III\textsubscript{$xy$} &  & \textsuperscript{1}A\textsubscript{1} &  & 2.50 &  & 0.018 &  & $\arrowvert H\rightarrow L+1\rangle(0.68531)$\tabularnewline
 &  & \textsuperscript{1}B\textsubscript{2} &  & 2.54 &  & 0.072 &  & $\arrowvert H-2\rightarrow L\rangle(0.65599)$\tabularnewline
IV\textsubscript{$y$} &  & \textsuperscript{1}B\textsubscript{2} &  & 2.85 &  & 0.124 &  & $\arrowvert H\rightarrow L+2\rangle(0.64371)$\tabularnewline
V\textsubscript{$x$} &  & \textsuperscript{1}A\textsubscript{1} &  & 3.18 &  & 0.103 &  & $\arrowvert H\rightarrow L+5\rangle(0.60398)$\tabularnewline
VI\textsubscript{$x$} &  & \textsuperscript{1}A\textsubscript{1} &  & 3.66 &  & 0.019 &  & $\arrowvert H\rightarrow L+9\rangle(0.66119)$\tabularnewline
VII\textsubscript{$y$} &  & \textsuperscript{1}B\textsubscript{2} &  & 4.32 &  & 0.308 &  & $\arrowvert H-1\rightarrow L+3\rangle(0.53077)$\tabularnewline
VIII\textsubscript{$x$} &  & \textsuperscript{1}A\textsubscript{1} &  & 4.55 &  & 0.138 &  & $\arrowvert H-2\rightarrow L+1\rangle(0.51314)$\tabularnewline
\hline 
\end{tabular}
\end{table}

\begin{table}[H]
\caption{Information about the excited states contributing to the peaks in
the linear optical absorption spectrum of model 8 of $C_{s}$ symmetry
{[}Fig. 8(d), main article{]}, computed using the first-principles
TDDFT method, and the B3LYP functional.}

\centering{}%
\begin{tabular}{ccccccccc}
\hline 
Peak & \hspace{1.7cm} & Symmetry & \hspace{1.7cm} & E (eV) & \hspace{1.7cm} & $f$ & \hspace{1.7cm} & Wave function\tabularnewline
\hline 
\hline 
I\textsubscript{$xy$} &  & \textsuperscript{1}A$'$ &  & 1.23 &  & 0.170 &  & $\arrowvert H\rightarrow L\rangle(0.71559)$\tabularnewline
II\textsubscript{$xy$} &  & \textsuperscript{1}A$'$ &  & 1.79 &  & 0.093 &  & $\arrowvert H-1\rightarrow L\rangle(0.62914)$\tabularnewline
III\textsubscript{$xy$} &  & \textsuperscript{1}A$'$ &  & 2.93 &  & 0.513 &  & $\arrowvert H-1\rightarrow L+1\rangle(0.66382)$\tabularnewline
IV\textsubscript{$xy$} &  & \textsuperscript{1}A\textsubscript{}$'$ &  & 3.36 &  & 0.032 &  & $\arrowvert H\rightarrow L+5\rangle(0.68594)$\tabularnewline
V\textsubscript{$xy$} &  & \textsuperscript{1}A\textsubscript{}$'$ &  & 3.56 &  & 0.032 &  & $\arrowvert H-3\rightarrow L\rangle(0.47538)$\tabularnewline
 &  & \textsuperscript{1}A$'$ &  & 3.62 &  & 0.033 &  & $\arrowvert H\rightarrow L+8\rangle(0.50498)$\tabularnewline
VI\textsubscript{$xy$} &  & \textsuperscript{1}A$'$ &  & 3.79 &  & 0.119 &  & $\arrowvert H\rightarrow L+9\rangle(0.44030)$\tabularnewline
 &  & \textsuperscript{1}A$'$ &  & 3.81 &  & 0.054 &  & $\arrowvert H-1\rightarrow L+2\rangle(0.55266)$\tabularnewline
VII\textsubscript{$xy$} &  & \textsuperscript{1}A$'$ &  & 4.11 &  & 0.056 &  & $\arrowvert H-1\rightarrow L+2\rangle(0.55266)$\tabularnewline
VIII\textsubscript{$xy$} &  & \textsuperscript{1}A$'$ &  & 4.35 &  & 0.149 &  & $\arrowvert H\rightarrow L+4\rangle(0.59709)$\tabularnewline
\hline 
\end{tabular}
\end{table}

\subsubsection{Doping with two separated BN-rings}

\begin{table}[H]
\caption{Information about the excited states contributing to the peaks in
the linear optical absorption spectrum of model 9 ($\uparrow-\uparrow$)
of $C_{2v}$ symmetry {[}Fig. 9(a), main article{]}, computed using
the first-principles TDDFT method, and the B3LYP functional.}

\centering{}%
\begin{tabular}{ccccccccc}
\hline 
Peak & \hspace{1.7cm} & Symmetry & \hspace{1.7cm} & E (eV) & \hspace{1.7cm} & $f$ & \hspace{1.7cm} & Wave function\tabularnewline
\hline 
\hline 
I\textsubscript{$y$} &  & \textsuperscript{1}B\textsubscript{2} &  & 2.75 &  & 0.393 &  & $\arrowvert H\rightarrow L\rangle(0.67093)$\tabularnewline
II\textsubscript{x} &  & \textsuperscript{1}A\textsubscript{1} &  & 3.62 &  & 0.168 &  & $\arrowvert H-2\rightarrow L\rangle(0.53655)$\tabularnewline
III\textsubscript{$y$} &  & \textsuperscript{1}B\textsubscript{2} &  & 3.98 &  & 0.247 &  & $\arrowvert H-1\rightarrow L+2\rangle(0.45296)$\tabularnewline
 &  & \textsuperscript{1}B\textsubscript{2} &  & 4.03 &  & 0.265 &  & $\arrowvert H-1\rightarrow L+1\rangle(0.44186)$\tabularnewline
IV\textsubscript{$x$} &  & \textsuperscript{1}A\textsubscript{1} &  & 4.51 &  & 0.009 &  & $\arrowvert H-1\rightarrow L+3\rangle(0.65830)$\tabularnewline
V\textsubscript{$y$} &  & \textsuperscript{1}B\textsubscript{2} &  & 4.80 &  & 0.486 &  & $\arrowvert H-2\rightarrow L+2\rangle(0.65482)$\tabularnewline
\hline 
\end{tabular}
\end{table}

\begin{table}[H]
\caption{Information about the excited states contributing to the peaks in
the linear optical absorption spectrum of model 9 ($\uparrow-\downarrow$)
of $D_{2h}$ symmetry {[}Fig. 9(d), main article{]}, computed using
the first-principles TDDFT method, and the B3LYP functional.}

\centering{}%
\begin{tabular}{ccccccccc}
\hline 
Peak & \hspace{1.7cm} & Symmetry & \hspace{1.7cm} & E (eV) & \hspace{1.7cm} & $f$ & \hspace{1.7cm} & Wave function\tabularnewline
\hline 
\hline 
I\textsubscript{$y$} &  & \textsuperscript{1}B\textsubscript{2u} &  & 2.79 &  & 0.399 &  & $\arrowvert H\rightarrow L\rangle(0.67381)$\tabularnewline
II\textsubscript{$y$} &  & \textsuperscript{1}B\textsubscript{2u} &  & 4.06 &  & 0.650 &  & $\arrowvert H-2\rightarrow L+1\rangle(0.62120)$\tabularnewline
III\textsubscript{$y$} &  & \textsuperscript{1}B\textsubscript{2u} &  & 4.32 &  & 0.069 &  & $\arrowvert H\rightarrow L+3\rangle(0.64725)$\tabularnewline
IV\textsubscript{$x$} &  & \textsuperscript{1}B\textsubscript{3u} &  & 4.72 &  & 0.010 &  & $\arrowvert H-7\rightarrow L\rangle(0.65575)$\tabularnewline
V\textsubscript{$x$} &  & \textsuperscript{1}B\textsubscript{3u} &  & 4.95 &  & 0.067 &  & $\arrowvert H\rightarrow L+5\rangle(0.57600)$\tabularnewline
VI\textsubscript{$x$} &  & \textsuperscript{1}B\textsubscript{3u} &  & 5.18 &  & 0.056 &  & $\arrowvert H-4\rightarrow L+1\rangle(0.64487)$\tabularnewline
\hline 
\end{tabular}
\end{table}

\begin{table}[H]
\caption{Information about the excited states contributing to the peaks in
the linear optical absorption spectrum of model 10 ($\uparrow-\uparrow$)
of $C_{2v}$ symmetry {[}Fig. 9(b), main article{]}, computed using
the first-principles TDDFT method, and the B3LYP functional.}

\centering{}%
\begin{tabular}{ccccccccc}
\hline 
Peak & \hspace{1.7cm} & Symmetry & \hspace{1.7cm} & E (eV) & \hspace{1.7cm} & $f$ & \hspace{1.7cm} & Wave function\tabularnewline
\hline 
\hline 
I\textsubscript{$y$} &  & \textsuperscript{1}B\textsubscript{2} &  & 1.27 &  & 0.348 &  & $\arrowvert H\rightarrow L\rangle(0.76497)$\tabularnewline
II\textsubscript{$x$} &  & \textsuperscript{1}A\textsubscript{1} &  & 1.66 &  & 0.010 &  & $\arrowvert H\rightarrow L+1\rangle(0.50669)$\tabularnewline
III\textsubscript{$x$} &  & \textsuperscript{1}A\textsubscript{1} &  & 2.38 &  & 0.006 &  & $\arrowvert H\rightarrow L+2\rangle(0.55592)$\tabularnewline
IV\textsubscript{$x$} &  & \textsuperscript{1}A\textsubscript{1} &  & 2.86 &  & 0.260 &  & $\arrowvert H-2\rightarrow L\rangle(0.54839)$\tabularnewline
V\textsubscript{$y$} &  & \textsuperscript{1}B\textsubscript{2} &  & 3.69 &  & 0.145 &  & $\arrowvert H-1\rightarrow L+2\rangle(0.45314)$\tabularnewline
VI\textsubscript{$x$} &  & \textsuperscript{1}A\textsubscript{1} &  & 4.27 &  & 0.019 &  & $\arrowvert H\rightarrow L+4\rangle(0.52350)$\tabularnewline
VII\textsubscript{$x$} &  & \textsuperscript{1}A\textsubscript{1} &  & 4.52 &  & 0.052 &  & $\arrowvert H-1\rightarrow L+4\rangle(0.68022)$\tabularnewline
\hline 
\end{tabular}
\end{table}

\begin{table}[H]
\caption{Information about the excited states contributing to the peaks in
the linear optical absorption spectrum of model 10 ($\uparrow-\downarrow$)
of $C_{2h}$ symmetry {[}Fig. 9(e), main article{]}, computed using
the first- principles TDDFT method, and the B3LYP functional.}

\centering{}%
\begin{tabular}{ccccccccc}
\hline 
Peak & \hspace{1.7cm} & Symmetry & \hspace{1.7cm} & E (eV) & \hspace{1.7cm} & $f$ & \hspace{1.7cm} & Wave function\tabularnewline
\hline 
\hline 
I\textsubscript{$xy$} &  & \textsuperscript{1}B\textsubscript{u} &  & 1.46 &  & 0.433 &  & $\arrowvert H\rightarrow L\rangle(0.74266)$\tabularnewline
II\textsubscript{$xy$} &  & \textsuperscript{1}B\textsubscript{u} &  & 2.28 &  & 0.041 &  & $\arrowvert H\rightarrow L+2\rangle(0.54892)$\tabularnewline
III\textsubscript{$xy$} &  & \textsuperscript{1}B\textsubscript{u} &  & 2.71 &  & 0.192 &  & $\arrowvert H-2\rightarrow L\rangle(0.53178)$\tabularnewline
IV\textsubscript{$xy$} &  & \textsuperscript{1}B\textsubscript{u} &  & 3.49 &  & 0.287 &  & $\arrowvert H-1\rightarrow L+1\rangle(0.67251)$\tabularnewline
V\textsubscript{$xy$} &  & \textsuperscript{1}B\textsubscript{u} &  & 3.73 &  & 0.101 &  & $\arrowvert H-3\rightarrow L\rangle(0.66967)$\tabularnewline
VI\textsubscript{$xy$} &  & \textsuperscript{1}B\textsubscript{u} &  & 4.22 &  & 0.081 &  & $\arrowvert H-2\rightarrow L+2\rangle(0.53063)$\tabularnewline
VII\textsubscript{$xy$} &  & \textsuperscript{1}B\textsubscript{u} &  & 4.49 &  & 0.095 &  & $\arrowvert H-6\rightarrow L\rangle(0.53501)$\tabularnewline
\hline 
\end{tabular}
\end{table}

\begin{table}[H]
\caption{Information about the excited states contributing to the peaks in
the linear optical absorption spectrum of model 11 ($\uparrow-\uparrow$)
of $C_{s}$ symmetry {[}Fig. 9(c), main article{]}, computed using
the first-principles TDDFT method, and the B3LYP functional.}

\centering{}%
\begin{tabular}{ccccccccc}
\hline 
Peak & \hspace{1.7cm} & Symmetry & \hspace{1.7cm} & E (eV) & \hspace{1.7cm} & $f$ & \hspace{1.7cm} & Wave function\tabularnewline
\hline 
\hline 
I\textsubscript{$xy$} &  & \textsuperscript{1}A$'$ &  & 1.59 &  & 0.354 &  & $\arrowvert H\rightarrow L\rangle(0.71146)$\tabularnewline
II\textsubscript{$xy$} &  & \textsuperscript{1}A$'$ &  & 2.68 &  & 0.102 &  & $\arrowvert H-1\rightarrow L\rangle(0.57517)$\tabularnewline
III\textsubscript{$xy$} &  & \textsuperscript{1}A$'$ &  & 3.01 &  & 0.029 &  & $\arrowvert H-2\rightarrow L\rangle(0.46478)$\tabularnewline
IV\textsubscript{$xy$} &  & \textsuperscript{1}A$'$ &  & 3.33 &  & 0.412 &  & $\arrowvert H-3\rightarrow L\rangle(0.41918)$\tabularnewline
V\textsubscript{$xy$} &  & \textsuperscript{1}A$'$ &  & 3.63 &  & 0.053 &  & $\arrowvert H\rightarrow L+5\rangle(0.52268)$\tabularnewline
VI\textsubscript{$xy$} &  & \textsuperscript{1}A$'$ &  & 3.88 &  & 0.065 &  & $\arrowvert H-4\rightarrow L\rangle(0.65638)$\tabularnewline
VII\textsubscript{$xy$} &  & \textsuperscript{1}A$'$ &  & 4.29 &  & 0.442 &  & $\arrowvert H-1\rightarrow L+1\rangle(0.59132)$\tabularnewline
VIII\textsubscript{$xy$} &  & \textsuperscript{1}A$'$ &  & 4.75 &  & 0.046 &  & $\arrowvert H-2\rightarrow L+2\rangle(0.63143)$\tabularnewline
\hline 
\end{tabular}
\end{table}

\begin{table}[H]
\caption{Information about the excited states contributing to the peaks in
the linear optical absorption spectrum of model 11 ($\uparrow-\downarrow$)
of $C_{2h}$ symmetry {[}Fig. 9(f), main article{]}, computed using
the first-principles TDDFT method, and the B3LYP functional.}

\centering{}%
\begin{tabular}{ccccccccc}
\hline 
Peak & \hspace{1.7cm} & Symmetry & \hspace{1.7cm} & E (eV) & \hspace{1.7cm} & $f$ & \hspace{1.7cm} & Wave function\tabularnewline
\hline 
\hline 
I\textsubscript{$xy$} &  & \textsuperscript{1}B\textsubscript{u} &  & 1.73 &  & 0.556 &  & $\arrowvert H\rightarrow L\rangle(0.73133)$\tabularnewline
II\textsubscript{$xy$} &  & \textsuperscript{1}B\textsubscript{u} &  & 2.63 &  & 0.021 &  & $\arrowvert H-2\rightarrow L\rangle(0.59122)$\tabularnewline
III\textsubscript{$xy$} &  & \textsuperscript{1}B\textsubscript{u} &  & 2.92 &  & 0.053 &  & $\arrowvert H\rightarrow L+2\rangle(0.58383)$\tabularnewline
IV\textsubscript{$xy$} &  & \textsuperscript{1}B\textsubscript{u} &  & 3.42 &  & 0.303 &  & $\arrowvert H\rightarrow L+3\rangle(0.56829)$\tabularnewline
V\textsubscript{$xy$} &  & \textsuperscript{1}B\textsubscript{u} &  & 4.11 &  & 0.434 &  & $\arrowvert H-1\rightarrow L+1\rangle(0.51329)$\tabularnewline
VI\textsubscript{$xy$} &  & \textsuperscript{1}B\textsubscript{u} &  & 4.61 &  & 0.353 &  & $\arrowvert H-3\rightarrow L+1\rangle(0.59960)$\tabularnewline
\hline 
\end{tabular}
\end{table}

\subsection{Orbital Projected Density of States}
\begin{figure}[H]
\begin{centering}
\includegraphics[scale=1.39]{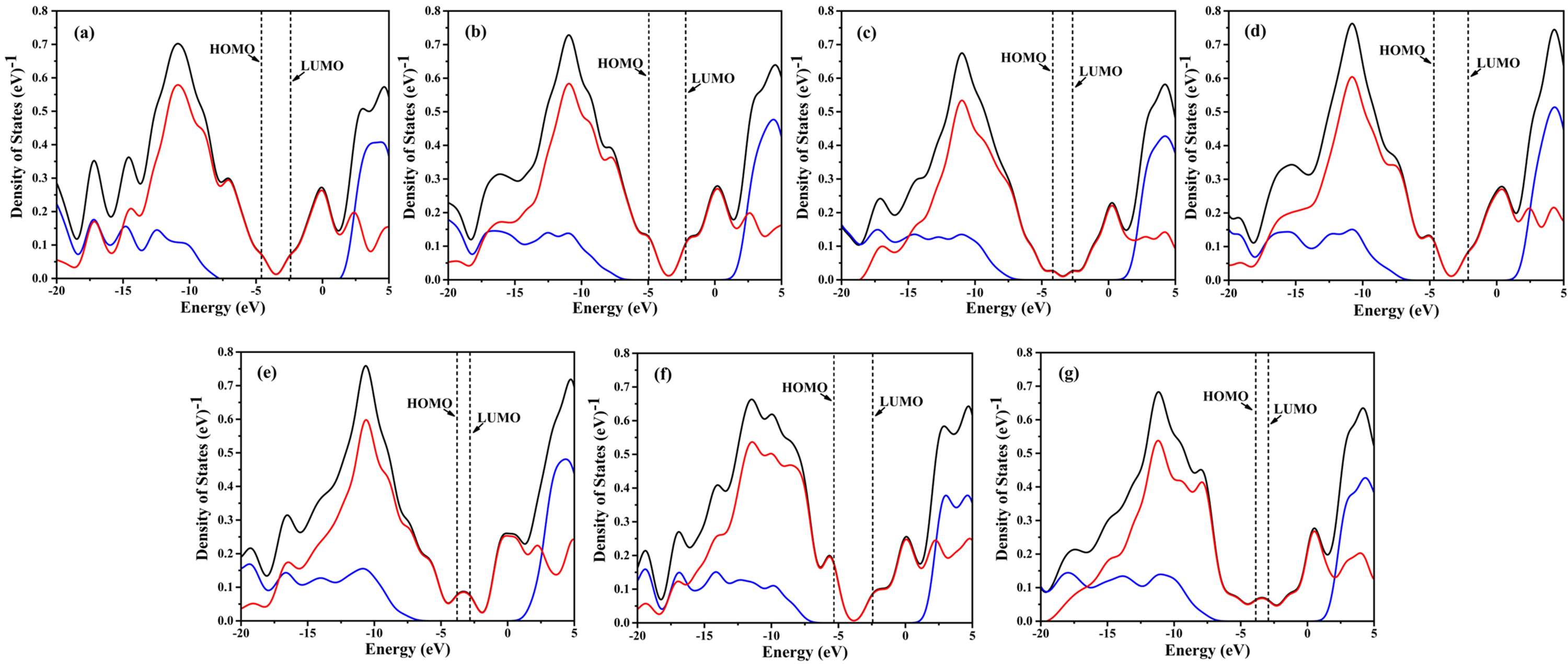}
\par\end{centering}
\caption{Total and orbital projected density of states (DOS) plots for: pristine
GQD {[}(a){]}, single BN-ring doping {[}(b) model 1, (c) model 3{]},
fused double BN-ring doping {[}(d) model 6, (e) model 7{]}, separated
double BN-ring doping {[}(f) Model 9 ($\uparrow-\downarrow$), (g)
model 10 ($\uparrow-\uparrow$){]}. Here black lines denote the total
density of states (TDOS), while red and blue lines represent the partial
DOS corresponding to the contributions of the $p$ and $s$ orbitals,
respectively. \protect\label{fig:orb-pdos}}
\end{figure}

\begin{quotation}
In Fig. \ref{fig:orb-pdos} we present the plots of both total and
the orbital-projected density of states (DOS) for the pristine GQD
as well as a few BN-GQDs. We note that for all the structures, the
$p$-orbitals, but more specifically $\pi/\pi^{*}$ orbitals, dominate
the total DOS near the Fermi level, consistent with their planar geometries. 
\end{quotation}

\subsection{Comparison between absorption spectra computed using different functionals}
\begin{center}
\begin{figure}[H]
\begin{centering}
\includegraphics[scale=1.29]{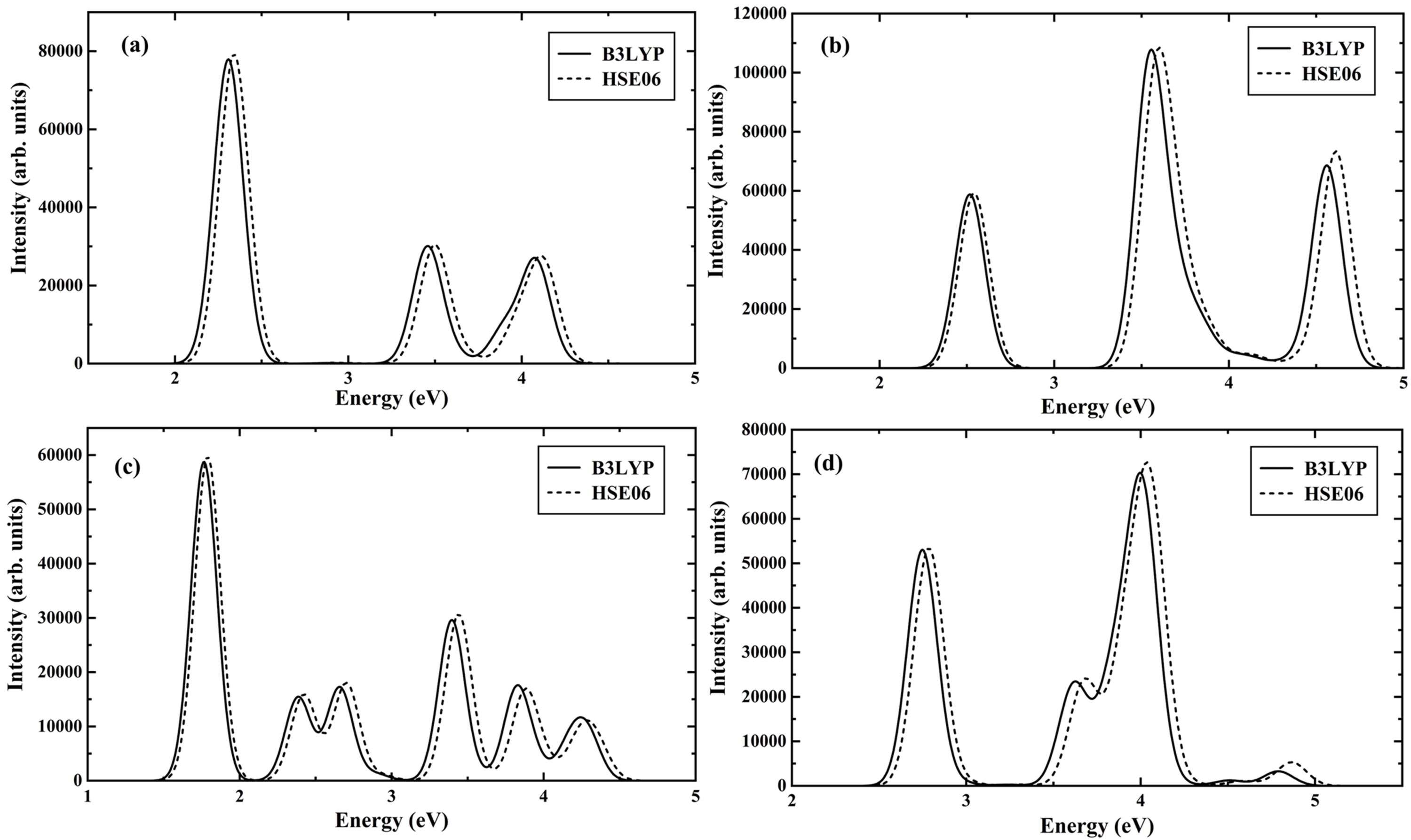}
\par\end{centering}
\caption{Comparison of TDDFT UV-Vis absorption spectra obtained with B3LYP
and HSE06 functionals with 20 excited states: (a) pristine GQD, (b)
single BN-ring doping (model 1), (c) fused double BN-ring doping (model
5), (d) separated double BN-ring doping (model 9 ($\uparrow-\uparrow$)).\protect\label{fig:Comparison-of-TDDFTfunctional}}
\end{figure}
\par\end{center}

\subsection{Absorption spectra calculated with different numbers of excited states}

UV-Vis absorption spectra were calculated with different numbers of
excited states to check convergence using the TDDFT-B3LYP level of
theory.
\begin{center}
\begin{figure}[H]
\begin{centering}
\includegraphics[scale=1.32]{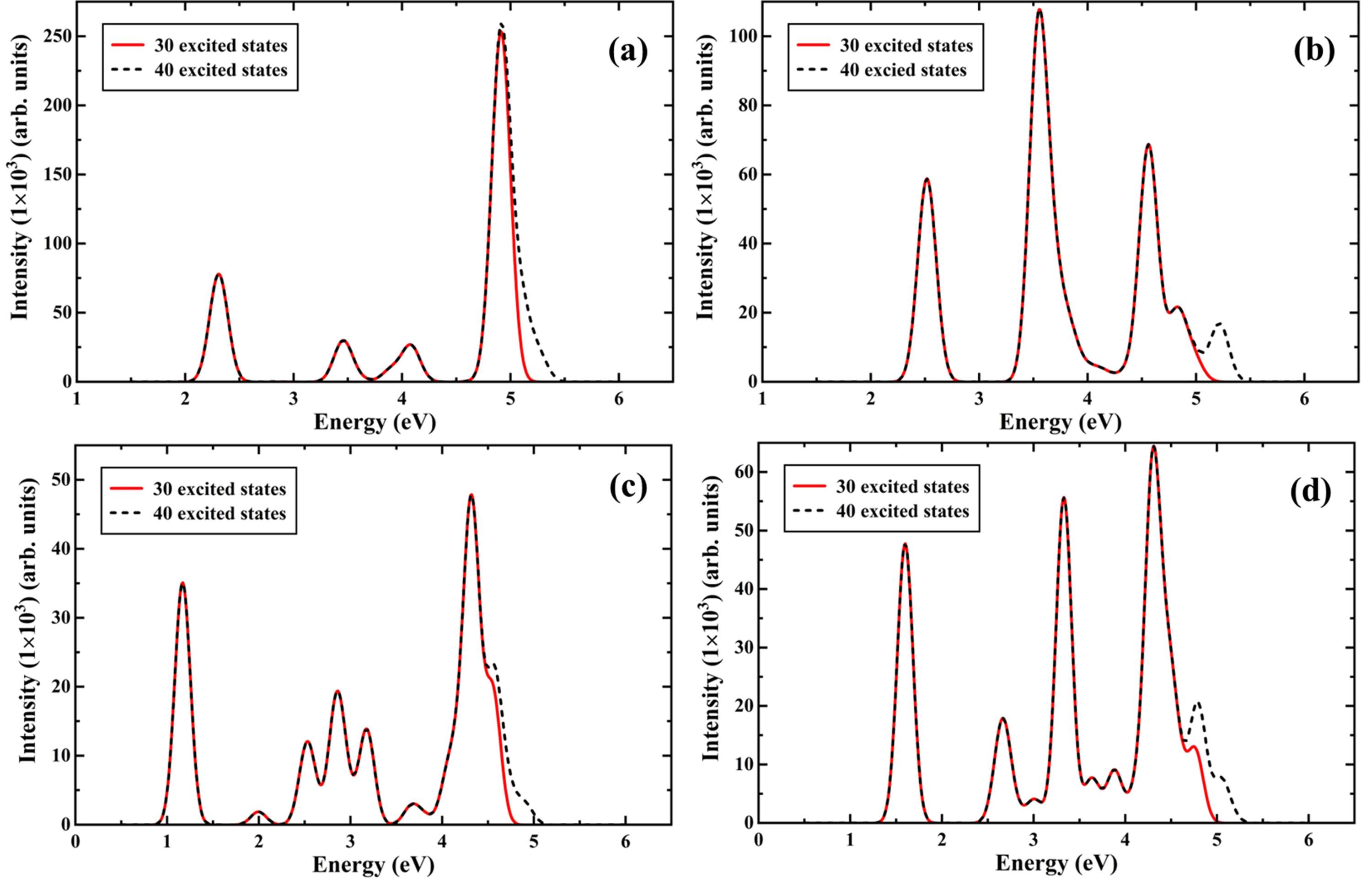}
\par\end{centering}
\caption{(a) pristine GQD; (b) single BN-ring doping (model 1); (c) fused double
BN-ring doping (model 7); (d) separated double BN-ring doping (model
11 ($\uparrow-\uparrow$)).\protect\label{fig:(a)-pristine-GQDdifferentstates=000020}}
\end{figure}
\par\end{center}

\subsection{Comparison of UV-Vis absorption spectra (TDDFT-B3LYP) for doped and
pristine GQDs}

\begin{figure}[H]
\centering{}\includegraphics[scale=0.94]{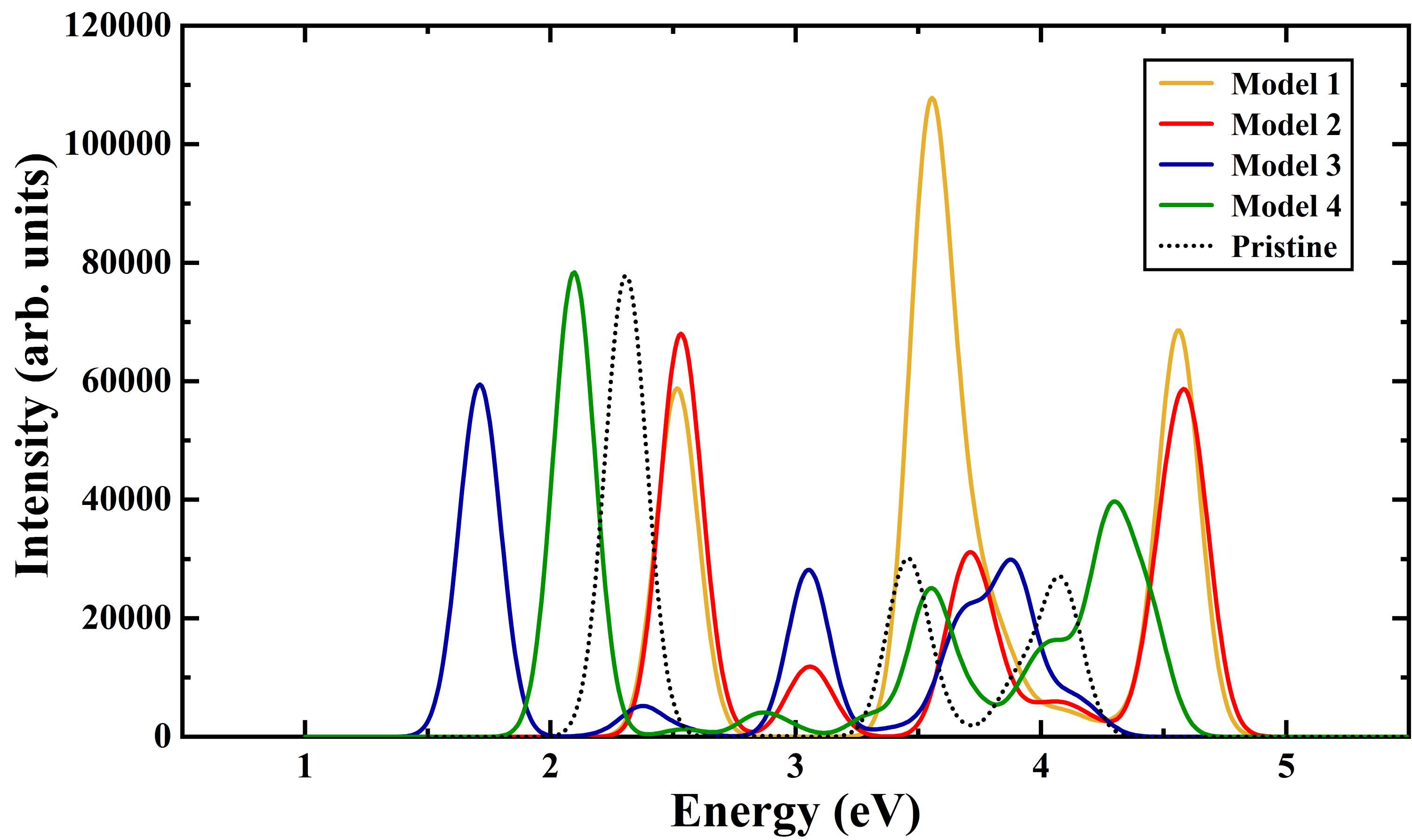}\caption{UV-Vis absorption spectra of single BN-ring doped BN-GQDs compared
to that of the pristine GQD, computed at the TDDFT-B3LYP level of
theory.\protect\label{fig:Variation-of-UV-Vis(1-4)}}
\end{figure}

\begin{center}
\begin{figure}[H]
\begin{centering}
\includegraphics[scale=0.95]{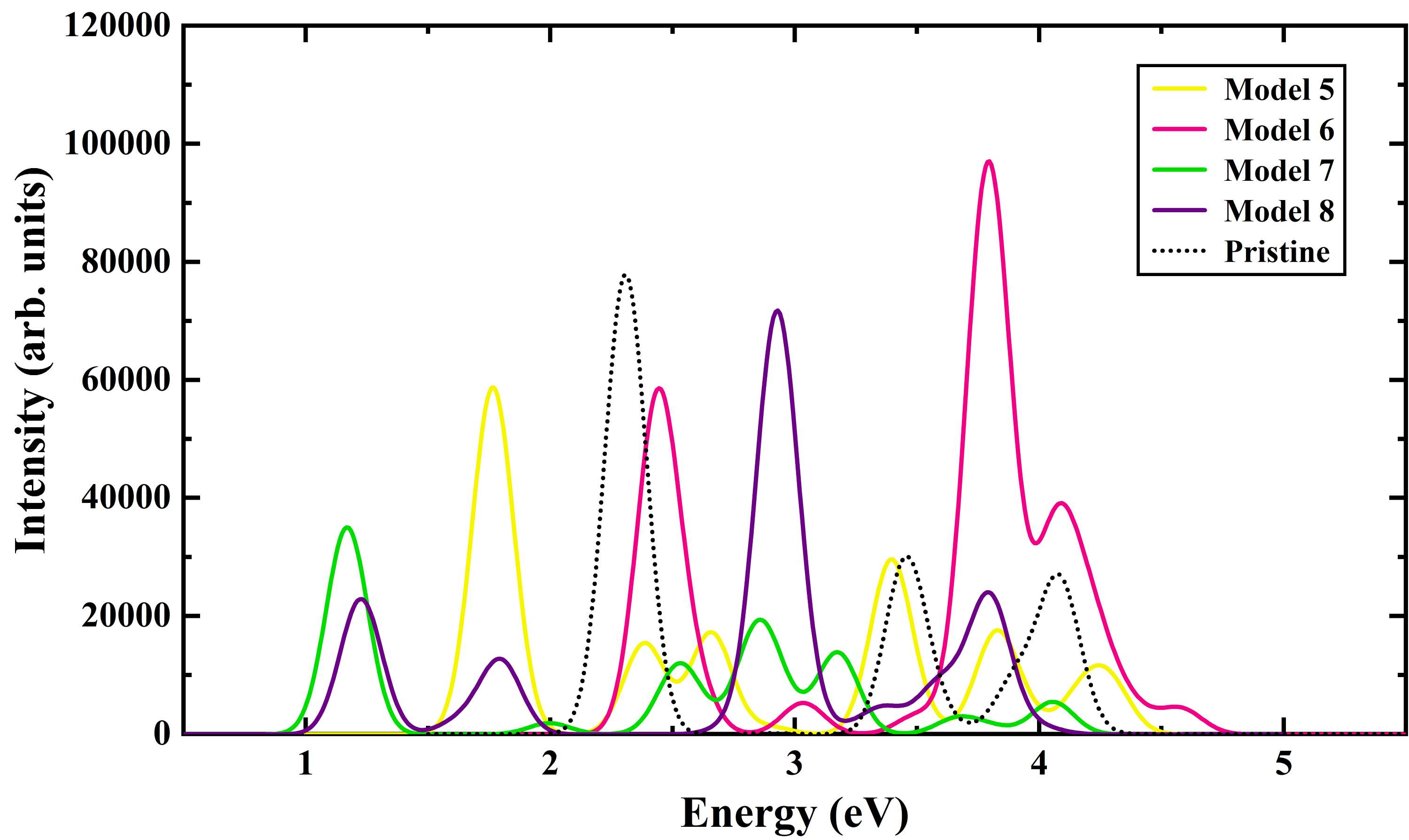}
\par\end{centering}
\caption{UV-Vis absorption spectra of BN-GQDs doped by two fused BN-rings,
compared to that of the pristine GQD, computed at the TDDFT-B3LYP
level of theory.\protect\label{fig:Variation-of-UV-Vis(5-8)}}
\end{figure}
\par\end{center}

\begin{figure}[H]
\begin{centering}
\includegraphics{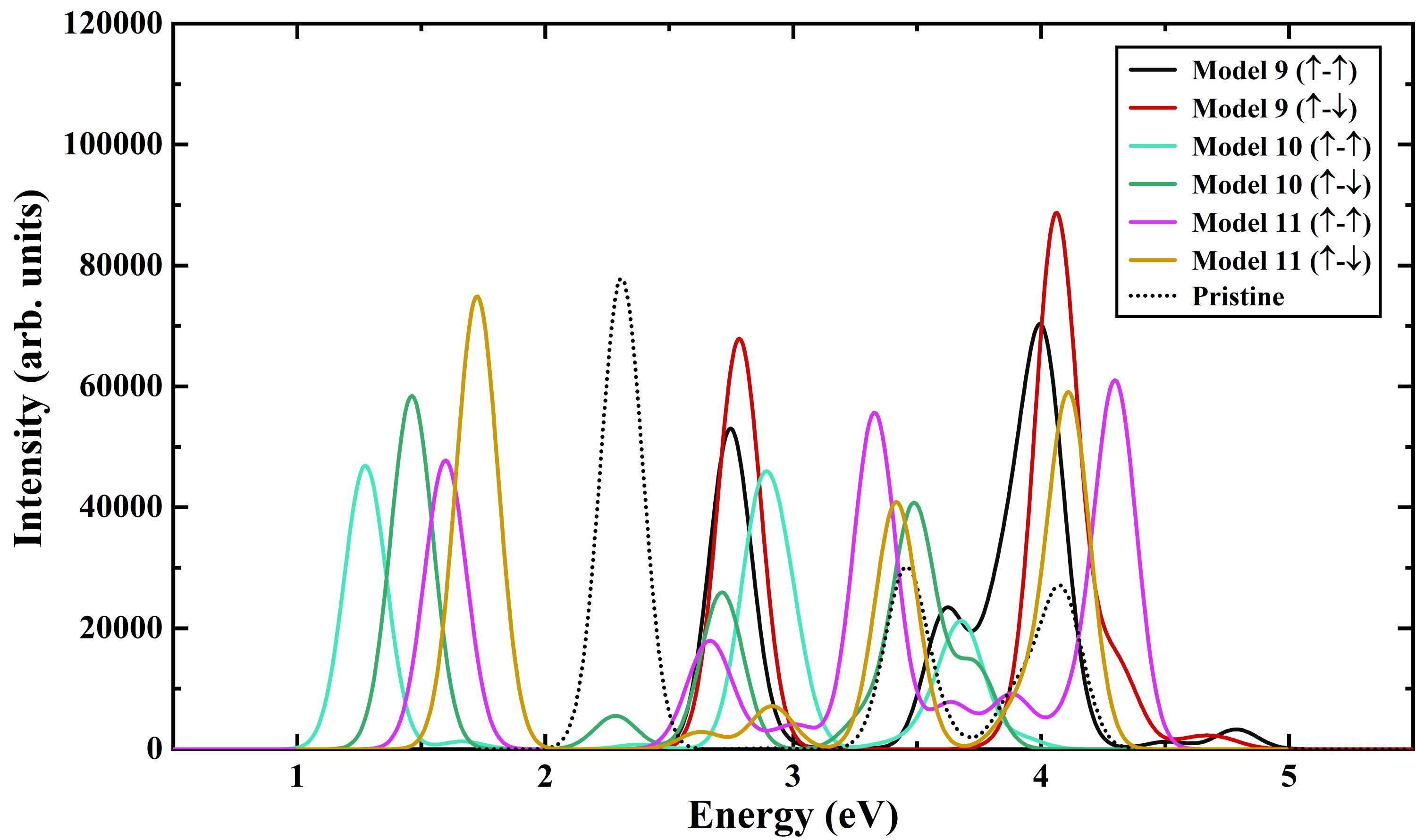}
\par\end{centering}
\caption{UV-Vis absorption spectra of BN-GQDs doped by two separated BN-rings,
compared to that of the pristine GQD, computed at the TDDFT-B3LYP
level of theory.\protect\label{fig:Variation-of-UV-Vis(9-11)}}
\end{figure}